\documentclass{aa}  

\usepackage{graphicx}
\usepackage{txfonts}
\usepackage{lscape}
\usepackage{booktabs}
\usepackage{multirow}
\usepackage[breaklinks=true]{hyperref}
\hypersetup{colorlinks=true, linkcolor=blue, citecolor=blue, filecolor=black, urlcolor=blue}

\begin{document} 

\title{The CARMENES search for exoplanets around M dwarfs}
\subtitle{Stellar atmospheric parameters of target stars with \textsc{SteParSyn}}
\author{E.~Marfil\inst{1,2},
        H.\,M.~Tabernero\inst{3,4},
        D.~Montes\inst{1},
        J.\,A.~Caballero\inst{2},
        F.\,J.~L\'{a}zaro\inst{1},
        J.\,I.~Gonz\'{a}lez~Hern\'{a}ndez\inst{5,6},
        E.~Nagel\inst{7,8},
        V.\,M.~Passegger\inst{7,9},
        A.~Schweitzer\inst{7},
        I.~Ribas\inst{10, 11},
        A.~Reiners\inst{12},
        A.~Quirrenbach\inst{13},
        P.\,J.~Amado\inst{14},
        C.~Cifuentes\inst{2},
        M.~Cort\'{e}s-Contreras\inst{2},
        S.~Dreizler\inst{12},
        C.~Duque-Arribas\inst{1},
        D.~Galadí-Enríquez\inst{15},
        Th.~Henning\inst{16},
        S.\,V.~Jeffers\inst{17,12},
        A.~Kaminski\inst{13},
        M.~K\"{u}rster\inst{16},
        M.~Lafarga\inst{10,11,18},
        \'{A}.~L\'{o}pez-Gallifa\inst{1},
        J.\,C.~Morales\inst{10,11},
        Y. Shan\inst{12},
        and
        M.~Zechmeister\inst{12}}
\titlerunning{Stellar atmospheric parameters of target stars with \textsc{SteParSyn}}
\authorrunning{E.~Marfil et al.}
\institute{Departamento de F{\'i}sica de la Tierra y Astrof{\'i}sica \& 
           IPARCOS-UCM (Instituto de F\'{i}sica de Part\'{i}culas y del Cosmos de la UCM), 
           Facultad de Ciencias F{\'i}sicas, Universidad Complutense de Madrid, 28040 Madrid, Spain\\
           \email{emigom01@ucm.es}
           \and
           Centro de Astrobiolog\'{i}a (CSIC-INTA), ESAC,
           Camino Bajo del Castillo s/n, 28691 Villanueva de la Ca\~{n}ada, Madrid, Spain
           \and
           Centro de Astrobiolog\'{i}a (CSIC-INTA), Carretera a Ajalvir km 4, 28850 Torrej\'{o}n de Ardoz, Madrid, Spain
           \and
           Instituto de Astrof{\'i}sica e Ci{\^e}ncias do Espa\c{c}o, Universidade do Porto, CAUP, 
           Rua das Estrelas, 4150-762 Porto, Portugal
           \and
           Instituto de Astrof\'{i}sica de Canarias,
           c/ V\'{i}a L\'{a}ctea s/n, 38205 La Laguna, Tenerife, Spain
           \and
           Universidad de La Laguna, Departamento de Astrof\'{i}sica,
           38206 La Laguna, Tenerife, Spain
           \and
           Hamburger Sternwarte,
           Gojenbergsweg 112, 21029 Hamburg, Germany
           \and
           Th\"{u}ringer Landessternwarte Tautenburg, Sternwarte 5, 07778 Tautenburg, Germany
           \and
           Homer L. Dodge Department of Physics and Astronomy, University of Oklahoma, 440 West Brooks Street, Norman, OK-73019 Oklahoma, United States of America
           \and
           Institut de Ci\`{e}ncies de l'Espai (CSIC), Campus UAB, c/ de Can Magrans s/n, 
           08193 Cerdanyola del Vall\`{e}s, Spain
           \and
           Institut d'Estudis Espacials de Catalunya (IEEC), c/ Gran Capit\`{a} 2-4, 
           08034 Barcelona, Spain
           \and
           Institut f\"{u}r Astrophysik, Georg-August-Universit\"{a}t-G\"{o}ttingen, 
           Friedrich-Hund-Platz 1, 37077 G\"{o}ttingen, Germany
           \and
           Landessternwarte, Zentrum f\"{u}r Astronomie der Universit\"{a}t Heidelberg,
           K\"{o}nigstuhl 12, 69117 Heidelberg, Germany
           \and
           Instituto de Astrof\'{i}sica de Andaluc\'{i}a (IAA-CSIC), 
           Glorieta de la Astronom\'{i}a s/n, 18008 Granada, Spain
           \and
           Observatorio de Calar Alto, Sierra de los Filabres, 04550 G\'{e}rgal, Almer\'{i}a, Spain
           \and
           Max-Planck-Institut f\"{u}r Astronomie, K\"{o}nigstuhl 17, 69117 Heidelberg, Germany
           \and
           Max-Planck-Institut f\"{u}r Sonnensystemforschung, Justus-von- Liebig-Weg 3, 37077 G\"{o}ttingen, Germany
           \and
           Department of Physics, University of Warwick, Gibbet Hill Road, Coventry CV4 7AL, United Kingdom}

\date{Received 06 August 2021 / Accepted 7 October 2021}

\abstract
{We determined effective temperatures, surface gravities, and metallicities for a sample of 343 M dwarfs observed with CARMENES, the double-channel, high-resolution spectrograph installed at the 3.5\,m telescope at Calar Alto Observatory. We employed {\sc SteParSyn}, a Bayesian spectral synthesis implementation particularly designed to infer the stellar atmospheric parameters of late-type stars following a Markov chain Monte Carlo approach. We made use of the BT-Settl model atmospheres and the radiative transfer code {\tt turbospectrum} to compute a grid of synthetic spectra around 75 magnetically insensitive \ion{Fe}{i} and \ion{Ti}{i} lines plus the TiO $\gamma$ and $\epsilon$ bands. To avoid any potential degeneracy in the parameter space, we imposed Bayesian priors on $T_{\rm eff}$ and $\log{g}$ based on the comprehensive, multi-band photometric data available for the sample. We find that this methodology is suitable down to M7.0\,V, where refractory metals such as Ti are expected to condense in the stellar photospheres. The derived $T_{\rm eff}$, $\log{g}$, and [Fe/H] range from 3000 to 4200\,K, 4.5 to 5.3\,dex, and $-$0.7 to 0.2\,dex, respectively. Although our $T_{\rm eff}$ scale is in good agreement with the literature, we report large discrepancies in the [Fe/H] scales, which might arise from the different methodologies and sets of lines considered. However, our [Fe/H] is in agreement with the metallicity distribution of FGK-type stars in the solar neighbourhood and correlates well with the kinematic membership of the targets in the Galactic populations. Lastly, excellent agreement in $T_{\rm eff}$ is found for M dwarfs with interferometric angular diameter measurements, as well as in the [Fe/H] between the components in the wide physical FGK+M and M+M systems included in our sample.}


\maketitle

\section{Introduction}
\label{sec:introduction}

M dwarfs are cool and faint stars that comprise about two-thirds of the stellar population in the solar neighbourhood \citep{Hen18, Rey21}. With main-sequence lifespans longer than the age of the Universe \citep{Lau97}, these stars are not only an excellent record of the structure and evolution of the Milky Way \citep{Boc07}, but have also become prime targets for exoplanet surveys owing to their low mass, low temperature, and ubiquity \citep{Bon13, Dre13, Rei18}; this combination favours detections of Earth-mass planets inside their habitable zone, where liquid water can be sustained \citep{Sca07}.

Installed at the 3.5\,m telescope at the Calar Alto Observatory, the CARMENES instrument \citep{Qui20} is, along with SPIRou \citep{Don20}, NIRPS \citep{Wil17}, HPF \citep{Mah12}, IRD \citep{Kot18}, MAROON-X \citep{Sei20}, NEID \citep{Schw16}, and GIARPS \citep{Cla18}, a new-generation spectrograph designed to detect Earth-mass planets around M dwarfs by means of the radial velocity technique \citep{Rei18}. It consists of  optical (hereafter VIS) and near-infrared (NIR) channels that cover the 5200--9600\,{\AA} and 9600--17100\,{\AA} wavelength regions with spectral resolutions of $R=94\,600$ and $R=80\,400$, respectively. As of mid-2020, the CARMENES exoplanet survey had collected more than 18\,500 VIS and 18\,000 NIR spectra for a sample of 365 M dwarfs \citep{Qui20} obtained as part of its 750-night guaranteed time observations (GTO) programme. In addition to detecting and characterising the orbits  of exoplanets \citep[e.g.][]{Tri18, Tri20, Luq19, Zec19} and their atmospheres \citep{Nor18, Alo19a, Yan19}, the survey represents a unique opportunity to gain insights into the nature of M dwarfs (namely, their photospheric parameters, chemical compositions, magnetic fields, and chromospheric activity levels) in a statistically relevant way \citep{Fuh18, Fuh20, Pas18, Pas19, Hin19, Hin20, Scho19, Sch19, Shu19, Abi20, Sha21}.

In the context of exoplanet research, a thorough account of the general properties of host stars directly affects the determination of the position of the habitable zone \citep{Sel07, Kop13}, the radius and mass of detected exoplanets, and their composition \citep{Mal20, Tak20}. In addition, planet formation theories attempt to explain the observed occurrence rate of exoplanets \citep{Joh10, How12, Buc12, Sab21} and selection effects \citep{Win15} in terms of the stellar parameters.

However, the computation of photospheric parameters for M dwarfs from stellar spectra, namely the effective temperature, $T_{\rm eff}$, surface gravity, $\log{g}$, and stellar metallicity, [Fe/H], is an arduous challenge. The low photospheric temperatures give rise to pervasive molecular bands that severely distort the stellar continuum \citep{Kir91, Tin98, Rei18}. This alone rules out classical techniques commonly adopted in the analysis of solar-type spectra, such as the equivalent width (EW) method \citep{Mar20}. Stellar convection in M dwarfs also calls into question some of the physical assumptions behind the model atmospheres of late-type stars, such as local thermodynamic equilibrium (LTE) and one-dimensional (1D) geometry for radiative transfer, which often represents a limitation to the interpretation at the resolution achieved by modern spectrographs \citep{Ber17, Ola21}. Furthermore, M dwarfs may exhibit strong magnetic fields that not only drive stellar activity \citep{Del98, Don08}, but may also distort magnetically sensitive lines via the Zeeman effect \citep{Lan04, Pas19}. It is, therefore, difficult to choose a set of lines to disentangle the impact of the magnetic field and the photospheric parameters on the stellar spectra \citep{Shu19}. To make matters worse, telluric absorption is ubiquitous in the NIR \citep{Rei18, Nagelthesis}, where the spectral energy distribution of M dwarfs peaks \citep{Cif20}.

Despite the above, several methods have proved fruitful for inferring $T_{\rm eff}$, $\log{g}$, and [Fe/H] in M dwarfs, including spectral synthesis \citep[e.g.][]{Pas18, Pas19, Raj18a, Sou20, Hej20}, pseudo-EWs \citep[e.g.][]{Man13a, Man14, Nev14, Mal15, Mal20}, and spectral indices \citep[e.g.][]{Roj12, Kha20}, all of which can be further investigated following a machine-learning approach \citep[e.g.][]{Sar18, Ant20, Pas20, Li21}. In general terms, spectral synthesis relies on a minimisation algorithm to find the synthetic spectrum that best matches the observed spectrum \citep{Val05, Bre16}, while pseudo-EW and spectral index approaches draw on the sensitivity of certain features to $T_{\rm eff}$ and [Fe/H], as well as on calibrations with wide physical binaries harbouring an FGK-type primary with known metallicity \citep[][and references therein]{Cas08, Nev12, Mon18}. Further studies have focused on ensuring consistency between these two approaches \citep{Vey17}.

Semi-empirically calibrated methods have been widely adopted in the literature. For example, based on observations in the photometric $Y$ band (9470--11210\,{\AA}) with NIRSPEC on Keck II, \citet{Vey17} used a method to derive $T_{\rm eff}$, [Fe/H], and [Ti/H] that rests on the approach described by \citet{Man13b} and a calibration sample of M dwarfs in common proper motion FGK+M binaries from \citet{Man13a}. The method relies on the measured FeH index in the Wing-Ford band and on the EWs of seven \ion{Fe}{i} and ten \ion{Ti}{i} lines found in the $Y$ band. More recently, \citet{Mal20} derived $T_{\rm eff}$, $\log{g}$, and [Fe/H] for a sample of 204 stars with spectral types from K7\,V to M4.0\,V with HARPS and HARPS-N data, assuming the photometric $M_K$-[Fe/H] relationship from \citet{Nev12} and the $T_{\rm eff}$ scale from \citet{Man13b}. 

M-dwarf studies based on spectral synthesis differ from one to another in terms of the synthetic grid employed, the spectral resolution of the data, and the features selected for comparison across different wavelength regions. For instance, \citet{Pas18} derived $T_{\rm eff}$, $\log{g}$, and [Fe/H] for 300 M dwarfs, 235 of which were observed with CARMENES. They adopted a $\chi^2$ minimisation procedure based on the PHOENIX-ACES library of synthetic spectra \citep{Hus13} and several spectral features in the VIS wavelength region covered by CARMENES, namely the TiO $\gamma$ band and several \ion{K}{i}, \ion{Fe}{i}, \ion{Ti}{i}, and \ion{Mg}{i} lines. To avoid unreliable values of $\log{g}$ and [Fe/H] for some stars in the sample caused by degeneracies in the parameter space, they constrained $\log{g}$ by using the evolutionary models of \citet{Bar98}. \citet{Pas19} expanded the analysis into the NIR wavelength region covered by CARMENES. For the first time, they directly compared the stellar parameters of 282 M dwarfs determined from simultaneous observations in the VIS and NIR wavelength regions. Despite some discrepancies, \citet{Pas19} conclude that it is best to consider both regions simultaneously to maximise the amount of spectral information and, thus, minimise the effects of imperfect modelling in M dwarfs. Furthermore, \citet{Pas19} lessened the impact of the Zeeman line broadening caused by the stellar magnetic field in the parameter computations in the NIR by selecting lines with low effective Land\'{e} factors. Adapting the method described by \citet{Raj12, Raj18a}, \citet{Raj18b} computed $T_{\rm eff}$, $\log{g}$, and [Fe/H] for 292 M dwarfs from the individual CARMENES spectra published by \citet{Rei18}, using a grid of BT-Settl models \citep{All12}, while keeping $\log{g}$ free throughout the minimisation process. \citet{Hej20} also employed BT-Settl models to derive the $T_{\rm eff}$, $\log{g}$, metallicity [M/H], and $\alpha$ element to iron abundance ratio, [$\alpha$/Fe], of 1544 M dwarfs in the solar neighbourhood from low- and medium-resolution spectra collected at the Michigan-Darmouth-MIT, Lick, Kitt Peak National, and Cerro-Tololo Inter-American observatories. \citet{Sou20} computed $T_{\rm eff}$, $\log{g}$, and [Fe/H] for 21 M dwarfs from mid-resolution ($R\approx22\,500$) APOGEE $H$-band spectra \citep{Wil10}, a grid of MARCS models \citep{Gus08}, and the {\tt turbospectrum} code \citep{Ple12} through the {\tt bacchus} wrapper \citep{Mas16} and find excellent agreement in [Fe/H] between components in the wide physical binaries included in their sample. More recently, \citet{Sar21} derived $T_{\rm eff}$, $\log{g}$, [M/H], and $\varv_{\rm mic}$ for a sample of 313 M dwarfs from APOGEE $H$-band spectra by means of a $\chi^2$ minimisation against a synthetic grid generated with {\tt turbospectrum}, {\tt ispec} \citep{Bla14}, and MARCS model atmospheres and find good agreement with the parameters obtained with the APOGEE Stellar Parameter and Chemical Abundances Pipeline \citep{Gar16}.

\begin{figure}
\centering
\includegraphics[width=0.49\textwidth]{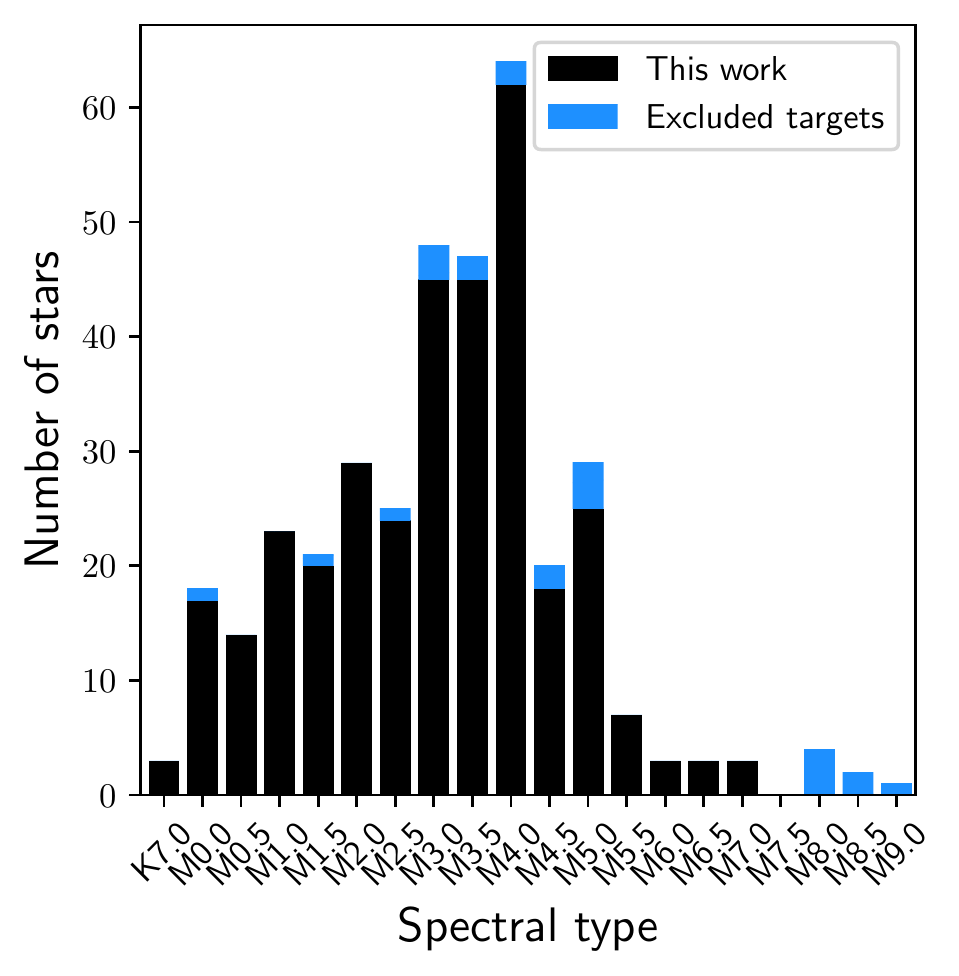}
\caption{\label{fig:histogram} Histogram of the M-dwarf sample analysed in this work (black) and the excluded CARMENES GTO targets (blue).}
\end{figure}

In this work we derived the stellar atmospheric parameters ($T_{\rm eff}$, $\log{g}$, and [Fe/H]) for a sample of 343 M dwarfs observed with CARMENES (i.e. 5200--17100\,{\AA} at $R\sim90\,000$) by means of the {\sc SteParSyn} code \citep{Tab20xx}, a Bayesian implementation of the spectral synthesis technique. This approach relied on the comparison of a synthetic grid against 75 magnetically insensitive \ion{Ti}{i} and \ion{Fe}{i} absorption lines along with the TiO $\gamma$ and $\epsilon$ bands. For the computation of the synthetic grid, we employed the {\tt turbospectrum} code \citep{Ple12} along with the BT-Settl model atmospheres \citep{All12}. In contrast to other works that disregard stellar activity \citep{Raj18a, Sar21}, our methodology allowed us to expand the analysis to M dwarfs that are strongly active, fast-rotating, or late-type (down to M7.0\,V), as a result of the careful selection of \ion{Ti}{i} and \ion{Fe}{i} lines according to their low sensitivity to chromospheric activity and the stellar magnetic field. In addition, the selected \ion{Ti}{i} and \ion{Fe}{i} lines are mostly free from non-LTE effects, which seem to affect lines of other elements widely used in other works \citep{Pas18, Pas19, Raj18a}, such as \ion{K}{i} \citep{Ola21}. 

This study is structured as follows. In Sect.~\ref{sec:sample} we describe the CARMENES GTO sample. We explain the derivation of the stellar atmospheric parameters in Sect.~\ref{sec:analysis}. Finally, we discuss the results and highlight the main conclusions in Sects.~\ref{sec:discussion} and~\ref{sec:summary}, respectively.

\section{Sample}
\label{sec:sample}

\begin{figure}
    \centering
    \includegraphics[width=0.49\textwidth]{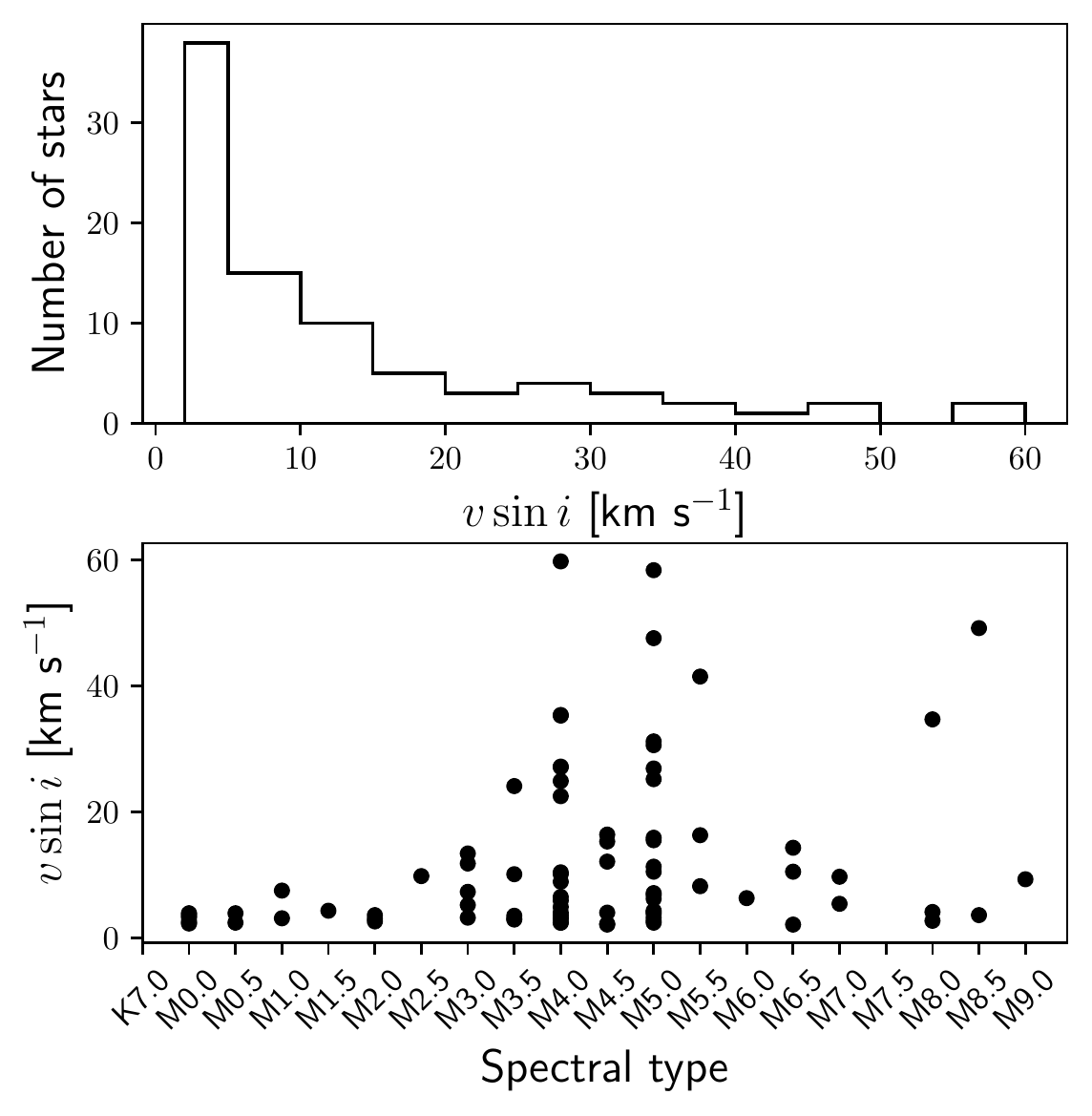}
    \caption{Distribution of projected rotational velocities larger than 2\,km\,s$^{-1}$ in the sample.}
    \label{fig:vsinihist}
\end{figure}

\begin{table}
\centering
\caption{\label{tab:excluded} GTO M dwarfs excluded from the analysis.}
\begin{tabular}{lll}
\hline\hline\noalign{\smallskip}
Karmn         & Name & Reason \\
\hline\noalign{\smallskip}
J00162$+$198W & EZ~Psc AB                    & (1) \\
J01056$+$284  & GJ~1029 AB                   & (1) \\
J02486$+$621  & 2MASS~J02483695$+$6211228    & (3) \\ 
J02573$+$765  & G~245-61                     & (3) \\ 
J04198$+$425  & LSR~J0419$+$4233             & (4) \\
J04219$+$213  & K2-155                       & (3) \\ 
J05394$+$406  & LSR~J0539$+$4038             & (4) \\
J05532$+$242  & Ross~59 AB                   & (1) \\
J07001$-$190  & 2MASS~J07000682$-$1901235 AB & (1) \\
J08536$-$034  & LP~666-009                   & (4) \\
J10182$-$204  & NLTT~23956 AB                & (1) \\
J10354$+$694  & LP~037-179 AB                & (1) \\
J14155$+$046  & GJ~1182 AB                   & (1) \\
J15412$+$759  & UU~UMi AB                    & (1) \\
J15474$-$108  & LP~743-031 ABC               & (2) \\
J16343$+$571  & CM~Dra Aab                   & (5) \\
J18356$+$329  & LSR~J1835$+$3259             & (4) \\
J19169$+$051S & V1298~Aql                    & (4) \\
J19255$+$096  & LSPM~J1925$+$0938            & (4) \\
J20198$+$229  & LP 395-008 AB                & (1) \\
J20556$-$140N & GJ~810~Aab                   & (1) \\
J23064$-$050  & 2MUCD~12171 (Trappist-1)     & (4) \\
J23585$+$076  & Wolf~1051 ABC                & (2) \\
\hline
\end{tabular}
\tablefoot{(1) Double-line spectroscopic (SB2) binary \citep{Bar18, Bar21}. (2) Triple-line spectroscopic triple systems \citep{Bar21}. (3) Conspicuous artefacts in the CARMENES template spectra. (4) Spectral type M8.0\,V or later. (5) Eclipsing binary \citep{Mor09}.}
\end{table}

\begin{table}
\centering
\caption{\label{tab:div}  Reference stars for the selection of \ion{Fe}{i} and \ion{Ti}{i} lines.}
\begin{tabular}{lcc}
\hline\hline\noalign{\smallskip}
Reference star   & SpT     & $T_{\rm eff}$\,[K] \\
\hline\noalign{\smallskip}
GX~And           & M1.0\,V & $\sim$3700 \\
Luyten's star    & M3.5\,V & $\sim$3400 \\
Teegarden's star & M7.0\,V & $\sim$2900 \\
\hline
\end{tabular}
\end{table}

\begin{figure*}
    \centering
    \includegraphics[width=\textwidth]{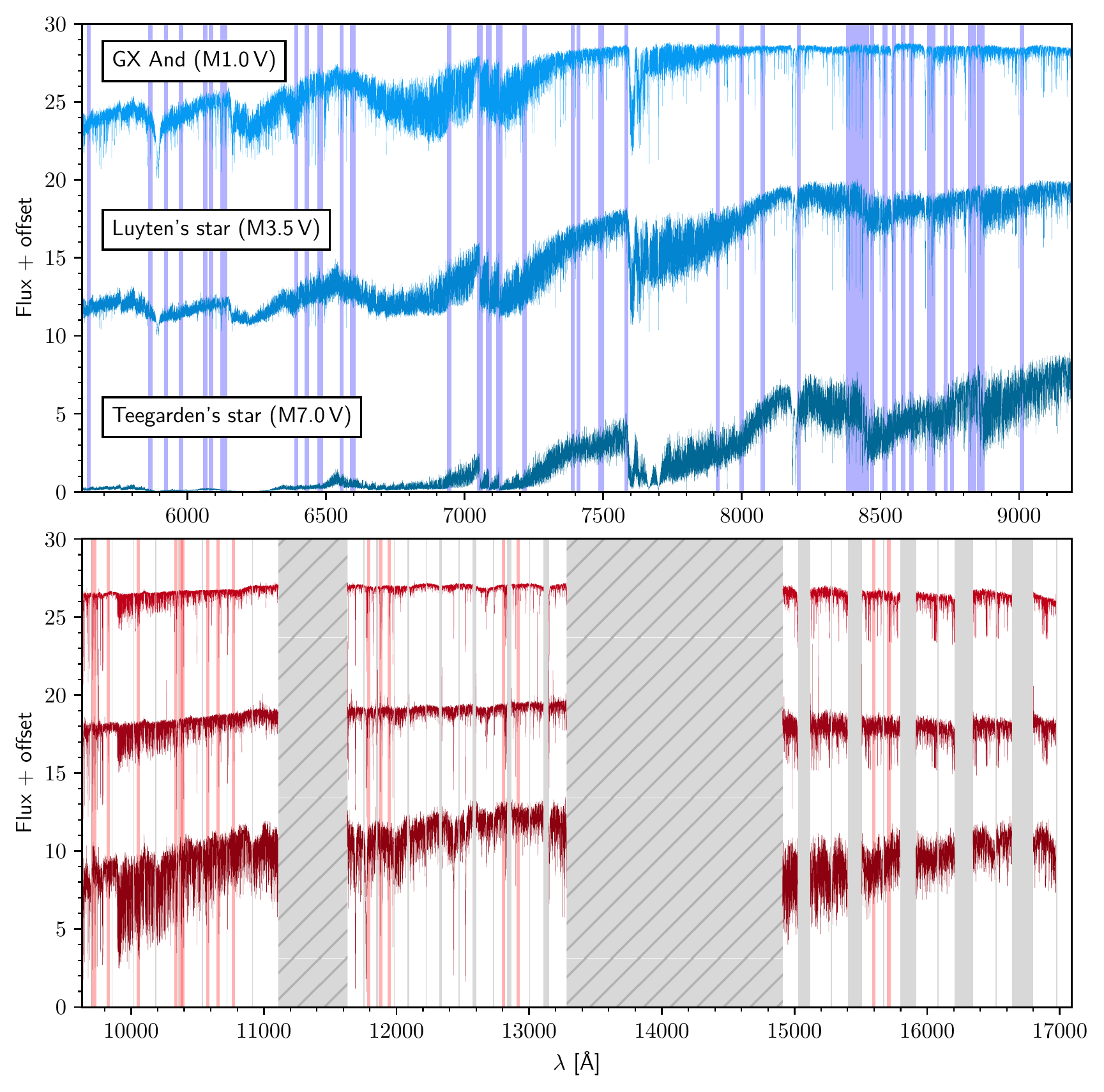}
    \caption{CARMENES template spectra of GX~And (M1.0\,V, J00183$+$440), Luyten's~star (M3.5\,V, J07274$+$052), and Teegarden's~star (M7.0\,V, J02530$+$168) in the VIS channel ({\it top panel}) and the NIR channel ({\it bottom panel}). Blue and red shaded regions denote the ranges synthesised in this work. The two wide spectral gaps in the NIR channel shown as hatched rectangles correspond to regions severely affected by telluric absorption \citep{Rei18}. Inter- and intra-order gaps in the NIR channel are also shown, as grey shaded regions.}
    \label{fig:refspectra}
\end{figure*}

Until mid-2020, the CARMENES M-dwarf radial velocity survey comprised 365 targets that cover the full M sequence from M0.0\,V to M9.0\,V \citep{Qui20}. The sample also includes three K7 dwarfs. Spectra for all stars were collected as part of the GTO programme of the instrument consortium from January 2016 to March 2021, equivalent to more than 5\,000 hours of observing time. 

A histogram of the M-dwarf sample analysed in this work is shown in Fig.~\ref{fig:histogram}, while all targets excluded from further analysis are listed in Table~\ref{tab:excluded}. Among these targets are one eclipsing binary \citep[CM~Dra;][]{Mor09}, ten double-line spectroscopic (SB2) binaries \citep{Bar18, Bar21}, and two triple-line spectroscopic triple systems \citep{Bar21}, for which no reliable parameter determinations could be made. We also discarded three targets because of conspicuous artefacts over multiple orders of their CARMENES spectra. Lastly, after performing the spectral processing presented in Sect.~\ref{subsec:processing}, seven M dwarfs with spectral types M8.0\,V or later were excluded due to limitations in our method. Altogether, our final sample includes 343 CARMENES GTO stars.

Table~\ref{tab:gtosample} displays the CARMENES identifiers, common names, spectral types from the CARMENES Cool dwarf Information and daTa Archive \citep[Carmencita;][]{Alo15, Cab16a}, radial velocities ($\varv_{\rm r}$), projected rotational velocities ($\varv\sin{i}$) adopted from \citet{Rei18} if available or computed following the same method otherwise, signal-to-noise ratios (S/N) over the entire spectra, and activity flags for the sample. Active stars are identified with an activity flag 1 if the pseudo-EW of the H$\alpha$ line satisfies ${\rm pEW'}({\rm H\alpha})<-0.3$\,{\AA}, following \citet{Scho19}. For most targets, radial velocities were adopted from \citet{Laf20}, who used cross-correlation with weighted binary masks. For targets that were not in \citet{Laf20}, we adopted the radial velocities provided by the CARMENES standard radial-velocity pipeline {\tt serval}\footnote{\url{www.github.com/mzechmeister/serval}} \citep{Zec18, Zec20}, based on least-squares fitting using high-S/N templates.

The distribution of $\varv\sin{i}$ in the sample is shown in Fig.~\ref{fig:vsinihist}. Most targets in our sample (268 stars) do not exhibit significant rotation, that is, $\varv\sin{i}<2$\,km\,s$^{-1}$, while targets with $\varv\sin{i}$ values higher than this upper limit (74 stars) are typically mid-type M dwarfs (M4.0--5.0\,V).

\section{Analysis}
\label{sec:analysis}

In the following we describe the processing of the CARMENES observations, the selection of the spectral features to be compared with the synthetic spectra, the computation of the synthetic grid by means of the BT-Settl models and {\tt turbospectrum}, and the analysis of the sample with the {\sc SteParSyn} code.

\subsection{Spectral processing}
\label{subsec:processing}

All M dwarfs considered have been observed with CARMENES at several epochs since January 2016 \citep{Rei18, Qui20}. The spectra were reduced following the standard CARMENES data flow. They were processed and calibrated in wavelength with the {\tt caracal} pipeline \citep{Cab16b}. The wavelength calibration relies on a combination of hollow cathode lamps and two temperature- and pressure-stabilised Fabry-P\'{e}rot units that ensure median uncertainties of about 1\,m~s$^{-1}$ in the VIS channel \citep{Tri18, Zec18}. As an additional step in the data flow, individual spectra were corrected for telluric absorption before the co-addition into template spectra as a way to increase the S/N. The modelling of the telluric spectra was done with the software package {\tt molecfit}\footnote{\url{http://www.eso.org/sci/software/pipelines/skytools/ molecfit}} \citep{Sme15, Kau15}, which incorporates the line-by-line radiative transfer code {\tt LBLRTM} \citep{Clo05} and the HITRAN molecular line list \citep{Rot09}. Further details on the removal of the telluric features can be found in \citet{Nagelthesis}. The radial velocity pipeline {\tt serval} \citep{Zec18} co-adds all individual spectra and provides one high-S/N template spectrum per star and instrument channel. We transformed the wavelengths onto the air scale following the International Astronomical Union standard \citep{Mor00} and corrected the spectra for the Doppler shift using the corresponding radial velocities of the targets.

\subsection{Selection of spectral features and line masks}
\label{subsec:selection}

\begin{figure}
    \centering
    \includegraphics[width=0.49\textwidth]{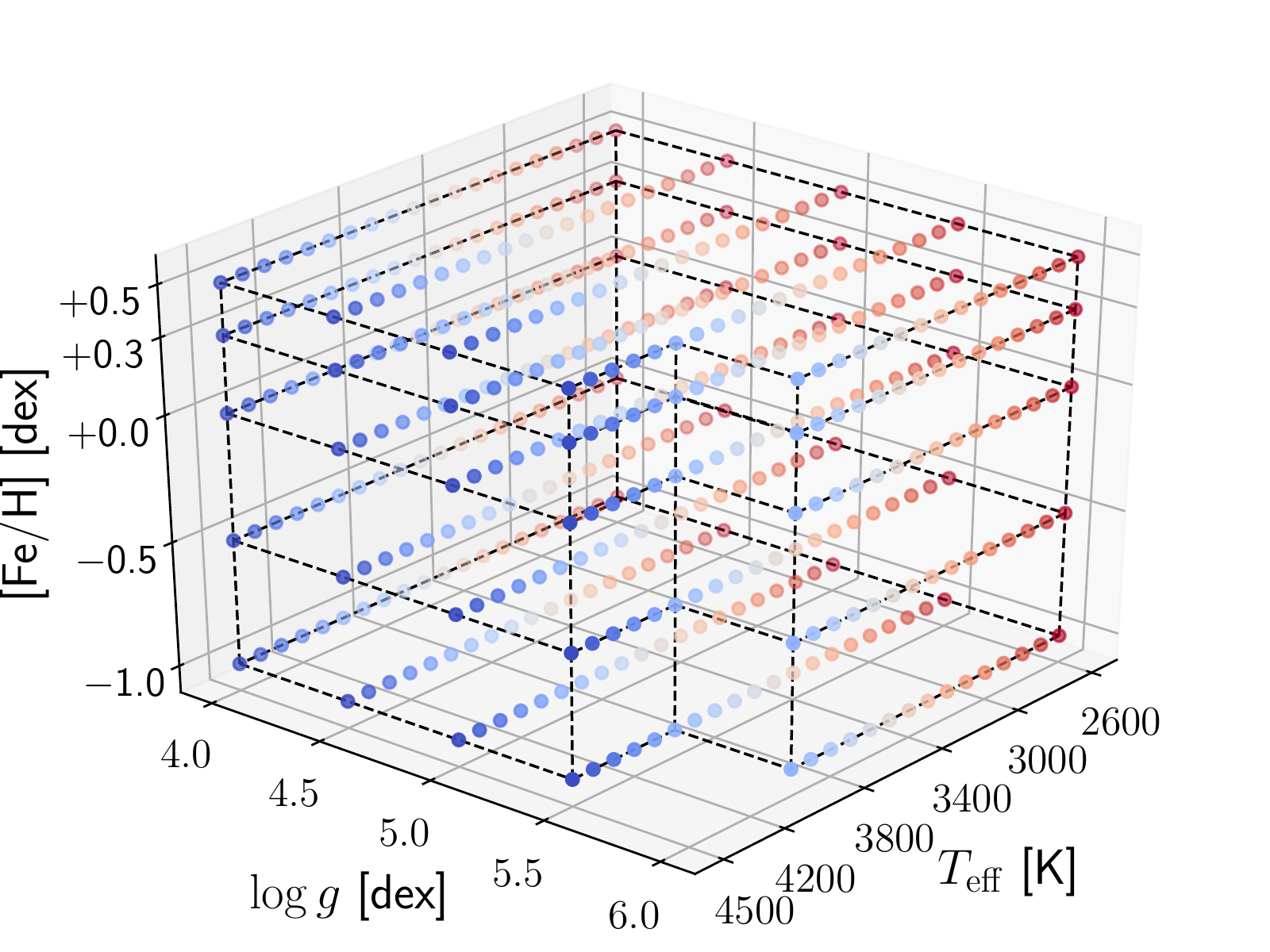}
    \caption{Coverage of our synthetic grid in $T_{\rm eff}$, $\log{g}$, and [Fe/H].}
    \label{fig:gridplot}
\end{figure}

The selection of \ion{Fe}{i} and \ion{Ti}{i} lines employed in this work was the result of a careful inspection of the template spectra of three representative early-, mid-, and late-type M dwarfs, namely GX~And, Luyten's~star, and Teegarden's~star (see Table~\ref{tab:div}). These M dwarfs show no significant rotation or activity, except for Teegarden's star, which shows relatively little activity for its spectral type \citep{Zec19}. In addition, their template spectra have a very high S/N. This choice of reference stars mirrors that of \citet{Rei18} for their spectral atlas. 

For this purpose, we first requested several line lists from the Vienna Atomic Line Database\footnote{\url{ http://vald.astro.uu.se}} \citep[VALD3;][]{Rya15} via the {\tt extract all} option, across the VIS and NIR wavelength regions covered by CARMENES. Following the prescriptions of \citet{Pas19}, we excluded all \ion{Fe}{i} and \ion{Ti}{i} lines in the NIR channel ($\lambda>9600$\,{\AA}) with large Land\'{e} factors (i.e. $g_{\rm eff} > 1.5$) to alleviate the impact of the stellar magnetic field on the line profiles. The visual inspection was complemented by a literature search to include some of the \ion{Fe}{i} and \ion{Ti}{i} line selections of \citet{Tab18} and \citet{Pas19}. The final list of 75 \ion{Fe}{i} and \ion{Ti}{i} lines along with the synthesised wavelength ranges can be found in Table~\ref{tab:linelist}. The ranges are wide enough (i.e. a few {\AA}) to circumvent potential problems posed by convolution at the edges. We also included the TiO $\gamma$ and $\epsilon$ bands for the present analysis (see Table~\ref{tab:bands}) given their high sensitivity to $T_{\rm eff}$ \citep{Raj14, Pas16}. In Fig.~\ref{fig:refspectra} we show the template spectra of the three reference stars along with the selected wavelength ranges (see also Figs.~\ref{fig:lineplot} and~\ref{fig:bandplot}). In contrast to some of the lines used in other works, such as \ion{K}{i} \citep{Pas18, Pas19, Raj18a}, non-LTE effects are negligible for \ion{Fe}{i} and \ion{Ti}{i} lines \citep{Ola21}.

We followed \citet{Tab20xx} to define the wavelength regions around the observed \ion{Fe}{i} and \ion{Ti}{i} line profiles in the template spectra to be compared with the synthetic grid (i.e. the line masks). We first performed a preliminary Gaussian fit to the \ion{Ti}{i} and \ion{Fe}{i} lines by means of the Levenberg-Marquardt algorithm through the Python {\tt scipy} package \citep{Vir20}. Next, we adjusted the line profiles assuming an initial width of 3$\sigma$ around their centre, avoiding adjacent spectral features. For targets with $\varv\sin{i}>4$~km\,s$^{-1}$ where the Gaussian approximation may no longer accurately reproduce the line profiles, we also broadened these line regions to account for rotation following the expression \citep{Tab20xx}:
\begin{equation}
\Delta\lambda = 2\lambda_{\rm line}\frac{\varv\sin{i}}{c},
\end{equation}
where $\Delta\lambda$ is the broadening, $\lambda_{\rm line}$ is the central wavelength of the line, and $c$ is the speed of light.

For active stars (as indicated by the H$\alpha$ flag), we specifically excluded some \ion{Fe}{i} and \ion{Ti}{i} lines that appear to be particularly sensitive to chromospheric activity and the stellar magnetic field. These lines are marked as magnetically sensitive in Table~\ref{tab:linelist} and will be discussed in a future publication \citep[{\color{blue} L\'{o}pez-Gallifa et al., in prep.}; for more details, see][]{Mon20}. This additional consideration is particularly relevant for M dwarfs showing magnetic fields that can be as high as several kilogauss, such as EV~Lac and YZ~CMi \citep{Shu19}. Overall, the line exclusion affected a total of 91 active stars in our sample and proved crucial in reaching convergence and, thus, avoiding getting unreliable parameters with {\sc SteParSyn}.

\begin{table}
\centering
\caption{\label{tab:bands} List of TiO bands and wavelength ranges synthesised.}
\begin{tabular}{lccc}
\hline\hline\noalign{\smallskip}
TiO band & \multicolumn{2}{c}{Range} & Band head \\
& $\lambda_{\rm min}$\,[\AA] & $\lambda_{\rm max}$\,[\AA] & $\lambda_{\rm head}$\,[\AA]\\
\hline\noalign{\smallskip}
\multirow{3}{*}{$\gamma$ bands}   & 7049.69 & 7058.85 & \textasciitilde 7054 \\
& 7082.49 & 7092.31 & \textasciitilde 7087 \\ 
& 7120.30 & 7129.69 & \textasciitilde 7125 \\
\hline\noalign{\smallskip}
\multirow{2}{*}{$\epsilon$ bands} & 8436.60 & 8446.51 & \textasciitilde 8440 \\
& 8854.87 & 8864.58 & \textasciitilde 8859 \\
\hline\noalign{\smallskip}
\end{tabular}
\end{table}

\subsection{Synthetic grid}
\label{subsec:grid}

After the line selection step, we computed a coarse grid of synthetic spectra based on a set of BT-Settl model atmospheres \citep{All12} and the 1D LTE spectrum synthesis code {\tt turbospectrum} \citep{Alv98, Ple12}. The grid is representative of the M-dwarf regime and ranges from 2600\,K to 4500\,K and 4.0\,dex to 6.0\,dex, when available (i.e. 5.5\,dex instead for models with $T_{\rm eff}>4000$\,K) in steps of 100\,K and 0.5\,dex in $T_{\rm eff}$ and $\log{g}$, respectively (see Fig.~\ref{fig:gridplot}). We set the metallicity, [Fe/H], to the following values: $-$1.0, $-$0.5, 0.0, $+$0.3, and $+$0.5\,dex. For all models, we adopted the same micro-turbulence velocities as in the PHOENIX-ACES library of synthetic spectra\footnote{\url{http://phoenix.astro.physik.uni-goettingen.de/}} \citep{Hus13}. We assumed the solar abundances as reported by \citet{Asp09} and scaled the abundances of the $\alpha$ elements (O, Ne, Mg, Si, S, Ar, Ca, and Ti) following the discussion of \citet{Gus08} for the standard grid of MARCS models, so that  [$\alpha$/Fe]\,=\,$-$0.4[Fe/H] for $-$1.0\,$\le$\,[Fe/H]$\,<$\,0.0, and [$\alpha$/Fe]\,=\,0.0 (i.e. no enhancement) for [Fe/H]\,$\ge$\,0.0. The synthetic grid was finally stored in a binary file by means of the Python module {\tt pickle} \citep{Van20}.

The atomic data were gathered via automatic email requests to the VALD3 database using the {\tt extract all} option for all the ranges listed in Tables~\ref{tab:bands} and \ref{tab:linelist}. The required molecular data were compiled from several sources and included H$_2$O \citep{Bar06}, FeH \citep{Dul03}, MgH \citep{Kur14}, CO \citep{Goo94}, SiH \citep{Kur14}, OH \citep{Kur14}, VO \citep{kem16}, CaH, ZrO, and TiO \citep[B. Plez priv. comm.; see][]{Hei21}.  To reduce computation times, we only considered the most relevant transitions in each molecular line list by applying the following Boltzmann cut \citep{Gra08}:
\begin{equation}
\log{({gf\cdot\lambda_i})}-\chi_i\cdot\theta > \max_i[ \log{({gf\cdot\lambda_i})}-\chi_i\cdot\theta]-5, 
\end{equation}
where $\lambda_i$ denotes the wavelength in {\AA} of the transition $i$, $\log{gf}$ is the oscillator strength, $\chi_i$ is the excitation potential in eV, and $\theta=5040\,{\rm K}/T$, with $T=3000$\,K.

As discussed in \citet{Tab20xx}, the {\sc SteParSyn} code allows the synthetic grid to be sequentially convolved on the fly (i.e. prior to the comparison with the observed spectra) to account for both the stellar rotation and the instrumental profile as the main sources of spectral broadening. As for rotation, we assumed the rotational profile with a limb darkening coefficient of 0.6 from \citet{Gra08}, and adopted the projected rotational velocities from Table~\ref{tab:gtosample}. Finally, to account for the instrumental profile, we convolved the spectra with a Voigt profile, assuming the corresponding median averaged Gaussian and Lorentzian components for the CARMENES VIS and NIR channels \citep{Nagelthesis}. 

\subsection{{\sc SteParSyn}}
\label{subsec:steparsyn}

\begin{figure}
\centering
\includegraphics[width=0.49\textwidth]{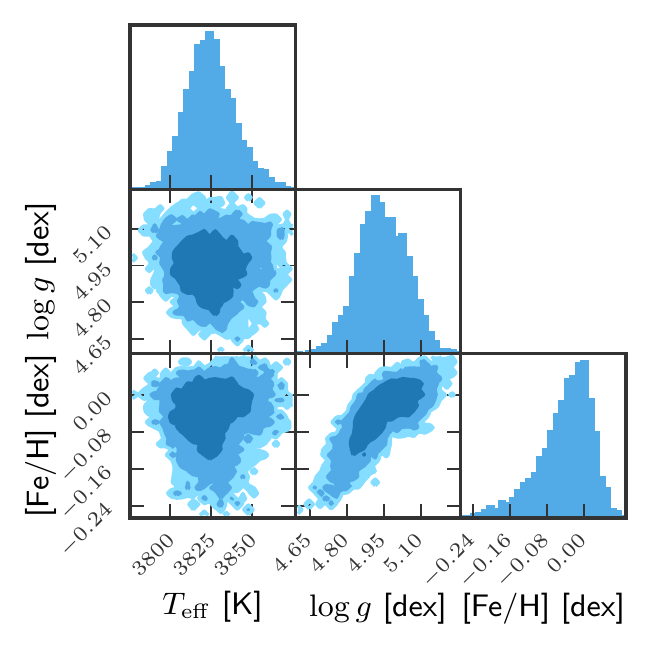}
\caption{\label{fig:cornerplot} Marginalised posterior distributions in $T_{\rm eff}$, $\log{g}$, and [Fe/H] for the M1.0\,V star HD~233153 (J05415+534). The colour shades denote the 1$\sigma$, 2$\sigma$, and 3$\sigma$ levels. The retrieved parameters are $T_{\rm eff}=3825\pm14$\,K, $\log{g}=4.94\pm0.10$\,dex, and ${\rm [Fe/H]}=-0.04\pm0.06$\,dex.}
\end{figure}

\begin{figure*}[ht!]
\centering
\includegraphics[width=0.95\textwidth]{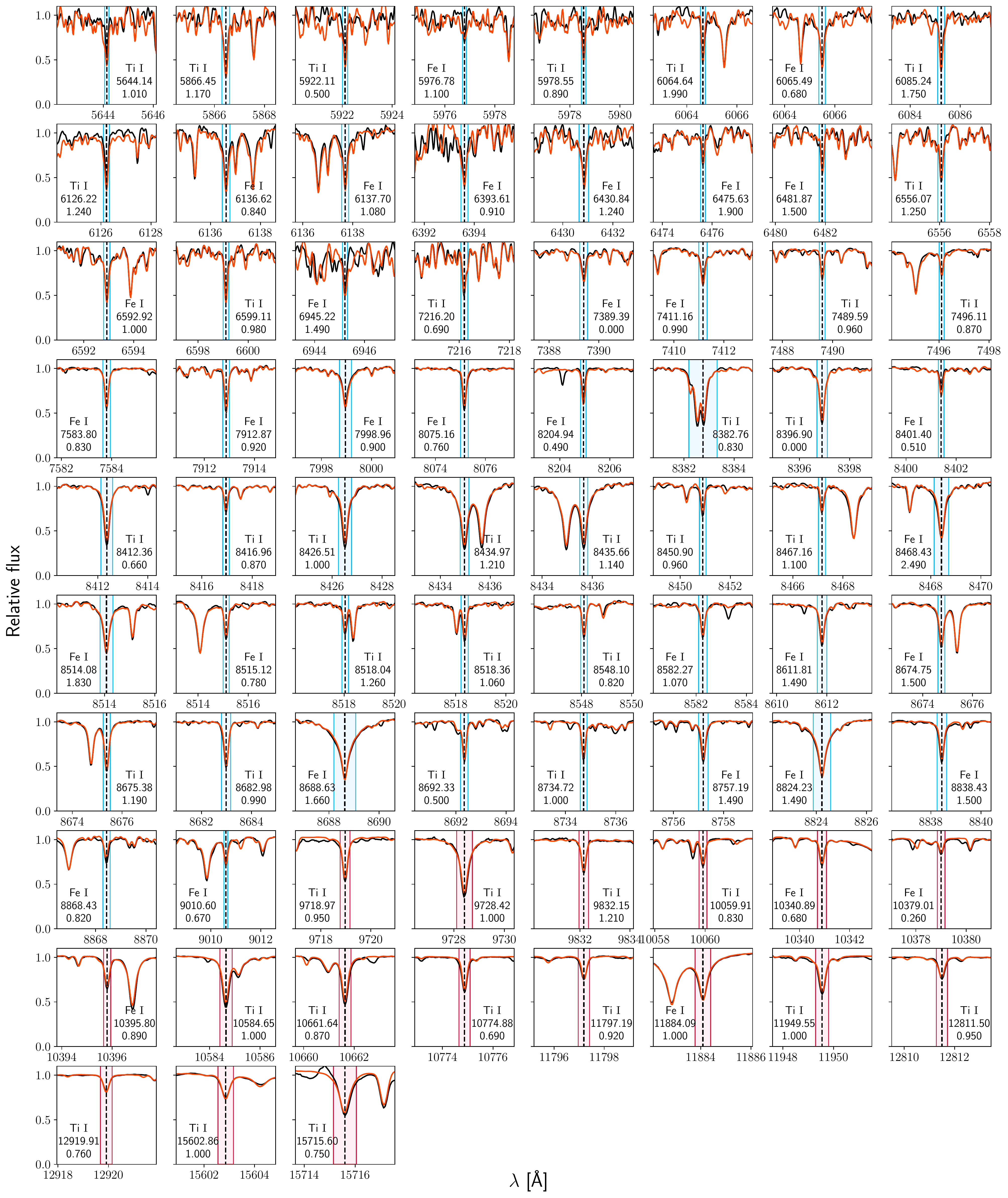}
\caption{\label{fig:lineplot} Atomic line fits for the M1.0\,V star HD~233153 (J05415$+$534). The solid black and orange lines are the observed and synthetic spectra, respectively. The blue and pink shaded areas are the regions of comparison between the observed and the synthetic spectra (i.e. line masks) for features in the VIS and NIR channels of CARMENES, respectively. The vertical dashed black lines mark the central wavelengths. The species, the central wavelength, and the effective Land\'{e}  factor, $g_{\rm eff}$, of the lines are indicated inside each box.}
\end{figure*}

\begin{figure}
\centering
\includegraphics[width=0.49\textwidth]{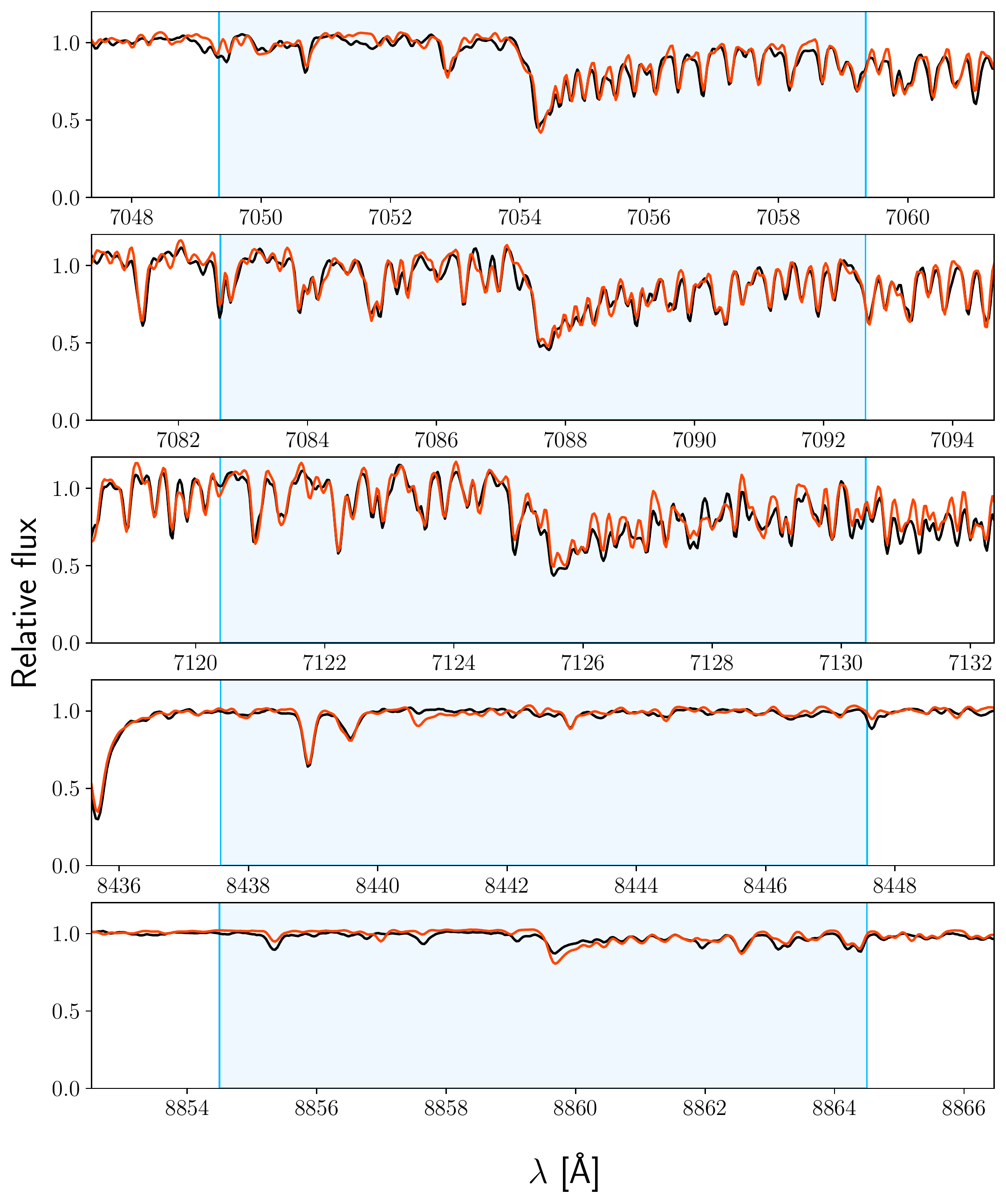}
\caption{\label{fig:bandplot} Same as Fig.~\ref{fig:lineplot}, but for the TiO $\gamma$ and $\epsilon$ bands.}
\end{figure}

\begin{figure}
\centering
\includegraphics[width=0.49\textwidth]{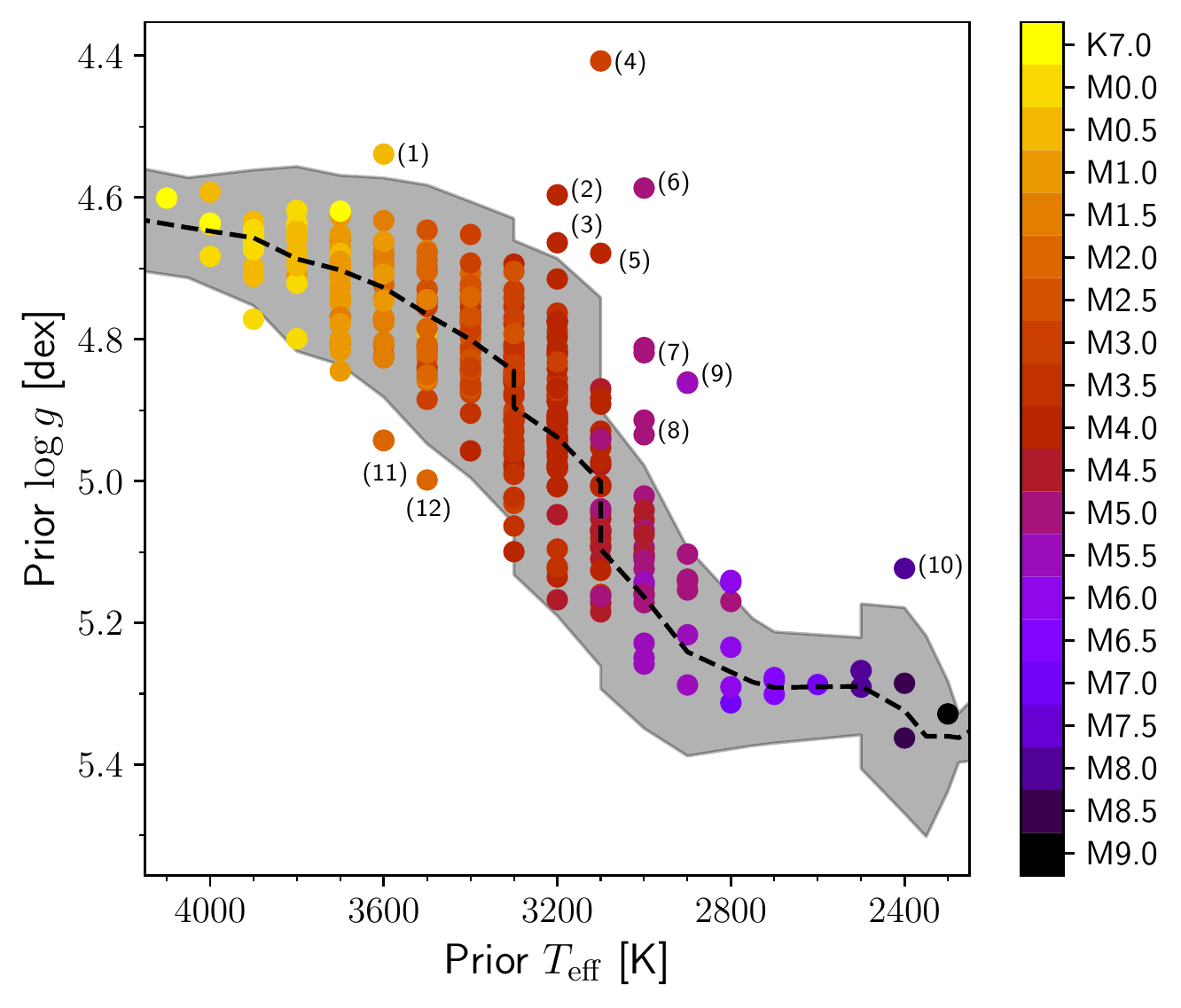}
\caption{Central prior $T_{\rm eff}$ and $\log{g}$ values for the sample following \citet{Cif20}. The spectral types are colour-coded. The dashed black line is the average $T_{\rm eff}$ and $\log{g}$ for K7\,V to M9.0\,V objects, with the corresponding uncertainties shaded in grey. The 15 outliers discussed in Sect.~\ref{subsec:steparsyn} are labelled and listed in Table~\ref{tab:young}. Two and three stars are superimposed under labels (7) and (11), respectively.}
\label{fig:prior}
\end{figure}

\begin{table*}
\centering
\caption{List of outliers in $\log{g}$.}
\label{tab:young}
\begin{tabular}{lllccl}
\hline\hline\noalign{\smallskip}
Name & Karmn & SpT & Prior $\log{g}$ & Label\,$^{(a)}$ & Remarks \\
     &       &     & [dex]           &                 &       \\
\hline\noalign{\smallskip}
Barta~161~12              & J01352$-$072 & M4.0\,V & 4.60 & (2)  & Member of $\beta$~Pictoris Moving Group  \\
RX~J0447.2+2038           & J04472$+$206 & M5.0\,V & 4.82 & (7)  & Member of IC~2391 Supercluster \\
1RXS~J050156.7$+$010845   & J05019$+$011 & M4.0\,V & 4.66 & (3)  & Member of $\beta$~Pictoris Moving Group  \\
RX~J0506.2$+$0439         & J05062$+$046 & M4.0\,V & 4.68 & (5)  & Member of $\beta$~Pictoris Moving Group  \\
2MASS~J05082729$-$2101444 & J05084$-$210 & M5.0\,V & 4.59 & (6)  & Member of Local Association \\
LP~205-044                & J06318$+$414 & M5.0\,V & 4.81 & (7)  & Member of Local Association \\
1RXS~J114728.8$+$664405   & J11474$+$667 & M5.0\,V & 4.91 & (8)  & Member of Castor Moving Group  \\
K2-33                     & J16102$-$193 & M3.0\,V & 4.41 & (4)  & Member of Local Association \\
1RXS~J173353.5$+$165515   & J17338$+$169 & M5.5\,V & 4.86 & (9)  & Member of Local Association \\
LSPM~J1925$+$0938         & J19255$+$096 & M8.0\,V & 5.12 & (10) & Spectral type M8.0\,V (see Table~\ref{tab:div}) \\
AU~Mic                    & J20451$-$313 & M1.0\,V & 4.54 & (1)  & Member of $\beta$~Pictoris Moving Group  \\
\noalign{\smallskip}
\hline
\noalign{\smallskip}
Ross~695                  & J12248$-$182 & M2.0\,V & 5.00 & (12) & Behaviour akin to a subdwarf \\
GJ~625                    & J16254$+$543 & M1.5\,V & 4.94 & (11) & Behaviour akin to a subdwarf \\
Ross~730                  & J19070$+$208 & M2.0\,V & 4.94 & (11) & Behaviour akin to a subdwarf \\
HD~349726                 & J19072$+$208 & M2.0\,V & 4.94 & (11) & Behaviour akin to a subdwarf \\
\hline
\end{tabular}
\tablefoot{Membership in young stellar kinematic groups following {\color{blue} Cort\'{e}s-Contreras et al. in prep.} $^{(a)}$Label column refers to Fig.~\ref{fig:prior}. Two and three stars are superimposed under labels (7) and (11), respectively.}
\end{table*}

{\sc SteParSyn} is a Bayesian code written in Python 3 designed to map the probability distributions of the stellar atmospheric parameters ($T_{\rm eff}$, $\log{g}$, [Fe/H], and $\varv\sin{i}$) from a given observed spectrum following an Markov chain Monte Carlo approach \citep{Tab20xx}. It has already been employed in many astrophysical contexts, including the study of stars in open clusters \citep{Loh18, Alo19}, stars in galaxies of the Local Group \citep{Tab18}, and exoplanet host stars \citep{Bor21, Dem21}.

In broad terms, the {\sc SteParSyn} code compares the observed spectrum with a grid of synthetic spectra computed around particular features. Prior decomposition of the synthetic grid following a principal component analysis \citep{Fra99} helps to assess any given point of the parameter space in a computationally inexpensive way. {\sc SteParSyn} finally returns the marginalised posterior distributions in the stellar atmospheric parameters, as shown in Fig.~\ref{fig:cornerplot} for a representative case (the M1.0\,V star \object{HD~233153}), along with the best synthetic fits for the atomic lines and molecular bands, as shown in Figs.~\ref{fig:lineplot} and \ref{fig:bandplot}, respectively, for the same star (see Figs.~\ref{fig:refGXAndcorner} to \ref{fig:refTeegardenbands} for the reference stars GX~And, Luyten's~star, and Teegarden's~star). The corner plot showing the marginalised posterior distribution obtained with {\sc SteParSyn} was made with the Python {\tt pygtc} package \citep{Boc16}. Further details about {\sc SteParSyn} can be found in \citet{Tab20xx}.

To avoid degeneracies in the M-dwarf parameter space, particularly between $\log{g}$ and [Fe/H], we assumed Gaussian prior probability distributions in $T_{\rm eff}$ and $\log{g}$ centred according to the $T_{\rm eff}$, stellar mass, and radius reported by \citet{Cif20} for each target when available\footnote{\url{https://github.com/ccifuentesr/CARMENES-V}}, or otherwise the averaged parameters as listed in Table~6 in \citet{Cif20}, with standard deviations of 200~K and 0.2~dex, respectively. $T_{\rm eff}$ and $\log{g}$ prior distributions for our sample are included in Table~\ref{tab:par_stars_stepar}. These $T_{\rm eff}$ values were derived from a thorough multi-band photometric analysis carried out with the Virtual Observatory Spectral energy distribution Analyser \citep[{\tt vosa};][]{Bay08}, whereas the stellar radii and masses were obtained from the Stefan-Boltzmann law and the mass-radius relation derived by \citet{Sch19}, respectively. 

In Fig.~\ref{fig:prior} we show a Kiel diagram (i.e. $\log{g}$ versus $T_{\rm eff}$) of the prior $T_{\rm eff}$ and $\log{g}$ values that we adopted for our sample. Some stars show low $\log{g}$ values, which can be explained with stellar age. These stars were already identified as young stars by \citet{Sch19} on account of their membership in young stellar kinematic groups {\color{blue} (Cort\'{e}s-Contreras et al. in prep.}). Among these are the Local Association, $\beta$~Pictoris Moving Group, the IC~2391 Supercluster, and Castor Moving Group, with estimated ages of 10--150~Ma \citep{Bel15}, 18.5\,Ma \citep{Mir20}, 50\,Ma~\citep{Bar04}, and 440\,Ma \citep{Mam13}, respectively. On the contrary, a few additional targets exhibit a high $\log{g}$, probably due to their behaviour akin to subdwarfs, following the discussion by \citet{Sch19}. All these outliers in $\log{g}$ are listed in Table~\ref{tab:young}.

\section{Results and discussion}\label{sec:discussion}

\begin{figure}
\centering
\includegraphics[width=0.49\textwidth]{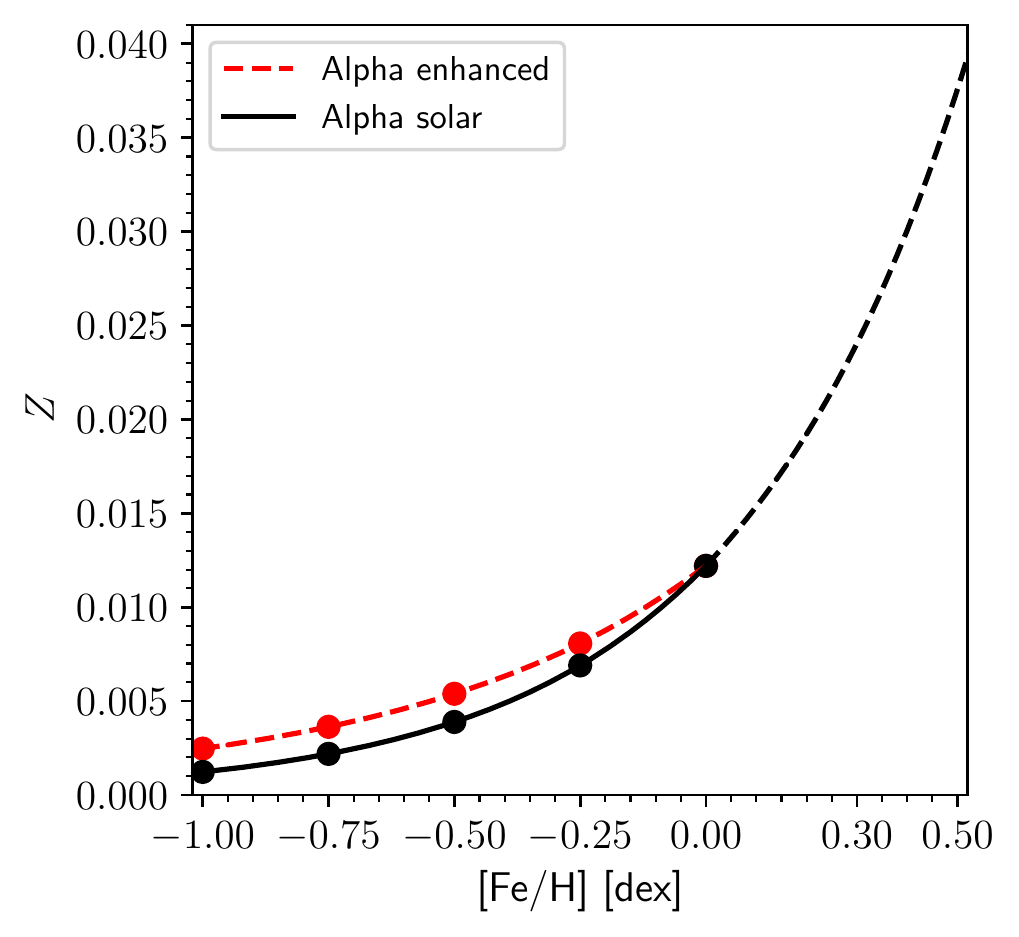}
\caption{\label{fig:alpha} Interpolation scheme between $Z$ and [Fe/H] in alpha-enhanced (red) and alpha-solar (black) models.}
\end{figure}

\begin{figure*}
\centering
\includegraphics[width=\textwidth]{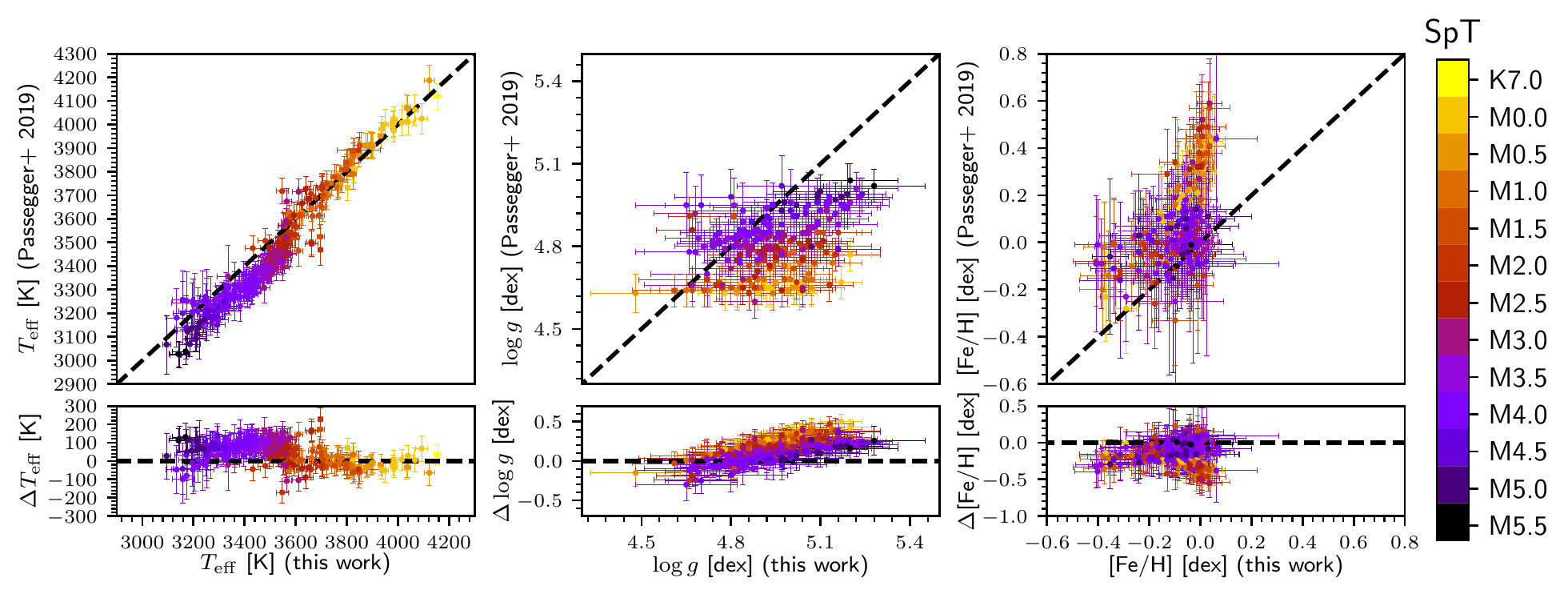}
\caption{\label{fig:referencecomparisons} Comparison between this work and the VIS+NIR analysis done by \citet{Pas19}. Spectral types are colour-coded. The dashed black lines indicate the 1:1 relationship.}
\end{figure*}

\begin{table*}
\centering
\caption{Summary of the Monte Carlo simulations on $T_{\rm eff}$, $\log{g}$, and [Fe/H] between different comparison samples.}
\label{tab:mc_par}
\begin{tabular}{lccccc}
\hline\hline\noalign{\smallskip}
Sample & Size & Parameter & Difference & $r_{\rm p}\pm\sigma_{\rm p}$ & $r_{\rm s}\pm\sigma_{\rm s}$ \\
\hline\noalign{\smallskip}
               &     & $T_{\rm eff}$       &    $64\pm95$\,K     & $0.807\pm0.016$ & $0.824\pm0.013$ \\
\citet{Raj18a} & 292 & $\log{g}$           & $-0.23\pm0.30$\,dex & $0.069\pm0.054$ & $0.051\pm0.054$ \\
               &     & [Fe/H]              & $-0.25\pm0.28$\,dex & $0.079\pm0.054$ & $0.084\pm0.053$ \\
\noalign{\smallskip}
               &     & $T_{\rm eff}$       &    $25\pm53$\,K     & $0.937\pm0.006$ & $0.940\pm0.005$ \\
\citet{Sch19}  & 287 & $\log{g}$           &  $0.03\pm0.20$\,dex & $0.253\pm0.040$ & $0.288\pm0.039$ \\
               &     & [Fe/H]              & $-0.19\pm0.17$\,dex & $0.397\pm0.018$ & $0.467\pm0.013$ \\
\noalign{\smallskip}
               &     & $T_{\rm eff}$       &    $48\pm61$\,K     & $0.931\pm0.007$ & $0.930\pm0.006$ \\
\citet{Pas19}  & 282 & $\log{g}$           &  $0.13\pm0.14$\,dex & $0.223\pm0.041$ & $0.259\pm0.042$ \\
               &     & [Fe/H]              & $-0.20\pm0.15$\,dex & $0.275\pm0.051$ & $0.288\pm0.049$ \\
\noalign{\smallskip}
               &     & $T_{\rm eff}$       &    $29\pm44$\,K     & $0.932\pm0.008$ & $0.945\pm0.005$ \\
\citet{Pas18}  & 235 & $\log{g}$           &  $0.02\pm0.15$\,dex & $0.182\pm0.045$ & $0.241\pm0.045$ \\
               &     & [Fe/H]              & $-0.10\pm0.12$\,dex & $0.288\pm0.051$ & $0.350\pm0.048$ \\
\noalign{\smallskip}
               &     & $T_{\rm eff}$       &    $29\pm80$\,K     & $0.883\pm0.014$ & $0.877\pm0.015$ \\
\citet{Mal20}  & 102 & $\log{g}$           &  $0.08\pm0.15$\,dex & $0.249\pm0.071$ & $0.285\pm0.072$ \\
               &     & [Fe/H]              & $-0.06\pm0.15$\,dex & $0.312\pm0.063$ & $0.364\pm0.065$ \\
\noalign{\smallskip}
               &     & $T_{\rm eff}$       &    $85\pm63$\,K     & $0.928\pm0.011$ & $0.924\pm0.011$ \\
\citet{Man15}  &  86 & $\log{g}$           &  $0.11\pm0.13$\,dex & $0.355\pm0.063$ & $0.427\pm0.060$ \\
               &     & [Fe/H]              & $-0.13\pm0.13$\,dex & $0.611\pm0.057$ & $0.648\pm0.046$ \\
\noalign{\smallskip}
               &     & $T_{\rm eff}$       &     $4\pm76$\,K     & $0.850\pm0.030$ & $0.860\pm0.023$ \\
\citet{Gai14}  &  63 & $\log{g}$           &  $0.09\pm0.15$\,dex & $0.088\pm0.084$ & $0.156\pm0.083$ \\
               &     & [Fe/H]              & $-0.12\pm0.11$\,dex & $0.643\pm0.064$ & $0.665\pm0.051$ \\
\noalign{\smallskip}
               &     & $T_{\rm eff}$       &    $23\pm107$\,K    & $0.917\pm0.017$ & $0.911\pm0.015$ \\
\citet{Pas20}  &  50 & $\log{g}$           &  $0.17\pm0.16$\,dex & $0.341\pm0.088$ & $0.398\pm0.098$ \\
               &     & [Fe/H]              & $-0.32\pm0.12$\,dex & $0.264\pm0.122$ & $0.245\pm0.138$ \\
\noalign{\smallskip}
               &     & $T_{\rm eff}$       &    $41\pm124$\,K    & $0.915\pm0.019$ & $0.886\pm0.021$ \\
\citet{Roj12}  &  40 & $\log{g}$\,$^{(a)}$ & \ldots              & \ldots          & \ldots          \\
               &     & [Fe/H]              & $-0.15\pm0.13$\,dex & $0.513\pm0.103$ & $0.533\pm0.099$ \\
\noalign{\smallskip}
               &     & $T_{\rm eff}$       &    $39\pm42$\,K     & $0.884\pm0.040$ & $0.883\pm0.037$ \\
\citet{Mal15}  &  21 & $\log{g}$           &  $0.12\pm0.13$\,dex & $0.213\pm0.169$ & $0.210\pm0.171$ \\
               &     & [Fe/H]              & $-0.07\pm0.14$\,dex & $0.576\pm0.116$ & $0.581\pm0.100$ \\
\hline
\end{tabular}
\tablefoot{We give the average difference on each parameter and the Pearson ($r_{\rm p}$) and the Spearman ($r_{\rm s}$) correlation coefficients (see Sect.~\ref{sec:discussion} for details). $^{(a)}$\citet{Roj12} fixed $\log{g}$ to 5.0\,dex.}
\end{table*}

\begin{figure}
    \centering
    \includegraphics[width=0.49\textwidth]{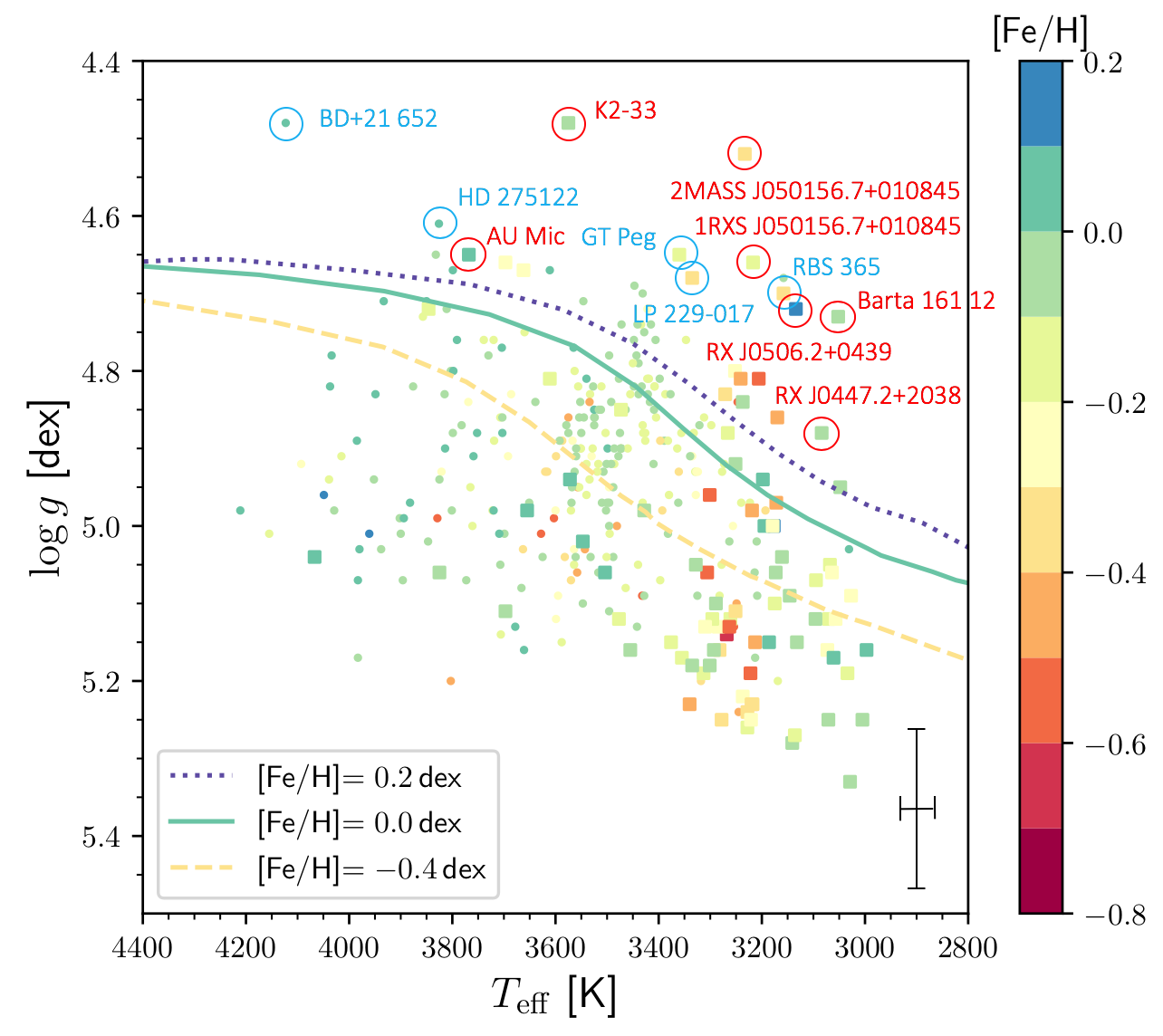}
    \caption{Kiel diagram of the sample showing the parameters computed with {\sc SteParSyn}. Squares depict H$\alpha$-active stars. The metallicity [Fe/H]$_{\rm corr}$ is colour-coded. Lines correspond to 5-Ga PARSEC isochrones with [Fe/H]\,=\,$-$0.4, 0.0, and 0.2\,dex \citep{Bre12}. Typical uncertainties in $\log{g}$ and $T_{\rm eff}$ are shown in the bottom-right corner. Outliers in $\log{g}$ are marked with red and blue circles (see Sect.~\ref{sec:discussion} for details).}
    \label{fig:HRdiagram}
\end{figure}

\begin{figure}
    \centering
    \includegraphics[width=0.49\textwidth]{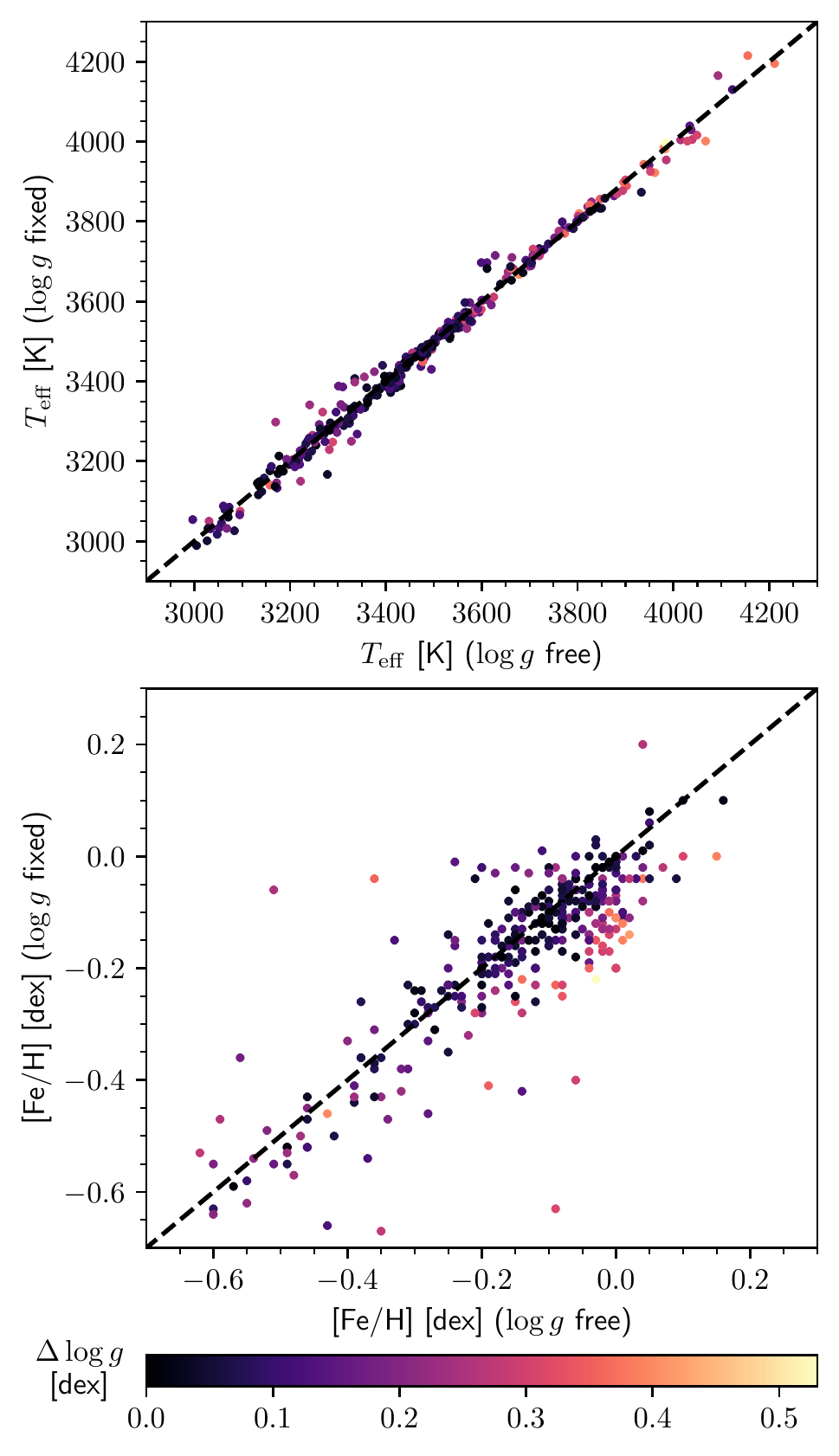}
    \caption{Comparison of $T_{\rm eff}$ ({\it top panel}) and [Fe/H] ({\it bottom panel}) between the runs with free and fixed $\log{g}$. The quantity $\Delta\log{g}$ is calculated as $\log{g}\,{\rm (prior)}-\log{g}\,{\rm (free)}$.}
    \label{fig:comparisonruns}
\end{figure}

\begin{figure}
    \centering
    \includegraphics[width=0.49\textwidth]{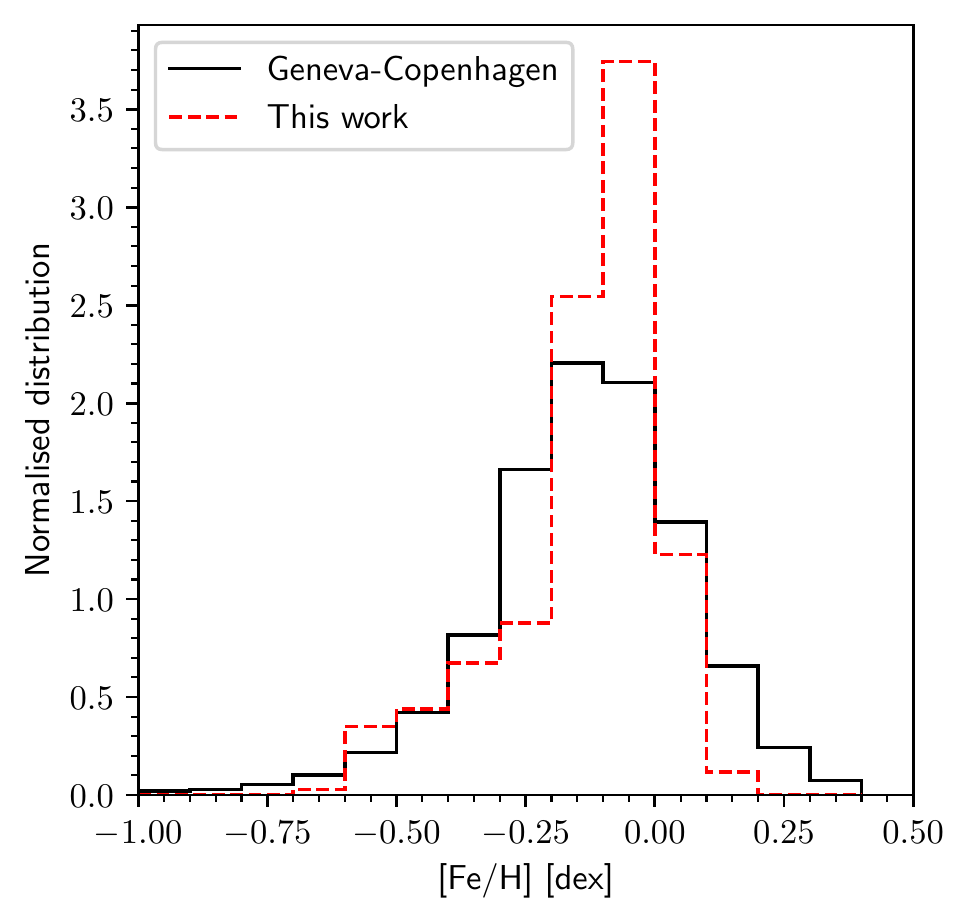}
    \caption{Metallicity distribution in this work and in the Geneva-Copenhagen survey \citep{Nor04}.}
    \label{fig:fehgalaxy}
\end{figure}

\begin{figure}
    \centering
    \includegraphics[width=0.49\textwidth]{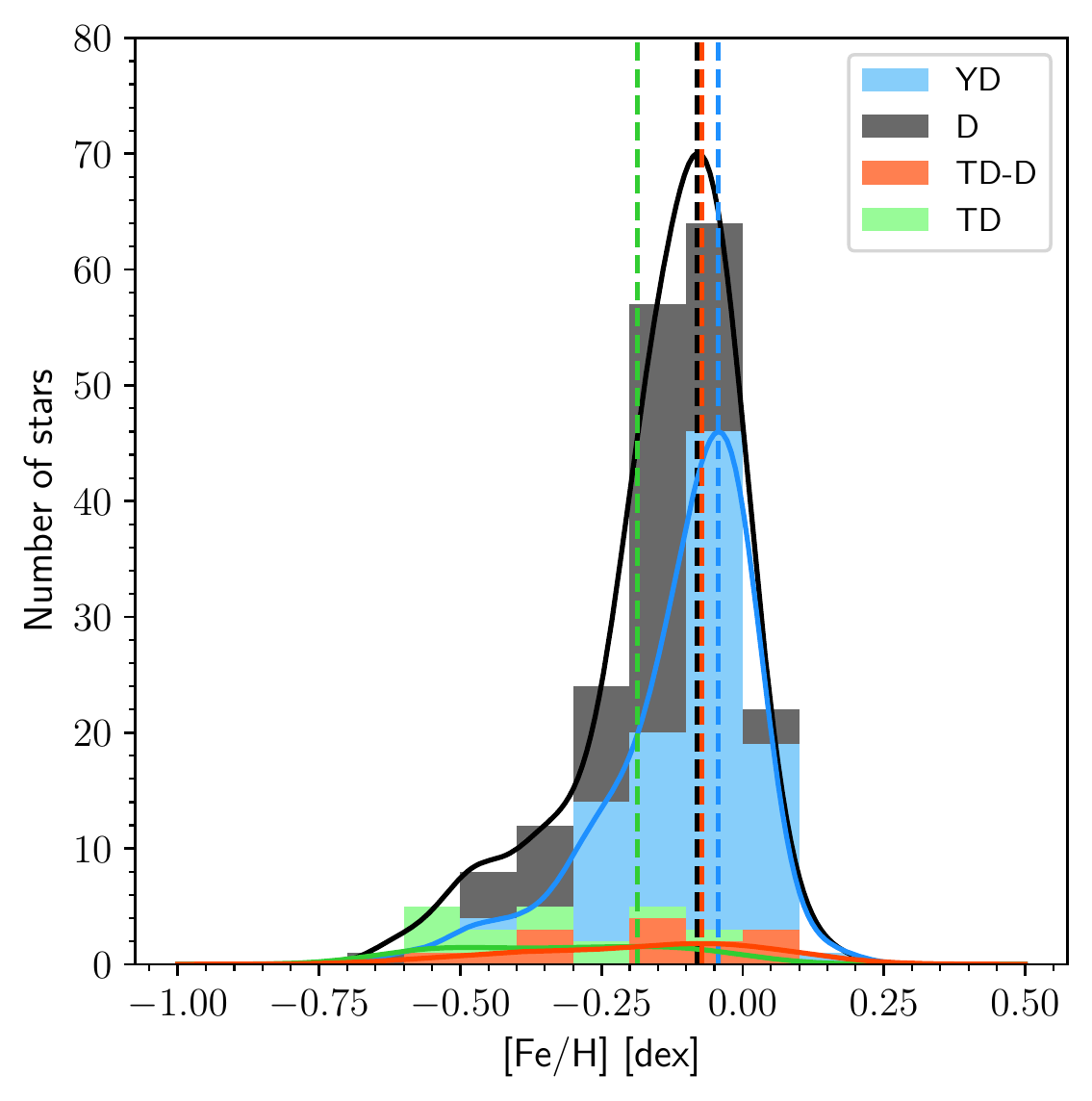}
    \caption{M-dwarf metallicity distributions separated by the kinematic membership of the targets in the thick disc (TD), the thick disc-thin disc transition (TD-D), the thin disc (D), and the young disc (YD), following {\color{blue} Cort\'{e}s-Contreras et al. in prep.} The vertical dashed lines indicate the peak of the distributions.}
    \label{fig:pop}
\end{figure}

The stellar atmospheric parameters of the sample computed with {\sc SteParSyn} ($T_{\rm eff}$, $\log{g}$, [Fe/H]) are given in Table~\ref{tab:par_stars_stepar}. Since some M-dwarf studies do not consider alpha enhancement in the computation of synthetic spectra \citep[see e.g.][]{Pas18}, we corrected our [Fe/H] scale into [Fe/H]$_{\rm corr}$ in order to compare our results more directly with that of the literature. We did this through a linear interpolation scheme via the {\tt scipy} package between the mass fraction of elements heavier than helium ($Z$) and the corresponding [Fe/H] value in alpha-enhanced and alpha-solar (i.e. no alpha enhancement) models, respectively, as shown in Fig.~\ref{fig:alpha}. The correspondence between $Z$ and [Fe/H] was adopted from the MARCS atmosphere models with standard and alpha-poor compositions\footnote{\url{https://marcs.astro.uu.se/}}. The interpolation scheme retrieves the [Fe/H]$_{\rm corr}$ value that corresponds to a given $Z$ defined by the original [Fe/H] scale.

We note the limitations of our methodology with respect to the analysis of M dwarfs with spectral types M8.0\,V or later due to the scarcity of useful \ion{Fe}{i} and \ion{Ti}{i} lines and the poor sensitivity of the TiO molecular bands to $T_{\rm eff}$ as a result of dust formation in the stellar photospheres \citep{Tsu96a, All01}. Furthermore, the version of the BT-Settl models used in this work does not allow for the formation of enough dust in very late M dwarfs \citep{All12}. However, we note that this limit in spectral type is similar to other studies based on spectral synthesis in M dwarfs \citep{Raj18a, Sch19}.

We compared our results against some recent M-dwarf studies, including 
\citet[][Fig.~\ref{fig:referencecomparisons}]{Pas19}, 
\citet[][Fig.~\ref{fig:compRA12}]{Roj12}, 
\citet[][Fig.~\ref{fig:compGM14}]{Gai14}, 
\citet[][Fig.~\ref{fig:compMald15}]{Mal15},
\citet[][Fig.~\ref{fig:compMann15}]{Man15}, 
\citet[][Fig.~\ref{fig:compPas18}]{Pas18},  
\citet[][Fig.~\ref{fig:compRaj18}]{Raj18a}, 
\citet[][Fig.~\ref{fig:compSch19}]{Sch19},
\citet[][Fig.~\ref{fig:compMald20}]{Mal20}, 
and \citet[][Fig.~\ref{fig:compPas20}]{Pas20}. The approaches to the computation of stellar parameters in the M-dwarf regime are manifold and generally dependent on the covered wavelength region $\Delta\lambda$, the spectral resolution $R$, and the S/N of the spectra. For instance, $T_{\rm eff}$ can be derived from the H$_2$O-K2 index \citep{Roj12}, the pseudo-EW of spectral features \citep{Mal15, Mal20}, and spectral fitting to a variety of synthetic grids, including PHOENIX-ACES \citep{Pas18, Sch19}, PHOENIX-SESAM \citep{Pas19}, and grids based on BT-Settl models \citep{Gai14, Man15, Raj18a}. As for $\log{g}$, in some studies it is left as a free parameter in the fitting procedure \citep{Raj18a}, while in others $\log{g}$ is kept fixed instead based on evolutionary models to break the observed degeneracy in the M-dwarf parameter space \citep{Roj12, Pas18, Pas19}. In other cases, $\log{g}$ is derived indirectly from the stellar masses and radii obtained from different approaches, primarily semi-empirical calibrations \citep{Gai14, Man15, Mal15, Mal20, Sch19}. Lastly, stellar metallicity is usually derived from semi-empirically calibrated relations established between the spectral features of M dwarfs that are companions of FGK-type stars with known metallicities \citep{Ter12, Man13a, Man14, Man15, New14, Nev14, Mal15, Mal20}, and the use of weights in the spectral fitting for spectral features that appear to be more sensitive to metallicity variations \citep{Pas16, Pas18, Pas19}.

To explore possible sources of systematic trends or offsets, we followed the same Monte Carlo method as in \citet{Tab18}. We generated 10\,000 artificial samples based on the stellar atmospheric parameters derived with {\sc SteParSyn} by assuming a normal distribution centred at the original measurements, and widths equal to the uncertainties in the parameters. The summary of these Monte Carlo simulations can be found in Table~\ref{tab:mc_par}. We computed the Pearson ($r_{\rm p}$) and the Spearman ($r_{\rm s}$) correlation coefficients, which quantify the degree of correlation between any two given variables. Our temperature scale appears to have an intrinsic systematic offset relative to the literature values in the range from 4\,K to 85\,K. The offset is smallest for \citet{Gai14} and largest for \citet{Man15}. In other words, all literature sources report on average slightly lower effective temperatures. However, we note that the offsets are compatible with zero in all instances but \citet{Man15}. Regarding $\log{g}$, our values appear to be higher (between 0.02 and 0.17\,dex on average with respect to the literature), except for \citet{Raj18a}, although even in that case the discrepancies are compatible with zero. As to metallicity, our [Fe/H] scale seems to agree best with the results by \citet{Roj12}, \citet{Man15}, and \citet{Mal15, Mal20}. In contrast, large discrepancies in metallicity can be seen when comparing our results with \citet{Pas18, Pas19}, and \citet{Sch19}. Among the plausible explanations are the use of different synthetic models, minimisation algorithms, and sets of lines in their analyses, especially \ion{K}{i} lines, which have been shown to be affected by non-LTE effects that can translate into metallicity corrections of the order of 0.2\,dex \citep{Ola21}.

Regarding the comparison between our results and the prior distributions, we find that the spectroscopic $T_{\rm eff}$ values are higher by $136\pm88$\,K, which is in agreement with the findings of \citet{Cif20} when comparing their estimations using photometry and {\tt vosa} to spectroscopic studies, namely \citet{Pas19} and \citet{Raj18a}. This difference is in turn more noticeable towards the mid-type and late M dwarfs, probably due to missing opacity sources. Regarding $\log{g}$, it also seems to be higher by $0.08\pm0.14$\,dex with respect to the prior values.

Similarly to Fig.~\ref{fig:prior}, Fig.~\ref{fig:HRdiagram} is a Kiel diagram of our sample showing the results obtained with {\sc SteParSyn} along with theoretical isochrones from the PAdova and TRieste Stellar Evolution Code \citep[PARSEC;][]{Bre12}. Since M dwarfs evolve very slowly once they reach the main sequence \citep{Bur97}, we selected a set of isochrones with an age of 5\,Ga and different compositions, namely [Fe/H]\,=\,$-$0.4, 0.0, and 0.2\,dex. Active and inactive stars in Fig~\ref{fig:HRdiagram} are shown with different symbols. First, we note that the fraction of active stars increases towards the late M-dwarf regime, as discussed by \citet{Tal18} and \citet{Rei18}. While some stars fall inside the region encompassed by the isochrones, we note that most of them deviate towards higher values of $\log{g}$. However, the trend established by the isochrones is preserved in the retrieved $\log{g}$ values throughout the diagram. Additionally, conspicuous outliers have been labelled and marked with red and blue circles. Red circles denote seven of the ten targets identified as young according to Table~\ref{tab:young}, whereas blue circles indicate other stars that also correspond to higher $\log{g}$ values. We note that most of them are active stars, with the exception of the early M dwarfs BD$+$21~652 and HD~275122. 

In light of such a large dispersion in $\log{g}$, we performed an additional run by fixing $\log{g}$ to the prior values. The results of this run ($T_{\rm eff,\,fixed}$, [Fe/H]$_{\rm fixed}$, and [Fe/H]$_{\rm corr,\,fixed}$) can also be found in Table~\ref{tab:par_stars_stepar}. In Fig.~\ref{fig:comparisonruns} we show the differences in $T_{\rm eff}$ and [Fe/H] between these two runs. The impact of fixing $\log{g}$ on $T_{\rm eff}$ is well below 100\,K for most targets, whereas the impact on [Fe/H] is between 0.1 and 0.2\,dex in most cases. In general terms, the larger the difference in $\log{g}$ between the runs, the larger the change in [Fe/H], as expected.

To check our metallicity scale, in Fig.~\ref{fig:fehgalaxy} we compare the M-dwarf metallicity distribution presented in this work to that from the Geneva-Copenhagen survey \citep{Nor04}, which comprises around 14\,000 F- and G-type dwarfs in the solar neighbourhood, and is volume-complete to a distance of \textasciitilde40\,pc. We note that the two distributions are very similar in the low-metallicity regime, whereas high-metallicity stars appear to be marginally underrepresented. There appears to be an excess of M dwarfs with solar metallicity and we retrieve no metallicity higher than 0.25\,dex. The scarcity of high-metallicity values is also noticeable when comparing our determinations with the literature, as shown in Figs.~\ref{fig:compRA12} \citep{Roj12}, \ref{fig:compGM14} \citep{Gai14}, and Fig.~\ref{fig:compMann15} \citep{Man15}.

We also investigated the M-dwarf metallicity distribution in terms of the kinematic membership of the stars in the Galactic populations, namely the thick disc (TD), the thick disc-thin disc transition (TD-D), the thin disc (D), and the young disc (YD) \citep[][{\color{blue} Cort\'{e}s-Contreras et al. in prep.}]{Mon01}, as shown in Fig.~\ref{fig:pop}. It can be seen that the mean metallicity of M dwarfs belonging to the TD is lower than that of those belonging to the D, in accordance with the results presented by \citet{Ben03, Ben05}.

\subsection{M dwarfs with interferometric angular measurements}

{
\begin{table*}
\centering
\tiny
\caption{\label{tab:teffinterf} M dwarfs with interferometric angular diameter measurements in common with the CARMENES GTO survey.}
\begin{tabular}{llccccccccc}
\hline\hline\noalign{\smallskip}
Name & Karmn & SpT & $R$         & $T_{\rm eff}$ & $L_{\rm bol}$      & $M$         & Ref.\,$^{(a)}$ & $T_{\rm eff}$\,$^{(b)}$ & $\log{g}$\,$^{(b)}$ & [Fe/H]$_{\rm corr}$\,$^{(b)}$ \\
     &       &     & [$R_\odot$] & [K]           & [$10^{-4}L_\odot$] & [$M_\odot$] &                & [K]                     & [dex]               & [dex]            \\
\hline\noalign{\smallskip}
HD~79210       & J09143$+$526  & M0.0\,V & $0.5773\pm0.0131$ & $3907\pm35$ & $697\pm21$    & 0.622 & B12 & $4015\pm16$ & $4.91\pm0.08$ & $-0.12\pm0.05$ \\
HD~79211       & J09144$+$526  & M0.0\,V & $0.5673\pm0.0137$ & $3867\pm37$ & $647\pm19$    & 0.600 & B12 & $3983\pm17$ & $5.17\pm0.07$ & $-0.03\pm0.04$ \\
GX~And         & J00183$+$440  & M1.0\,V & $0.3874\pm0.0023$ & $3563\pm11$ & $217.3\pm2.1$ & 0.423 & B12 & $3603\pm24$ & $4.99\pm0.14$ & $-0.52\pm0.11$ \\
BD$+$44~2051~A & J11054$+$435  & M1.0\,V & $0.3982\pm0.0091$ & $3497\pm39$ & $212.9\pm2.6$ & 0.403 & B12 & $3628\pm19$ & $5.01\pm0.13$ & $-0.56\pm0.09$ \\
BD$+$25~3173   & J16581$+$257  & M1.0\,V & $0.5387\pm0.0157$ & $3590\pm45$ & $432\pm13$    & 0.54  & v14 & $3782\pm17$ & $4.87\pm0.09$ & $-0.05\pm0.05$ \\
HD~199305      & J20533$+$621  & M1.0\,V & $0.5472\pm0.0067$ & $3692\pm22$ & $499.0\pm6.2$ & 0.573 & B12 & $3815\pm16$ & $4.83\pm0.14$ & $-0.10\pm0.08$ \\
HD~36395       & J05314$-$036  & M1.5\,V & $0.5735\pm0.0044$ & $3801\pm9$  & $616.3\pm8.8$ & 0.615 & B12 & $3850\pm27$ & $4.71\pm0.18$ &  $0.05\pm0.08$ \\
Lalande~21185  & J11033$+$359  & M1.5\,V & $0.3921\pm0.0037$ & $3465\pm17$ & $198.9\pm1.2$ & 0.403 & B12 & $3557\pm26$ & $4.95\pm0.14$ & $-0.49\pm0.10$ \\
HD~119850      & J13457$+$148  & M1.5\,V & $0.4840\pm0.0084$ & $3618\pm31$ & $360\pm5$     & 0.520 & B12 & $3620\pm24$ & $4.93\pm0.12$ & $-0.34\pm0.07$ \\
HD~216899      & J22565$+$165  & M1.5\,V & $0.5477\pm0.0048$ & $3713\pm11$ & $511.2\pm7.4$ & 0.569 & B12 & $3798\pm21$ & $4.80\pm0.12$ &  $0.03\pm0.06$ \\
HD~285968      & J04429$+$189  & M2.0\,V & $0.4525\pm0.0221$ & $3679\pm77$ & $337\pm18$    & 0.45  & v14 & $3678\pm16$ & $5.13\pm0.08$ &  $0.00\pm0.03$ \\
Ross~905       & J11421$+$267  & M2.5\,V & $0.4546\pm0.0182$ & $3416\pm53$ & $252.5\pm1.2$ & 0.472 & B12 & $3533\pm26$ & $4.83\pm0.11$ & $-0.12\pm0.10$ \\
HO~Lib         & J15194$-$077  & M3.0\,V & $0.2990\pm0.0100$ & $3442\pm54$ & $113\pm8$     & 0.297 & B12 & $3500\pm26$ & $4.97\pm0.11$ & $-0.08\pm0.07$ \\
BD$+$68~946    & J17364$+$683  & M3.0\,V & $0.4183\pm0.0070$ & $3413\pm28$ & $212.8\pm2.3$ & 0.413 & B12 & $3389\pm81$ & $4.89\pm0.15$ & $-0.06\pm0.13$ \\
HD~173739      & J18427$+$526N & M3.0\,V & $0.3561\pm0.0039$ & $3407\pm15$ & $153.1\pm1.8$ & 0.318 & B12 & $3473\pm34$ & $4.90\pm0.11$ & $-0.31\pm0.12$ \\
Barnard's~star & J17578$+$046  & M3.5\,V & $0.1867\pm0.0012$ & $3224\pm10$ & $33.8\pm0.2$  & 0.150 & B12 & $3254\pm32$ & $5.13\pm0.12$ & $-0.57\pm0.10$ \\
HD~173740      & J18427$+$596S & M3.5\,V & $0.3232\pm0.0061$ & $3104\pm28$ & $87.1\pm1.2$  & 0.235 & B12 & $3393\pm48$ & $4.98\pm0.12$ & $-0.38\pm0.18$ \\
IL~Aqr         & J22532$-$142  & M4.0\,V & $0.3761\pm0.0059$ & $3129\pm19$ & $122\pm2$     & 0.37  & v14 & $3421\pm40$ & $4.83\pm0.10$ & $-0.06\pm0.10$ \\
\hline
\end{tabular}
\tablefoot{$^{(a)}$Stellar radii, effective temperatures, luminosities, and stellar masses from \citet[B12]{Boy12} and \citet[v14]{vBra14}. The estimates of the stellar mass from these works conservatively assume 10\% and 5\% uncertainties, respectively. $^{(b)}$This work.}
\end{table*}
}

\begin{figure}
    \centering
    \includegraphics[width=0.49\textwidth]{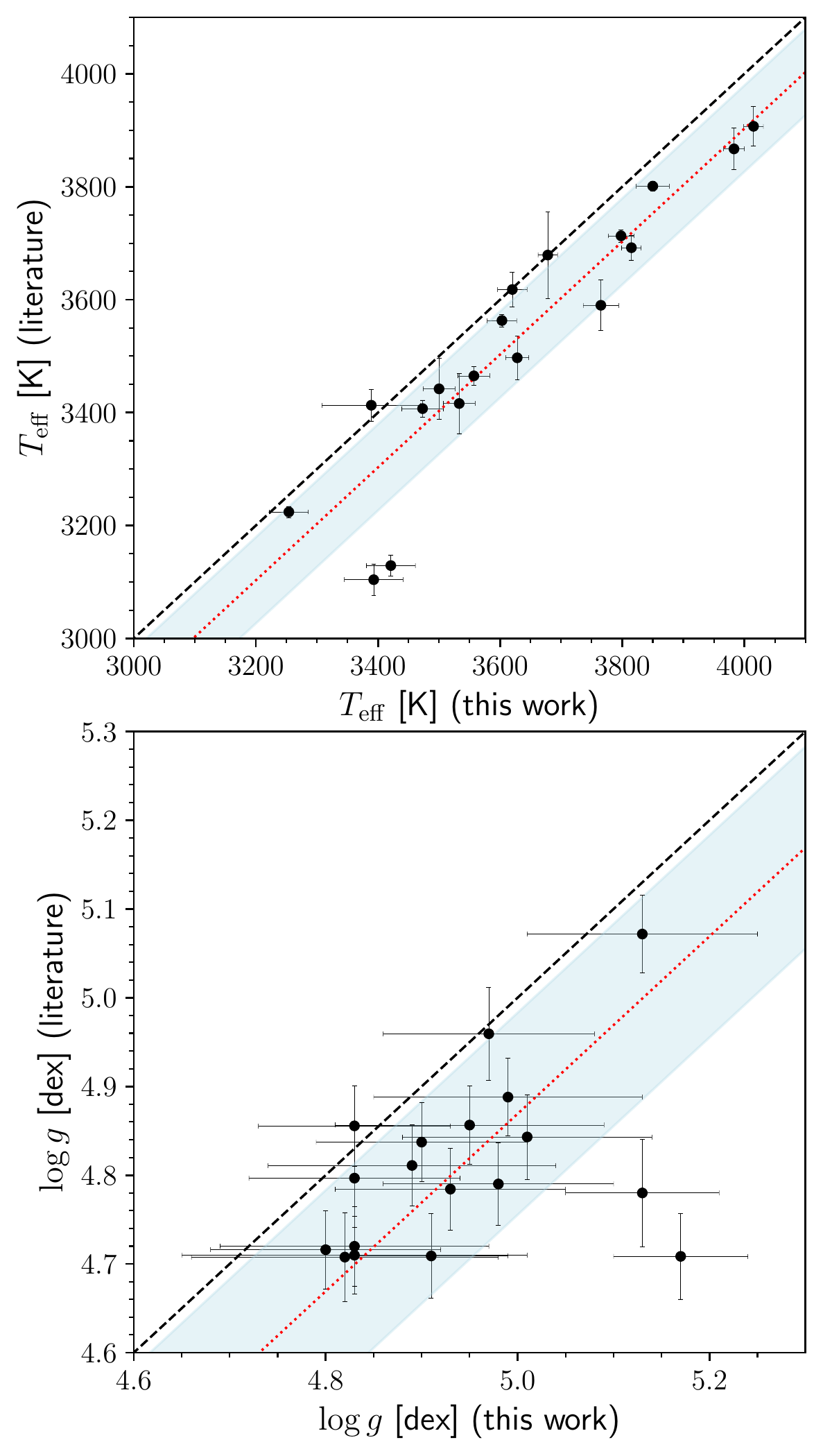}
    \caption{Comparison in $T_{\rm eff}$ ({\it upper panel}) and $\log{g}$ ({\it bottom panel}) for M dwarfs with interferometric angular measurements (see Table~\ref{tab:teffinterf}). The dashed black lines indicate the 1:1 relationship, whereas the dotted red lines show the 1:1 relationship, shifted following the average differences in each parameter. The blue shaded region denotes the 1$\sigma$ level.}
    \label{fig:interf}
\end{figure}

In Table~\ref{tab:teffinterf} we compile the 18 M dwarfs in our sample with interferometrically measured radii from \citet{Boy12} and \citet{vBra14}, which may be used along with other information (e.g. trigonometric parallax, bolometric flux, isochrone fitting, and mass-luminosity relations) to derive $T_{\rm eff}$ and $\log{g}$ independently from spectroscopy. For this reason, this set of stars is particularly helpful as a benchmark test for any method aimed at determining stellar atmospheric parameters \citep{Cas08, Man15, Bir20}.

Figure~\ref{fig:interf} shows the comparison in $T_{\rm eff}$ and $\log{g}$ for these M dwarfs between \citet{Boy12}, \citet{vBra14}, and this work. The $\log{g}$ value was computed from the stellar mass and radius using the standard definition (i.e. $g=GM/R^2$, where $G$ is the gravitational constant). In general, we find good agreement, with systematic offsets of $72\pm67$\,K in $T_{\rm eff}$, and $0.10\pm0.09$\,dex in $\log{g}$. However, we note two outliers in $T_{\rm eff}$, namely Il~Aqr and HD~173740, which show a difference around 300\,K between the interferometric and the spectroscopic approaches. Interestingly enough, HD~173740 also belongs to a wide binary system together with HD~173739, as discussed in Sect.~\ref{subsec:mmsystems}. Although they exhibit similar chemical compositions and differ by only one spectral subtype, the $T_{\rm eff}$ derived from interferometry appears to be irreconcilable between the two components. In contrast, we derived very similar $T_{\rm eff}$ with {\sc SteParSyn} for both components. However, accounting for this difference is beyond the scope of this paper.

\subsection{M dwarfs in FGK+M systems}

\begin{table*}
\centering
\caption{\label{tab:mcomp} M dwarfs in wide physical binaries with FGK-type primaries in common with the CARMENES GTO survey.}
\begin{tabular}{llclllccc}
\hline\hline\noalign{\smallskip}
\multicolumn{3}{c}{Primary} & \multicolumn{6}{c}{Secondary} \\
\cmidrule(lr){1-3} \cmidrule(lr){4-9}
Name & SpT & [Fe/H]\,[dex] & Name & Karmn & SpT & $T_{\rm eff}$\,[K] & $\log{g}$\,[dex] & [Fe/H]$_{\rm corr}$\,[dex] \\
\hline\noalign{\smallskip}
V538 Aur       & K1\,V   &  $0.04\pm0.02$ & HD~233153      & J05415$+$534 & M1.0\,V & $3825\pm14$ & $4.94\pm0.10$ & $-0.02\pm0.06$ \\
V869~Mon       & K0/2\,V & $-0.11\pm0.03$ & GJ~282~C       & J07361$-$031 & M1.0\,V & $3825\pm12$ & $5.06\pm0.08$ & $-0.04\pm0.06$ \\
HD~154363          & K5\,V   & $-0.62\pm0.05$ & HD~154363~B    & J17052$-$050 & M1.5\,V & $3587\pm14$ & $4.89\pm0.10$ & $-0.39\pm0.07$ \\
$\theta$~Boo~A & F7\,V   & $-0.09\pm0.01$ & $\theta$~Boo~B & J14251$+$518 & M2.5\,V & $3551\pm22$ & $4.90\pm0.10$ & $-0.14\pm0.06$ \\
HD~16160       & K3\,V   & $-0.20\pm0.02$ & BX~Cet         & J02362$+$068 & M4.0\,V & $3335\pm45$ & $4.91\pm0.10$ & $-0.24\pm0.12$ \\
$o^{02}$~Eri~A & K0.5\,V & $-0.37\pm0.02$ & $o^{02}$~Eri~C & J04153$-$076 & M4.5\,V & $3179\pm61$ & $5.00\pm0.18$ & $-0.30\pm0.17$ \\
$\rho$~Cnc     & G8\,V   &  $0.29\pm0.04$ & $\rho$~Cnc~B   & J08526$+$283 & M4.5\,V & $3321\pm37$ & $4.87\pm0.08$ & $-0.10\pm0.11$ \\
\hline
\end{tabular}
\tablefoot{Metallicities of the FGK-type primary stars from \citet{Mon18}, except for $\theta$~Boo~A, which is from \citet{Tab20xx}. Stellar atmospheric parameters of the M-dwarf secondaries ($T_{\rm eff}$, $\log{g}$, and [Fe/H]) as computed in this work.}
\end{table*}

\begin{figure}
    \centering
    \includegraphics[width=0.49\textwidth]{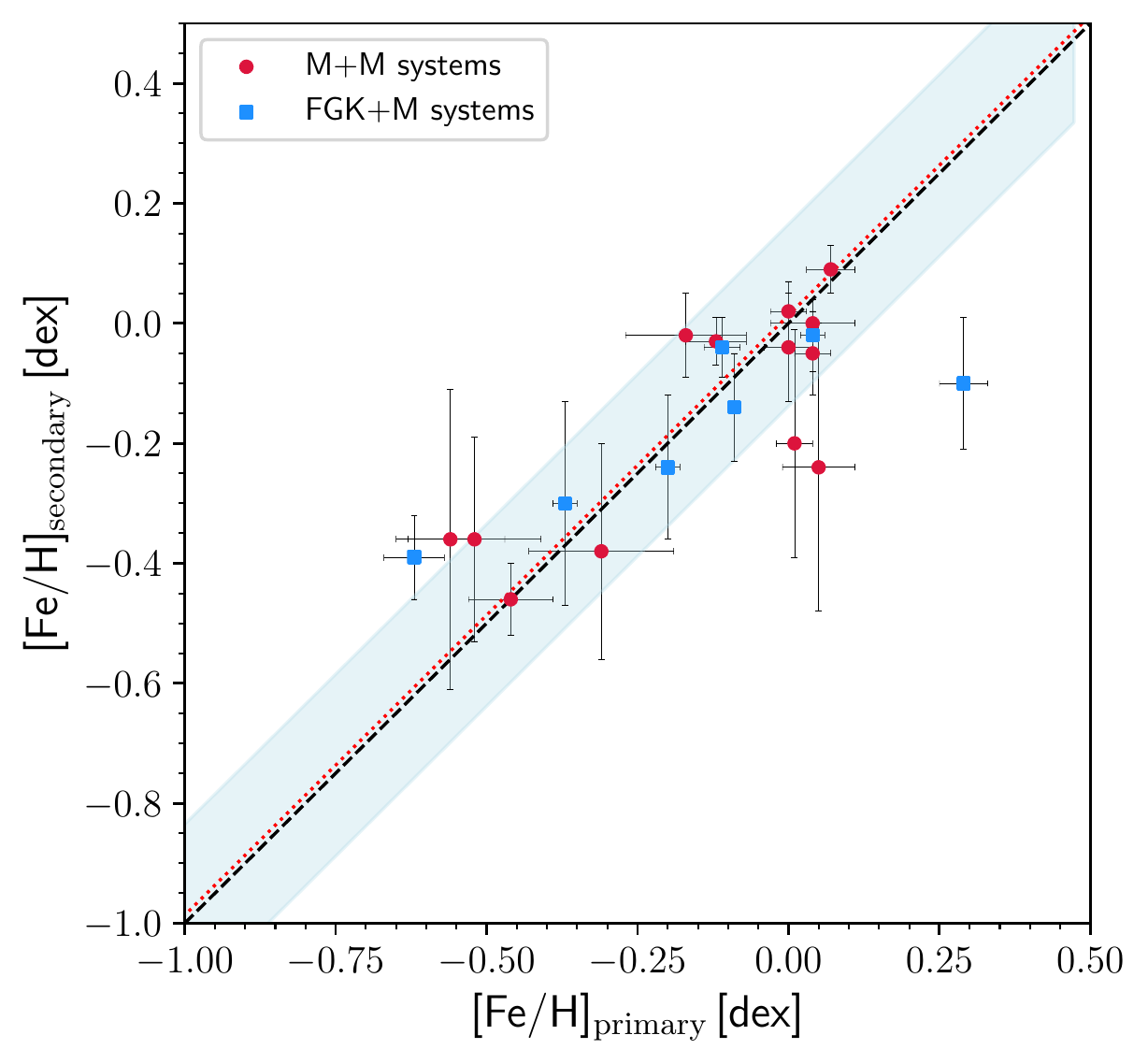}
    \caption{Same as Fig.~\ref{fig:interf}, but for the comparison in [Fe/H] between the components in FGK+M (blue) and M+M (red) systems (see Tables~\ref{tab:mcomp} and \ref{tab:mmsystems2}).}
    \label{fig:mbinary}
\end{figure}

The study of multi-star systems serves as a benchmark test for [Fe/H] given that all components are thought to be formed from the same molecular cloud, be coeval, and have roughly similar chemical compositions \citep{Des06, One12, Sou20, Tak20}. In Table~\ref{tab:mcomp} we compile all M dwarfs in our sample that are bound in wide physical binaries with an FGK-type primary \citep{Mon18}. For the FGK-type primaries, we adopted the metallicities from \citet{Mon18}, who employed the {\sc StePar} code \citep{Tab19} as the preferred implementation of the EW method. For the fast-rotating F7\,V star $\theta$~Boo~A \citep[{$\varv\sin{i}=30.4$\,km\,s$^{-1}$,}][]{Luc17}, we adopted the metallicity from \citet{Tab20xx}.

In Fig.~\ref{fig:mbinary} we show the comparison in [Fe/H] between the components in these FGK+M systems. We find good agreement between our [Fe/H] determinations for the M-dwarf secondaries, with an internal scatter of $-0.01\pm0.19$~dex, which is similar to what is found in other works \citep{Sou20, Ant20}. The largest discrepancy in metallicity between components is found for $\rho$~Cnc~B. If we assume a homogeneous chemical composition in FGK+M systems, this star also exhibits a C/O ratio that is larger than expected for an M dwarf, since ${\rm C/O}=1.12$ for the primary star \citep{Del10}. We note that the pseudo-continuum level is very sensitive to C/O and alterations in C/O can explain a large fraction of the variation in the strength of the atomic metal lines in M-dwarf spectra \citep{Vey16, Sou20}. However, disentangling the effect of C/O on [Fe/H] computations is beyond the scope of this paper.

\subsection{M dwarfs in M+M systems}
\label{subsec:mmsystems}

Similarly to FGK+M systems, wide physical M+M binaries can also prove useful as a benchmark test for [Fe/H]. In this case, both components are analysed with the same methodology (i.e. spectral synthesis). In Table~\ref{tab:mmsystems2} we list the astrometric parameters (parallax $\pi$, proper motion $\mu$, and radial velocity $\varv_{\rm r}$) along with the stellar atmospheric parameters ($T_{\rm eff}$, $\log{g}$, and [Fe/H]) of the 32 M dwarfs in our sample bound in M+M systems. These include two SB2 systems (EZ~Psc and GJ~810) and the M8.0\,V star V1298~Aql (vB~10), for which no reliable parameter determinations could be made (see Table~\ref{tab:excluded}). Parallaxes and proper motions were compiled from the {\it Gaia} DR2 and EDR3 releases \citep{GaiaDR2, GaiaDR3}. 

In Fig.~\ref{fig:mbinary} we compare the metallicity computed in this work for both components, in the same way as for the FGK+M wide binaries. The mean difference in [Fe/H] between components is $0.10\pm0.08$~dex, which is in agreement with other works \citep[$0.06\pm0.07$\,dex,][]{Ant20}. However, we note that the differences become larger (i.e. up to \textasciitilde0.2\,dex) in cases where the components differ by more than one spectral type.

\subsection{Choice of model atmospheres}
\label{subsec:grids}
\begin{table*}
    \centering\
    \caption{Parameters derived for the reference stars with different synthetic grids.}
    \label{tab:models}
    \begin{tabular}{llccc}
    \hline\hline\noalign{\smallskip}
    Reference star & Grid & $T_{\rm eff}$ & $\log{g}$ & [Fe/H]$_{\rm corr}$ \\
         &       & [K]           & [dex]     & [dex]  \\
    \hline\noalign{\smallskip}
                     & BT-Settl     & $3603\pm24$ & $4.99\pm0.14$ & $-0.52\pm0.11$ \\
    GX~And           & MARCS        & $3675\pm30$ & $4.90\pm0.10$ & $-0.45\pm0.08$ \\
                     & PHOENIX-ACES & $3664\pm37$ & $4.84\pm0.12$ & $-0.54\pm0.13$ \\
    \noalign{\smallskip}
                     & BT-Settl     & $3380\pm43$ & $4.96\pm0.11$ & $-0.11\pm0.11$ \\
    Luyten's~star    & MARCS        & $3429\pm38$ & $4.88\pm0.10$ & $-0.08\pm0.09$ \\
                     & PHOENIX-ACES & $3412\pm50$ & $4.81\pm0.14$ & $-0.19\pm0.12$ \\
    \noalign{\smallskip}
                     & BT-Settl     & $3034\pm45$ & $5.19\pm0.20$ & $-0.11\pm0.28$ \\
    Teegarden's~star & MARCS        & $3072\pm51$ & $5.08\pm0.16$ & $-0.13\pm0.24$ \\
                     & PHOENIX-ACES & $3081\pm57$ & $5.10\pm0.17$ & $-0.24\pm0.25$ \\
    \hline
    \end{tabular}
\end{table*}

Recent studies have shown that different parameter determination methods can lead to significant variations in the results for the same stars \citep{Bla19, Pas20xx}. These differences can be attributed to a number of factors, including the choice of model atmospheres, atomic and molecular data, and radiative transfer code. For this reason, we performed a test on the reference stars (i.e. GX~And, Luyten's star, and Teegarden's star) with the aim of evaluating the impact of different sets of models on our parameter computations with {\sc SteParSyn}. 

We first computed a synthetic grid based on MARCS models by following the same procedure as described in Sect.~\ref{subsec:grid}. We set the MARCS grid from 2500\,K to 3900\,K, 4.0\,dex to 5.5\,dex, and $-$1.0\,dex to $+$1.0\,dex, in steps of 100\,K, 0.5\,dex, and 0.25\,dex in $T_{\rm eff}$, $\log{g}$, and [Fe/H], respectively. Similarly, we adopted the PHOENIX-ACES grid used by \citet{Pas20xx}, which ranges from 2700\,K to 4500\,K, 4.2\,dex to 5.5\,dex, and $-$1.0\,dex to $+$0.8\,dex, in steps of 100\,K, 0.1\,dex, and 0.1\,dex in $T_{\rm eff}$, $\log{g}$, and [Fe/H], respectively. While the MARCS grid is based on atmosphere models with standard composition and, thus, follows the same scaled $\rm [\alpha/Fe]$ pattern as the BT-Settl grid, the PHOENIX-ACES grid has $\rm [\alpha/Fe]=0$ since models with $\rm [\alpha/Fe]\neq0$ are only available for $T_{\rm eff}>3500$\,K \citep{Hus13, Pas18}. Therefore, metallicities derived with the PHOENIX-ACES grid do not need to be corrected with the interpolation scheme described in Sect.~\ref{sec:discussion}.

In Table~\ref{tab:models} we list the parameters obtained for the reference stars and the considered grids (i.e. BT-Settl, MARCS, and PHOENIX-ACES). The derived $\log{g}$ and [Fe/H] values agree within the uncertainties for all models, while the average differences in $T_{\rm eff}$, $\log{g}$, and [Fe/H] between models are around 32\,K, 0.11\,dex, and 0.06\,dex, respectively. In addition, the derived $T_{\rm eff}$ values appear to be slightly higher when using MARCS and PHOENIX-ACES models. In contrast, $\log{g}$ values are slightly lower. The differences in these results might arise from the synthetic gap \citep{Pas20}, that is, the fact that synthetic spectra do not perfectly match the observed spectra as a result of the underlying chemo-physical assumptions (e.g. different equations of state). However, a more detailed study is required to further explore the origin of these small discrepancies.

\section{Summary and conclusions}\label{sec:summary}

We have computed the stellar atmospheric parameters ($T_{\rm eff}$, $\log{g}$, [Fe/H], and [Fe/H]$_{\rm corr}$) of 343 M dwarfs observed with CARMENES with the {\sc SteParSyn} code as a Bayesian implementation of the spectral synthesis technique, along with a grid of BT-Settl models, and the radiative transfer code {\tt turbospectrum}. We excluded one eclipsing binary, ten SB2 binaries, two triple-line spectroscopic triple systems, and three targets with low-quality spectra to prevent unreliable parameter determinations. To avoid any potential degeneracy in the M-dwarf parameter space, we imposed prior probability distributions in $T_{\rm eff}$ and $\log{g}$ based on the comprehensive, multi-band photometric data available for the sample \citep{Cif20}. Our method is suited for the M sequence from M0.0\,V to M7.0\,V, but not beyond, due to the scarcity of \ion{Ti}{i} and \ion{Fe}{i} lines at M8.0\,V and later spectral types, as well as the insensitivity of TiO bands to $T_{\rm eff}$ as a result of dust formation \citep{Tsu96a, All01}.

We selected 75 \ion{Fe}{i} and \ion{Ti}{i} lines in the VIS and NIR wavelength regions covered by CARMENES. Following the prescriptions of \citet{Pas19}, NIR lines with low Landé factors (i.e. $g_{\rm eff}<1.5$) were selected to minimise the impact of the stellar magnetic field on parameter computations. Even so, some \ion{Fe}{i} and \ion{Ti}{i} lines appear to be highly sensitive to chromospheric activity and the stellar magnetic field and were excluded in the analysis of the active stars in our sample based on the H$\alpha$ flag. Such lines will be discussed in a future publication ({\color{blue} L\'{o}pez-Gallifa et al. in prep.}).

We also compared our derived parameters with some recent M-dwarf studies in the literature \citep{Roj12, Mal15, Mal20, Man15,  Raj18a, Sch19, Pas18, Pas19, Pas20}, finding similar $T_{\rm eff}$ but disparate metallicity scales. The metallicity determinations that correlate best with the literature are the [Fe/H]$_{\rm corr}$ values since they are corrected from the $\alpha$ enhancement of the synthetic grid. The best agreement in metallicity is found when comparing with \citet{Man15}, \citet{Gai14}, and \citet{Roj12}. The M-dwarf metallicity distribution for our sample is also in agreement with both the Geneva-Copenhagen survey of FG-type stars in the solar neighbourhood \citep{Nor04} and the kinematic membership of the targets in the Galactic populations ({\color{blue} Cortés-Contreras et al. in prep.}). Regarding $\log{g}$, we find larger values than the ones reported in the literature, although fixing $\log{g}$ to the prior values does not have a strong impact on either the derived $T_{\rm eff}$ or the [Fe/H] scales. As a benchmark test, we placed special emphasis on the M dwarfs with interferometrically measured radii \citep{Boy12, vBra14}, wide physical FGK+M binaries \citep{Mon18}, and M+M systems. Despite systematic offsets that are inherent to any methodology, we find that our parameter determinations are fairly consistent with the literature values in all these cases.

\begin{acknowledgements}
We thank the anonymous referee for the insightful comments and suggestions that improved the manuscript of the paper. CARMENES is an instrument for the Centro Astron\'{o}mico Hispano en Andaluc\'{i}a at Calar Alto (CAHA). CARMENES is funded by the German Max-Plank Gesellschaft (MPG), the Spanish Consejo Superior de Investigaciones Cient\'{i}ficas (CSIC), the European Union through FEDER/ERF FICTS-2011-02 funds, and the members of the CARMENES Consortium (Max-Plank-Institut f\"{u}r Astronomie, Instituto de Astrof\'{i}sica de Andaluc\'{i}a, Landessternwarte K\"{o}nigstuhl, Institut de Ci\`{e}nces de l'Espai, Institut f\"{u}r Astrophysik G\"{o}ttingen, Universidad Complutense de Madrid, Th\"{u}ringer Landessternwarte Tautenberg, Instituto de Astrof\'{i}sica de Canarias, Hamburger Sternwarte, Centro de Astrobiolog\'{i}a and Centro Astron\'{o}mico Hispano-Andaluz), with additional contributions by the Spanish Ministry of Economy, the German Science Foundation through the Major Research Instrumentation Programme and DFG Research Unit FOR2544 "Blue Planets around Red Stars", the Klaus Tschira Stiftung, the states of Baden-W\"{u}rttemberg and Niedersachsen, and by the Junta de Andaluc\'{i}a. The authors acknowledge financial support from the Funda\c{c}\~{a}o para a Ci\^{e}ncia e a Tecnologia (FCT) through the research grants UID/FIS/04434/2019, UIDB/04434/2020 and UIDP/04434/2020, national funds PTDC/FIS-AST/28953/2017, by FEDER (Fundo Europeu de Desenvolvimento Regional) through COMPETE2020 - Programa Operacional Competitividade e Internacionaliza\c{c}\~{a}o (POCI-01-0145-FEDER-028953), and the Spanish {Ministerio de Ciencia, Innovaci\'{o}n y Universidades}, {Ministerio de Econom\'ia y Competitividad}, {the Universidad Complutense de Madrid}, and the Fondo Europeo de Desarrollo Regional (FEDER/ERF) through fellowship FPU15/01476, and projects AYA2016-79425-C3-1/2/3-P, PID2019-109522GB-C5[1:4]/AEI/10.13039/501100011033, AYA2014-56359-P, BES-2017-080769, and RYC-2013-14875. The authors also acknowledge financial support from the Centre of Excellence "Severo Ochoa" and "Mar\'{i}a de Maeztu" awards to the Instituto de Astrof\'{i}sica de Canarias (SEV-2015-0548), Instituto de Astrof\'{i}sica de Andaluc\'{i}a (SEV-2017-0709), and Centro de Astrobiolog\'{i}a (MDM-2017-0737), and the Generalitat de Catalunya/CERCA programme. This work has made use of the VALD database, operated at Uppsala University, the Institute of Astronomy RAS in Moscow, and the University of Vienna, and of data from the European Space Agency (ESA) mission {\it Gaia} (\url{https://www.cosmos.esa.int/gaia}), processed by the {\it Gaia} Data Processing and Analysis Consortium (DPAC, \url{https://www.cosmos.esa.int/web/gaia/dpac/consortium}). Funding for the DPAC has been provided by national institutions, in particular the institutions participating in the {\it Gaia} Multilateral Agreement. V.~M.~P. acknowledges financial support from NASA through grant NNX17AG24G. S.~V.~J. acknowledges the support of the DFG priority program SPP 1992 “Exploring the Diversity of Extrasolar Planets" (JE 701/5-1). E.~M. would also like to warmly thank the staff at the Hamburger Sternwarte for their hospitality during his stay funded by project EST18/00162. Based on data from the CARMENES data archive at CAB (INTA-CSIC).
\end{acknowledgements}

\begin{sidewaystable*}
\centering
\caption{Astrometric and stellar atmospheric parameters of the M dwarfs bound in M+M systems in the CARMENES GTO sample.}
\label{tab:mmsystems2}
\begin{tabular}{lllccccccc}
\hline\hline\noalign{\smallskip}
Karmn & Name & SpT & $\pi$ & $\mu_{\alpha}\cos{\delta}$ & $\mu_{\delta}$  & $\varv_{\rm r}$       & $T_{\rm eff}$ & $\log{g}$ & [Fe/H]$_{\rm corr}$ \\
      &      &     & [mas] & [mas\,a$^{-1}$]            & [mas\,a$^{-1}$] & [km\,s$^{-1}$] & [K]           & [dex]     & [dex] \\
\hline\noalign{\smallskip}
J00162$+$198E & LP~404$-$062         & M4.0\,V &  $65.05\pm0.04$ &   $708.14\pm0.05$ &  $-748.88\pm0.04$ &   $-1.62\pm0.02$ & $3329\pm30$ & $4.93\pm0.07$ & $-0.11\pm0.09$ \\
J00162$+$198W & EZ~Psc~AB\,$^{(a)}$  & M4.0\,V &  $65.11\pm0.04$ &   $714.64\pm0.05$ &  $-761.97\pm0.04$ &   $-0.50\pm0.04$ & \ldots      & \ldots        & \ldots         \\
\noalign{\smallskip}
J00183$+$440  & GX~And               & M1.0\,V & $280.71\pm0.02$ &  $2891.52\pm0.02$ &   $411.83\pm0.01$ &  $+11.48\pm0.02$ & $3603\pm24$ & $4.99\pm0.14$ & $-0.52\pm0.11$ \\
J00184$+$440  & GQ~And               & M3.5\,V & $280.69\pm0.03$ &  $2862.80\pm0.02$ &   $336.43\pm0.02$ &  $+10.68\pm0.03$ & $3318\pm53$ & $5.20\pm0.11$ & $-0.36\pm0.17$ \\
\noalign{\smallskip}
J01026$+$623  & BD+61~195            & M1.5\,V & $101.42\pm0.02$ &   $731.09\pm0.01$ &    $90.53\pm0.02$ &   $-6.30\pm0.02$ & $3791\pm19$ & $4.76\pm0.11$ &  $0.05\pm0.06$ \\
J01033$+$623  & V388~Cas             & M5.0\,V & $101.37\pm0.05$ &   $730.40\pm0.04$ &    $85.97\pm0.05$ &   $-6.52\pm0.08$ & $3057\pm49$ & $5.12\pm0.18$ & $-0.24\pm0.24$ \\
\noalign{\smallskip}
J02489$-$145W & PM~J02489$-$1432W    & M2.0\,V &  $26.57\pm0.02$ &   $164.07\pm0.03$ &    $46.55\pm0.03$ &  $+32.27\pm0.06$ & $3655\pm25$ & $4.98\pm0.10$ &  $0.04\pm0.05$ \\
J02489$-$145E & PM~J02489$-$1432E    & M2.5\,V &  $26.54\pm0.02$ &   $174.43\pm0.03$ &    $45.47\pm0.03$ &  $+32.40\pm0.06$ & $3572\pm27$ & $4.94\pm0.12$ &  $0.00\pm0.04$ \\
\noalign{\smallskip}
J05365$+$113  & V2689~Ori            & M0.0\,V &  $87.53\pm0.02$ &    $-2.82\pm0.03$ &   $-56.35\pm0.02$ &  $+21.51\pm0.02$ & $4067\pm14$ & $5.04\pm0.07$ &  $0.01\pm0.03$ \\
J05366$+$112  & PM~J05366$+$1117     & M4.0\,V &  $87.35\pm0.03$ &    $-3.32\pm0.04$ &   $-60.84\pm0.03$ &  $+21.53\pm0.02$ & $3355\pm23$ & $5.17\pm0.16$ & $-0.20\pm0.10$ \\
\noalign{\smallskip}
J09143$+$526  & HD~79210             & M0.0\,V & $157.89\pm0.02$ & $-1545.79\pm0.02$ &  $-569.05\pm0.02$ &  $+10.69\pm0.02$ & $4015\pm16$ & $4.91\pm0.08$ & $-0.12\pm0.05$ \\
J09144$+$526  & HD~79211             & M0.0\,V & $157.88\pm0.02$ & $-1573.04\pm0.02$ &  $-659.91\pm0.02$ &  $+11.90\pm0.02$ & $3983\pm12$ & $5.17\pm0.07$ & $-0.03\pm0.04$ \\
\noalign{\smallskip}
J09425$+$700  & GJ~360               & M2.0\,V &  $84.33\pm0.02$ &  $-671.78\pm0.02$ &  $-271.16\pm0.02$ &   $+6.70\pm0.02$ & $3547\pm23$ & $5.02\pm0.12$ &  $0.00\pm0.03$ \\
J09428$+$700  & GJ~362               & M3.0\,V &  $84.30\pm0.02$ &  $-670.96\pm0.02$ &  $-265.46\pm0.02$ &   $+6.35\pm0.03$ & $3504\pm30$ & $5.06\pm0.12$ &  $0.02\pm0.05$ \\
\noalign{\smallskip}
J11054$+$435  & BD+44~2051~A         & M1.0\,V & $203.89\pm0.03$ & $-4406.47\pm0.03$ &   $938.53\pm0.03$ &  $+68.58\pm0.02$ & $3628\pm19$ & $5.01\pm0.13$ & $-0.56\pm0.09$ \\
J11055$+$435  & WX~UMa               & M5.5\,V & $203.83\pm0.05$ & $-4339.85\pm0.05$ &   $960.70\pm0.04$ &  $+69.16\pm0.02$ & $3278\pm86$ & $5.25\pm0.20$ & $-0.36\pm0.25$ \\
\noalign{\smallskip}
J11110$+$304E & HD~97101~A           & K7.0\,V &  $84.18\pm0.03$ &   $591.60\pm0.02$ &  $-197.12\pm0.03$ &  $-16.00\pm0.03$ & $4211\pm13$ & $4.98\pm0.07$ &  $0.04\pm0.03$ \\
J11110$+$304W & HD~97101~B           & M2.0\,V &  $84.16\pm0.02$ &   $604.91\pm0.02$ &  $-206.05\pm0.02$ &  $-15.69\pm0.02$ & $3730\pm20$ & $4.78\pm0.13$ & $-0.05\pm0.07$ \\
\noalign{\smallskip}
J14257$+$236W & BD+24~2733~A         & M0.0\,V &  $61.24\pm0.02$ &   $792.55\pm0.01$ & $-1116.41\pm0.02$ &   $+8.88\pm0.02$ & $3985\pm13$ & $4.89\pm0.08$ &  $0.07\pm0.04$ \\
J14257$+$236E & BD+24~2733~B         & M0.5\,V &  $61.20\pm0.02$ &   $793.44\pm0.01$ & $-1118.91\pm0.02$ &   $+7.96\pm0.02$ & $3933\pm12$ & $4.71\pm0.11$ &  $0.09\pm0.04$ \\
\noalign{\smallskip}
J16167$+$672S & HD~147379            & M0.0\,V &  $92.88\pm0.01$ &  $-497.92\pm0.02$ &    $84.05\pm0.02$ &  $-19.36\pm0.02$ & $4034\pm17$ & $4.78\pm0.09$ &  $0.00\pm0.04$ \\
J16167$+$672N & EW~Dra               & M3.0\,V &  $92.90\pm0.02$ &  $-483.01\pm0.02$ &    $89.05\pm0.02$ &  $-18.78\pm0.03$ & $3569\pm32$ & $4.97\pm0.11$ & $-0.02\pm0.06$ \\
\noalign{\smallskip}
J16554$-$083N & GJ~643               & M3.5\,V & $153.88\pm0.05$ &  $-817.58\pm0.05$ &  $-898.60\pm0.04$ &  $+15.50\pm0.03$ & $3397\pm33$ & $5.07\pm0.12$ & $-0.17\pm0.10$ \\
J16555$-$083  & vB~8                 & M7.0\,V & $153.97\pm0.06$ &  $-813.04\pm0.06$ &  $-870.61\pm0.04$ &  $+14.41\pm0.03$ & $3005\pm21$ & $5.25\pm0.18$ & $-0.04\pm0.09$ \\
\noalign{\smallskip}
J18427$+$596N & HD~173739            & M3.0\,V & $283.84\pm0.02$ & $-1311.68\pm0.03$ &  $1792.33\pm0.03$ &   $-0.75\pm0.08$ & $3473\pm34$ & $4.90\pm0.11$ & $-0.31\pm0.12$ \\
J18427$+$596S & HD~173740            & M3.5\,V & $283.84\pm0.03$ & $-1400.26\pm0.04$ &  $1862.53\pm0.03$ &   $+1.02\pm0.08$ & $3493\pm48$ & $4.98\pm0.12$ & $-0.38\pm0.18$ \\
\noalign{\smallskip}
J19070$+$208  & Ross~730             & M2.0\,V & $113.25\pm0.03$ &  $-478.27\pm0.02$ &  $-349.09\pm0.03$ &  $+32.06\pm0.02$ & $3543\pm21$ & $5.03\pm0.12$ & $-0.46\pm0.07$ \\
J19072$+$208  & HD~349726            & M2.0\,V & $113.22\pm0.02$ &  $-480.75\pm0.02$ &  $-332.50\pm0.02$ &  $+31.83\pm0.02$ & $3558\pm19$ & $5.06\pm0.10$ & $-0.46\pm0.06$ \\
\noalign{\smallskip}
J19169$+$051N & V1428~Aql            & M2.5\,V & $169.06\pm0.02$ &  $-579.08\pm0.03$ & $-1332.87\pm0.02$ &  $+35.61\pm0.03$ & $3575\pm25$ & $4.88\pm0.12$ & $-0.08\pm0.08$ \\
J19169$+$051S & V1298~Aql$^{(b)}$    & M8.0\,V & $168.95\pm0.07$ &  $-598.76\pm0.07$ & $-1366.06\pm0.07$ &  $+35.73\pm0.03$ & \ldots      & \ldots        & \ldots         \\
\noalign{\smallskip}
J20556$-$140N & GJ~810~Aab\,$^{(a)}$ & M4.0\,V &  $84.86\pm0.47$ &  $1416.62\pm0.44$ &  $-472.38\pm0.33$ & $-142.10\pm0.02$ & \ldots      & \ldots        & \ldots         \\
J20556$-$140S & GJ~810~B             & M5.0\,V &  $80.20\pm0.04$ &  $1420.76\pm0.04$ &  $-472.27\pm0.03$ & $-142.04\pm0.02$ & $3193\pm23$ & $4.98\pm0.10$ & $-0.36\pm0.11$ \\
\hline\noalign{\smallskip}
\end{tabular}
\tablefoot{Parallaxes ($\pi$) and proper motions ($\mu_{\alpha}\cos{\delta}$, $\mu_{\delta}$) from {\it Gaia} DR2 and DR3 \citep{GaiaDR2, GaiaDR3}. Radial velocities ($\varv_{\rm r}$) from \citet{Laf20}, except for J00162$+$198W and J20556$-$140N, which are from \cite{Bar18}, and J02489$-$145E, J02489$-$145W, J11110$+$304E, and J18427$+$596S, which were computed with {\tt serval} \citep{Zec18, Zec20}. $^{(a)}$Double-line spectroscopic (SB2) binary \citep{Bar18}. $^{(b)}$Spectral type M8.0\,V or later.}
\end{sidewaystable*}

\bibliographystyle{aa_url}
\bibliography{41980corr_arXiv}

\begin{thebibliography}{153}
\expandafter\ifx\csname natexlab\endcsname\relax\def\natexlab#1{#1}\fi

\bibitem[{{Abia} {et~al.}(2020){Abia}, {Tabernero}, {Korotin}, {Montes},
  {Marfil}, {Caballero}, {Straniero}, {Prantzos}, {Ribas}, {Reiners},
  {Quirrenbach}, {Amado}, {B{\'e}jar}, {Cort{\'e}s-Contreras}, {Dreizler},
  {Henning}, {Jeffers}, {Kaminski}, {K{\"u}rster}, {Lafarga},
  {L{\'o}pez-Gallifa}, {Morales}, {Nagel}, {Passegger}, {Pedraz},
  {Rodr{\'\i}guez L{\'o}pez}, {Schweitzer}, \& {Zechmeister}}]{Abi20}
{Abia}, C., {Tabernero}, H.~M., {Korotin}, S.~A., {et~al.} 2020,
  \href{http://dx.doi.org/10.1051/0004-6361/202039032}{\color{blue}\aap},
  \href{https://ui.adsabs.harvard.edu/abs/2020A&A...642A.227A}{642, A227}

\bibitem[{{Allard} {et~al.}(2001){Allard}, {Hauschildt}, {Alexander},
  {Tamanai}, \& {Schweitzer}}]{All01}
{Allard}, F., {Hauschildt}, P.~H., {Alexander}, D.~R., {Tamanai}, A., \&
  {Schweitzer}, A. 2001,
  \href{http://dx.doi.org/10.1086/321547}{\color{blue}\apj},
  \href{https://ui.adsabs.harvard.edu/abs/2001ApJ...556..357A}{556, 357}

\bibitem[{{Allard} {et~al.}(2012){Allard}, {Homeier}, \& {Freytag}}]{All12}
{Allard}, F., {Homeier}, D., \& {Freytag}, B. 2012,
  \href{http://dx.doi.org/10.1098/rsta.2011.0269}{\color{blue}Philosophical
  Transactions of the Royal Society of London Series A},
  \href{https://ui.adsabs.harvard.edu/abs/2012RSPTA.370.2765A}{370, 2765}

\bibitem[{{Alonso-Floriano} {et~al.}(2015){Alonso-Floriano}, {Morales},
  {Caballero}, {Montes}, {Klutsch}, {Mundt}, {Cort{\'e}s-Contreras}, {Ribas},
  {Reiners}, \& {Amado}}]{Alo15}
{Alonso-Floriano}, F.~J., {Morales}, J.~C., {Caballero}, J.~A., {et~al.} 2015,
  \href{http://dx.doi.org/10.1051/0004-6361/201525803}{\color{blue}\aap},
  \href{https://ui.adsabs.harvard.edu/abs/2015A&A...577A.128A}{577, A128}

\bibitem[{{Alonso-Floriano} {et~al.}(2019){Alonso-Floriano},
  {S{\'a}nchez-L{\'o}pez}, {Snellen}, {L{\'o}pez-Puertas}, {Nagel}, {Amado},
  {Bauer}, {Caballero}, {Czesla}, {Nortmann}, {Pall{\'e}}, {Salz}, {Reiners},
  {Ribas}, {Quirrenbach}, {Aceituno}, {Anglada-Escud{\'e}}, {B{\'e}jar},
  {Guenther}, {Henning}, {Kaminski}, {K{\"u}rster}, {Lamp{\'o}n}, {Lara},
  {Montes}, {Morales}, {Tal-Or}, {Schmitt}, {Zapatero Osorio}, \&
  {Zechmeister}}]{Alo19a}
{Alonso-Floriano}, F.~J., {S{\'a}nchez-L{\'o}pez}, A., {Snellen}, I.~A.~G.,
  {et~al.} 2019,
  \href{http://dx.doi.org/10.1051/0004-6361/201834339}{\color{blue}\aap},
  \href{https://ui.adsabs.harvard.edu/abs/2019A&A...621A..74A}{621, A74}

\bibitem[{{Alonso-Santiago} {et~al.}(2019){Alonso-Santiago}, {Negueruela},
  {Marco}, {Tabernero}, {Gonz{\'a}lez-Fern{\'a}ndez}, \& {Castro}}]{Alo19}
{Alonso-Santiago}, J., {Negueruela}, I., {Marco}, A., {et~al.} 2019,
  \href{http://dx.doi.org/10.1051/0004-6361/201936109}{\color{blue}\aap},
  \href{https://ui.adsabs.harvard.edu/abs/2019A&A...631A.124A}{631, A124}

\bibitem[{{Alvarez} \& {Plez}(1998)}]{Alv98}
{Alvarez}, R. \& {Plez}, B. 1998, \aap,
  \href{http://adsabs.harvard.edu/abs/1998A%26A...330.1109A}{330, 1109}

\bibitem[{{Antoniadis-Karnavas} {et~al.}(2020){Antoniadis-Karnavas}, {Sousa},
  {Delgado-Mena}, {Santos}, {Teixeira}, \& {Neves}}]{Ant20}
{Antoniadis-Karnavas}, A., {Sousa}, S.~G., {Delgado-Mena}, E., {et~al.} 2020,
  \href{http://dx.doi.org/10.1051/0004-6361/201937194}{\color{blue}\aap},
  \href{https://ui.adsabs.harvard.edu/abs/2020A&A...636A...9A}{636, A9}

\bibitem[{{Asplund} {et~al.}(2009){Asplund}, {Grevesse}, {Sauval}, \&
  {Scott}}]{Asp09}
{Asplund}, M., {Grevesse}, N., {Sauval}, A.~J., \& {Scott}, P. 2009,
  \href{http://dx.doi.org/10.1146/annurev.astro.46.060407.145222}{\color{blue}\araa},
  \href{https://ui.adsabs.harvard.edu/abs/2009ARA&A..47..481A}{47, 481}

\bibitem[{{Baraffe} {et~al.}(1998){Baraffe}, {Chabrier}, {Allard}, \&
  {Hauschildt}}]{Bar98}
{Baraffe}, I., {Chabrier}, G., {Allard}, F., \& {Hauschildt}, P.~H. 1998, \aap,
  \href{https://ui.adsabs.harvard.edu/abs/1998A&A...337..403B}{337, 403}

\bibitem[{{Barber} {et~al.}(2006){Barber}, {Tennyson}, {Harris}, \&
  {Tolchenov}}]{Bar06}
{Barber}, R.~J., {Tennyson}, J., {Harris}, G.~J., \& {Tolchenov}, R.~N. 2006,
  \href{http://dx.doi.org/10.1111/j.1365-2966.2006.10184.x}{\color{blue}\mnras},
  \href{https://ui.adsabs.harvard.edu/abs/2006MNRAS.368.1087B}{368, 1087}

\bibitem[{{Baroch} {et~al.}(2021){Baroch}, {Morales}, {Ribas}, {B{\'e}jar},
  {Reffert}, {Cardona Guill{\'e}n}, {Reiners}, {Caballero}, {Quirrenbach},
  {Amado}, {Anglada-Escud{\'e}}, {Colom{\'e}}, {Cort{\'e}s-Contreras},
  {Dreizler}, {Galad{\'\i}-Enr{\'\i}quez}, {Hatzes}, {Jeffers}, {Henning},
  {Herrero}, {Kaminski}, {K{\"u}rster}, {Lafarga}, {Lodieu},
  {L{\'o}pez-Gonz{\'a}lez}, {Montes}, {Pall{\'e}}, {Perger}, {Pollacco},
  {Rodr{\'\i}guez-L{\'o}pez}, {Rodr{\'\i}guez}, {Rosich}, {Sch{\"o}fer},
  {Schweitzer}, {Shan}, {Tal-Or}, \& {Zechmeister}}]{Bar21}
{Baroch}, D., {Morales}, J.~C., {Ribas}, I., {et~al.} 2021,
  \href{http://dx.doi.org/10.1051/0004-6361/202141031}{\color{blue}\aap},
  \href{https://ui.adsabs.harvard.edu/abs/2021A&A...653A..49B}{653, A49}

\bibitem[{{Baroch} {et~al.}(2018){Baroch}, {Morales}, {Ribas}, {Tal-Or},
  {Zechmeister}, {Reiners}, {Caballero}, {Quirrenbach}, {Amado}, {Dreizler},
  {Lalitha}, {Jeffers}, {Lafarga}, {B{\'e}jar}, {Colom{\'e}},
  {Cort{\'e}s-Contreras}, {D{\'{\i}}ez-Alonso},
  {Galad{\'{\i}}-Enr{\'{\i}}quez}, {Guenther}, {Hagen}, {Henning}, {Herrero},
  {K{\"u}rster}, {Montes}, {Nagel}, {Passegger}, {Perger}, {Rosich},
  {Schweitzer}, \& {Seifert}}]{Bar18}
{Baroch}, D., {Morales}, J.~C., {Ribas}, I., {et~al.} 2018,
  \href{http://dx.doi.org/10.1051/0004-6361/201833440}{\color{blue}\aap},
  \href{https://ui.adsabs.harvard.edu/abs/2018A%26A...619A..32B}{619, A32}

\bibitem[{{Barrado y Navascu{\'e}s} {et~al.}(2004){Barrado y Navascu{\'e}s},
  {Stauffer}, \& {Jayawardhana}}]{Bar04}
{Barrado y Navascu{\'e}s}, D., {Stauffer}, J.~R., \& {Jayawardhana}, R. 2004,
  \href{http://dx.doi.org/10.1086/423485}{\color{blue}\apj},
  \href{https://ui.adsabs.harvard.edu/abs/2004ApJ...614..386B}{614, 386}

\bibitem[{{Bayo} {et~al.}(2008){Bayo}, {Rodrigo}, {Barrado Y Navascu{\'e}s},
  {Solano}, {Guti{\'e}rrez}, {Morales-Calder{\'o}n}, \& {Allard}}]{Bay08}
{Bayo}, A., {Rodrigo}, C., {Barrado Y Navascu{\'e}s}, D., {et~al.} 2008,
  \href{http://dx.doi.org/10.1051/0004-6361:200810395}{\color{blue}\aap},
  \href{https://ui.adsabs.harvard.edu/abs/2008A&A...492..277B}{492, 277}

\bibitem[{{Bell} {et~al.}(2015){Bell}, {Mamajek}, \& {Naylor}}]{Bel15}
{Bell}, C. P.~M., {Mamajek}, E.~E., \& {Naylor}, T. 2015,
  \href{http://dx.doi.org/10.1093/mnras/stv1981}{\color{blue}\mnras},
  \href{https://ui.adsabs.harvard.edu/abs/2015MNRAS.454..593B}{454, 593}

\bibitem[{{Bensby} {et~al.}(2003){Bensby}, {Feltzing}, \&
  {Lundstr{\"o}m}}]{Ben03}
{Bensby}, T., {Feltzing}, S., \& {Lundstr{\"o}m}, I. 2003,
  \href{http://dx.doi.org/10.1051/0004-6361:20031213}{\color{blue}\aap},
  \href{https://ui.adsabs.harvard.edu/abs/2003A&A...410..527B}{410, 527}

\bibitem[{{Bensby} {et~al.}(2005){Bensby}, {Feltzing}, {Lundstr{\"o}m}, \&
  {Ilyin}}]{Ben05}
{Bensby}, T., {Feltzing}, S., {Lundstr{\"o}m}, I., \& {Ilyin}, I. 2005,
  \href{http://dx.doi.org/10.1051/0004-6361:20040332}{\color{blue}\aap},
  \href{https://ui.adsabs.harvard.edu/abs/2005A&A...433..185B}{433, 185}

\bibitem[{{Bergemann} {et~al.}(2017){Bergemann}, {Collet}, {Sch{\"o}nrich},
  {Andrae}, {Kovalev}, {Ruchti}, {Hansen}, \& {Magic}}]{Ber17}
{Bergemann}, M., {Collet}, R., {Sch{\"o}nrich}, R., {et~al.} 2017,
  \href{http://dx.doi.org/10.3847/1538-4357/aa88b5}{\color{blue}\apj},
  \href{https://ui.adsabs.harvard.edu/abs/2017ApJ...847...16B}{847, 16}

\bibitem[{{Birky} {et~al.}(2020){Birky}, {Hogg}, {Mann}, \&
  {Burgasser}}]{Bir20}
{Birky}, J., {Hogg}, D.~W., {Mann}, A.~W., \& {Burgasser}, A. 2020,
  \href{http://dx.doi.org/10.3847/1538-4357/ab7004}{\color{blue}\apj},
  \href{https://ui.adsabs.harvard.edu/abs/2020ApJ...892...31B}{892, 31}

\bibitem[{{Blanco-Cuaresma}(2019)}]{Bla19}
{Blanco-Cuaresma}, S. 2019,
  \href{http://dx.doi.org/10.1093/mnras/stz549}{\color{blue}\mnras},
  \href{https://ui.adsabs.harvard.edu/abs/2019MNRAS.486.2075B}{486, 2075}

\bibitem[{{Blanco-Cuaresma} {et~al.}(2014){Blanco-Cuaresma}, {Soubiran},
  {Heiter}, \& {Jofr{\'e}}}]{Bla14}
{Blanco-Cuaresma}, S., {Soubiran}, C., {Heiter}, U., \& {Jofr{\'e}}, P. 2014,
  {iSpec: Stellar atmospheric parameters and chemical abundances}, Astrophysics
  Source Code Library

\bibitem[{{Bochanski} {et~al.}(2007){Bochanski}, {Munn}, {Hawley}, {West},
  {Covey}, \& {Schneider}}]{Boc07}
{Bochanski}, J.~J., {Munn}, J.~A., {Hawley}, S.~L., {et~al.} 2007,
  \href{http://dx.doi.org/10.1086/522053}{\color{blue}\aj},
  \href{https://ui.adsabs.harvard.edu/abs/2007AJ....134.2418B}{134, 2418}

\bibitem[{Bocquet \& Carter(2016)}]{Boc16}
Bocquet, S. \& Carter, F.~W. 2016,
  \href{http://dx.doi.org/10.21105/joss.00046}{\color{blue}The Journal of Open
  Source Software}, 1, 1

\bibitem[{{Bonfils} {et~al.}(2013){Bonfils}, {Delfosse}, {Udry}, {Forveille},
  {Mayor}, {Perrier}, {Bouchy}, {Gillon}, {Lovis}, {Pepe}, {Queloz}, {Santos},
  {S{\'e}gransan}, \& {Bertaux}}]{Bon13}
{Bonfils}, X., {Delfosse}, X., {Udry}, S., {et~al.} 2013,
  \href{http://dx.doi.org/10.1051/0004-6361/201014704}{\color{blue}\aap},
  \href{https://ui.adsabs.harvard.edu/abs/2013A&A...549A.109B}{549, A109}

\bibitem[{{Borsa} {et~al.}(2021){Borsa}, {Allart}, {Casasayas-Barris},
  {Tabernero}, {Zapatero Osorio}, {Cristiani}, {Pepe}, {Rebolo}, {Santos},
  {Adibekyan}, {Bourrier}, {Demangeon}, {Ehrenreich}, {Pall{\'e}}, {Sousa},
  {Lillo-Box}, {Lovis}, {Micela}, {Oshagh}, {Poretti}, {Sozzetti}, {Allende
  Prieto}, {Alibert}, {Amate}, {Benz}, {Bouchy}, {Cabral}, {Dekker},
  {D'Odorico}, {Di Marcantonio}, {Figueira}, {Genova Santos}, {Gonz{\'a}lez
  Hern{\'a}ndez}, {Lo Curto}, {Manescau}, {Martins}, {M{\'e}gevand}, {Mehner},
  {Molaro}, {Nunes}, {Riva}, {Su{\'a}rez Mascare{\~n}o}, {Udry}, \&
  {Zerbi}}]{Bor21}
{Borsa}, F., {Allart}, R., {Casasayas-Barris}, N., {et~al.} 2021,
  \href{http://dx.doi.org/10.1051/0004-6361/202039344}{\color{blue}\aap},
  \href{https://ui.adsabs.harvard.edu/abs/2021A&A...645A..24B}{645, A24}

\bibitem[{{Boyajian} {et~al.}(2012){Boyajian}, {von Braun}, {van Belle},
  {McAlister}, {ten Brummelaar}, {Kane}, {Muirhead}, {Jones}, {White},
  {Schaefer}, {Ciardi}, {Henry}, {L{\'o}pez-Morales}, {Ridgway}, {Gies}, {Jao},
  {Rojas-Ayala}, {Parks}, {Sturmann}, {Sturmann}, {Turner}, {Farrington},
  {Goldfinger}, \& {Berger}}]{Boy12}
{Boyajian}, T.~S., {von Braun}, K., {van Belle}, G., {et~al.} 2012,
  \href{http://dx.doi.org/10.1088/0004-637X/757/2/112}{\color{blue}\apj},
  \href{https://ui.adsabs.harvard.edu/abs/2012ApJ...757..112B}{757, 112}

\bibitem[{{Bressan} {et~al.}(2012){Bressan}, {Marigo}, {Girardi}, {Salasnich},
  {Dal Cero}, {Rubele}, \& {Nanni}}]{Bre12}
{Bressan}, A., {Marigo}, P., {Girardi}, L., {et~al.} 2012,
  \href{http://dx.doi.org/10.1111/j.1365-2966.2012.21948.x}{\color{blue}\mnras},
  \href{https://ui.adsabs.harvard.edu/abs/2012MNRAS.427..127B}{427, 127}

\bibitem[{{Brewer} {et~al.}(2016){Brewer}, {Fischer}, {Valenti}, \&
  {Piskunov}}]{Bre16}
{Brewer}, J.~M., {Fischer}, D.~A., {Valenti}, J.~A., \& {Piskunov}, N. 2016,
  \href{http://dx.doi.org/10.3847/0067-0049/225/2/32}{\color{blue}\apjs},
  \href{https://ui.adsabs.harvard.edu/abs/2016ApJS..225...32B}{225, 32}

\bibitem[{{Buchhave} {et~al.}(2012){Buchhave}, {Latham}, {Johansen},
  {Bizzarro}, {Torres}, {Rowe}, {Batalha}, {Borucki}, {Brugamyer}, {Caldwell},
  {Bryson}, {Ciardi}, {Cochran}, {Endl}, {Esquerdo}, {Ford}, {Geary},
  {Gilliland}, {Hansen}, {Isaacson}, {Laird}, {Lucas}, {Marcy}, {Morse},
  {Robertson}, {Shporer}, {Stefanik}, {Still}, \& {Quinn}}]{Buc12}
{Buchhave}, L.~A., {Latham}, D.~W., {Johansen}, A., {et~al.} 2012,
  \href{http://dx.doi.org/10.1038/nature11121}{\color{blue}\nat},
  \href{https://ui.adsabs.harvard.edu/abs/2012Natur.486..375B}{486, 375}

\bibitem[{{Burrows} {et~al.}(1997){Burrows}, {Marley}, {Hubbard}, {Lunine},
  {Guillot}, {Saumon}, {Freedman}, {Sudarsky}, \& {Sharp}}]{Bur97}
{Burrows}, A., {Marley}, M., {Hubbard}, W.~B., {et~al.} 1997,
  \href{http://dx.doi.org/10.1086/305002}{\color{blue}\apj},
  \href{https://ui.adsabs.harvard.edu/abs/1997ApJ...491..856B}{491, 856}

\bibitem[{{Caballero} {et~al.}(2016{\natexlab{a}}){Caballero},
  {Cort{\'e}s-Contreras}, {Alonso-Floriano}, {Montes}, {Quirrenbach}, {Amado},
  {Ribas}, {Reiners}, {Abellan}, {B{\'e}jar}, {Brinkm{\"o}ller}, {Czesla},
  {Dorda}, {Gallardo}, {Gonz{\'a}lez-{\'A}lvarez}, {Hidalgo}, {Holgado},
  {Jeffers}, {Kim}, {Klutsch}, {Lamert}, {Llamas}, {L{\'o}pez-Santiago},
  {Mart{\'\i}nez-Rodr{\'\i}guez}, {Morales}, {Mundt}, {Passegger},
  {Sch{\"o}fer}, {Seifert}, \& {Zechmeister}}]{Cab16a}
{Caballero}, J.~A., {Cort{\'e}s-Contreras}, M., {Alonso-Floriano}, F.~J.,
  {et~al.} 2016{\natexlab{a}}, in 19th Cambridge Workshop on Cool Stars,
  Stellar Systems, and the Sun (CS19),
  \href{https://ui.adsabs.harvard.edu/abs/2016csss.confE.148C}{148}

\bibitem[{{Caballero} {et~al.}(2016{\natexlab{b}}){Caballero}, {Gu{\`a}rdia},
  {L{\'o}pez del Fresno}, {Zechmeister}, {de Juan}, {Alonso-Floriano}, {Amado},
  {Colom{\'e}}, {Cort{\'e}s-Contreras}, {Garc{\'\i}a-Piquer}, {Gesa}, {de
  Guindos}, {Hagen}, {Helmling}, {Hern{\'a}ndez Casta{\~n}o}, {K{\"u}rster},
  {L{\'o}pez-Santiago}, {Montes}, {Morales Mu{\~n}oz}, {Pavlov}, {Quirrenbach},
  {Reiners}, {Ribas}, {Seifert}, \& {Solano}}]{Cab16b}
{Caballero}, J.~A., {Gu{\`a}rdia}, J., {L{\'o}pez del Fresno}, M., {et~al.}
  2016{\natexlab{b}}, in Society of Photo-Optical Instrumentation Engineers
  (SPIE) Conference Series, Vol. 9910,
  \href{https://ui.adsabs.harvard.edu/abs/2016SPIE.9910E..0EC}{99100E}

\bibitem[{{Casagrande} {et~al.}(2008){Casagrande}, {Flynn}, \&
  {Bessell}}]{Cas08}
{Casagrande}, L., {Flynn}, C., \& {Bessell}, M. 2008,
  \href{http://dx.doi.org/10.1111/j.1365-2966.2008.13573.x}{\color{blue}\mnras},
  \href{https://ui.adsabs.harvard.edu/abs/2008MNRAS.389..585C}{389, 585}

\bibitem[{{Cifuentes} {et~al.}(2020){Cifuentes}, {Caballero},
  {Cort{\'e}s-Contreras}, {Montes}, {Abell{\'a}n}, {Dorda}, {Holgado},
  {Zapatero Osorio}, {Morales}, {Amado}, {Passegger}, {Quirrenbach}, {Reiners},
  {Ribas}, {Sanz-Forcada}, {Schweitzer}, {Seifert}, \& {Solano}}]{Cif20}
{Cifuentes}, C., {Caballero}, J.~A., {Cort{\'e}s-Contreras}, M., {et~al.} 2020,
  \href{http://dx.doi.org/10.1051/0004-6361/202038295}{\color{blue}\aap},
  \href{https://ui.adsabs.harvard.edu/abs/2020A&A...642A.115C}{642, A115}

\bibitem[{{Claudi} {et~al.}(2018){Claudi}, {Benatti}, {Carleo}, {Ghedina},
  {Guerra}, {Ghinassi}, {Harutyunyan}, {Micela}, {Molinari}, {Oliva}, {Rainer},
  {Tozzi}, {Baffa}, {Baruffolo}, {Biliotti}, {Buchschacher}, {Cecconi},
  {Cosentino}, {Falcini}, {Fantinel}, {Fini}, {Giani}, {Gonzalez-Alvarez},
  {Gonzalez}, {Gonzalez}, {Gratton}, {Hernandez}, {Iuzzolino}, {Lodi},
  {Malavolta}, {Maldonado}, {Origlia}, {Puglisi}, {Sanna}, {San Juan
  G{\'o}mez}, {Scuderi}, {Seemann}, {Sozzetti}, {Sozzi}, {Perez Ventura},
  {Hernandez Diaz}, {Galli}, {Riverol}, \& {Riverol}}]{Cla18}
{Claudi}, R., {Benatti}, S., {Carleo}, I., {et~al.} 2018, in Society of
  Photo-Optical Instrumentation Engineers (SPIE) Conference Series, Vol. 10702,
  \href{https://ui.adsabs.harvard.edu/abs/2018SPIE10702E..0ZC}{107020Z}

\bibitem[{{Clough} {et~al.}(2005){Clough}, {Shephard}, {Mlawer}, {Delamere},
  {Iacono}, {Cady-Pereira}, {Boukabara}, \& {Brown}}]{Clo05}
{Clough}, S.~A., {Shephard}, M.~W., {Mlawer}, E.~J., {et~al.} 2005,
  \href{http://dx.doi.org/10.1016/j.jqsrt.2004.05.058}{\color{blue}\jqsrt},
  \href{https://ui.adsabs.harvard.edu/abs/2005JQSRT..91..233C}{91, 233}

\bibitem[{{Delfosse} {et~al.}(1998){Delfosse}, {Forveille}, {Perrier}, \&
  {Mayor}}]{Del98}
{Delfosse}, X., {Forveille}, T., {Perrier}, C., \& {Mayor}, M. 1998, \aap,
  \href{https://ui.adsabs.harvard.edu/abs/1998A&A...331..581D}{331, 581}

\bibitem[{{Delgado Mena} {et~al.}(2010){Delgado Mena}, {Israelian},
  {Gonz{\'a}lez Hern{\'a}ndez}, {Bond}, {Santos}, {Udry}, \& {Mayor}}]{Del10}
{Delgado Mena}, E., {Israelian}, G., {Gonz{\'a}lez Hern{\'a}ndez}, J.~I.,
  {et~al.} 2010,
  \href{http://dx.doi.org/10.1088/0004-637X/725/2/2349}{\color{blue}\apj},
  \href{https://ui.adsabs.harvard.edu/abs/2010ApJ...725.2349D}{725, 2349}

\bibitem[{{Demangeon} {et~al.}(2021){Demangeon}, {Zapatero Osorio}, {Alibert},
  {Barros}, {Adibekyan}, {Tabernero}, {Antoniadis-Karnavas}, \&
  {Camacho}}]{Dem21}
{Demangeon}, O.~D.~S., {Zapatero Osorio}, M.~R., {Alibert}, Y., {et~al.} 2021,
  \href{http://dx.doi.org/https://doi.org/10.1051/0004-6361/202140728}{\color{blue}\aap}

\bibitem[{{Desidera} {et~al.}(2006){Desidera}, {Gratton}, {Lucatello}, \&
  {Claudi}}]{Des06}
{Desidera}, S., {Gratton}, R.~G., {Lucatello}, S., \& {Claudi}, R.~U. 2006,
  \href{http://dx.doi.org/10.1051/0004-6361:20064896}{\color{blue}\aap},
  \href{https://ui.adsabs.harvard.edu/abs/2006A&A...454..581D}{454, 581}

\bibitem[{{Donati} {et~al.}(2020){Donati}, {Kouach}, {Moutou}, {Doyon},
  {Delfosse}, {Artigau}, {Baratchart}, {Lacombe}, {Barrick}, {H{\'e}brard},
  {Bouchy}, {Saddlemyer}, {Par{\`e}s}, {Rabou}, {Micheau}, {Dolon}, {Reshetov},
  {Challita}, {Carmona}, {Striebig}, {Thibault}, {Martioli}, {Cook},
  {Fouqu{\'e}}, {Vermeulen}, {Wang}, {Arnold}, {Pepe}, {Boisse}, {Figueira},
  {Bouvier}, {Ray}, {Feugeade}, {Morin}, {Alencar}, {Hobson}, {Castilho},
  {Udry}, {Santos}, {Hernandez}, {Benedict}, {Vall{\'e}e}, {Gallou}, {Dupieux},
  {Larrieu}, {Perruchot}, {Sottile}, {Moreau}, {Usher}, {Baril}, {Wildi},
  {Chazelas}, {Malo}, {Bonfils}, {Loop}, {Kerley}, {Wevers}, {Dunn}, {Pazder},
  {Macdonald}, {Dubois}, {Carri{\'e}}, {Valentin}, {Henault}, {Yan}, \&
  {Steinmetz}}]{Don20}
{Donati}, J.~F., {Kouach}, D., {Moutou}, C., {et~al.} 2020,
  \href{http://dx.doi.org/10.1093/mnras/staa2569}{\color{blue}\mnras},
  \href{https://ui.adsabs.harvard.edu/abs/2020MNRAS.498.5684D}{498, 5684}

\bibitem[{{Donati} {et~al.}(2008){Donati}, {Morin}, {Petit}, {Delfosse},
  {Forveille}, {Auri{\`e}re}, {Cabanac}, {Dintrans}, {Fares}, {Gastine},
  {Jardine}, {Ligni{\`e}res}, {Paletou}, {Ramirez Velez}, \&
  {Th{\'e}ado}}]{Don08}
{Donati}, J.~F., {Morin}, J., {Petit}, P., {et~al.} 2008,
  \href{http://dx.doi.org/10.1111/j.1365-2966.2008.13799.x}{\color{blue}\mnras},
  \href{https://ui.adsabs.harvard.edu/abs/2008MNRAS.390..545D}{390, 545}

\bibitem[{{Dressing} \& {Charbonneau}(2013)}]{Dre13}
{Dressing}, C.~D. \& {Charbonneau}, D. 2013,
  \href{http://dx.doi.org/10.1088/0004-637X/767/1/95}{\color{blue}\apj},
  \href{https://ui.adsabs.harvard.edu/abs/2013ApJ...767...95D}{767, 95}

\bibitem[{{Dulick} {et~al.}(2003){Dulick}, {Bauschlicher}, {Burrows}, {Sharp},
  {Ram}, \& {Bernath}}]{Dul03}
{Dulick}, M., {Bauschlicher}, C.~W., J., {Burrows}, A., {et~al.} 2003,
  \href{http://dx.doi.org/10.1086/376791}{\color{blue}\apj},
  \href{https://ui.adsabs.harvard.edu/abs/2003ApJ...594..651D}{594, 651}

\bibitem[{{Francis} \& {Wills}(1999)}]{Fra99}
{Francis}, P.~J. \& {Wills}, B.~J. 1999, in Astronomical Society of the Pacific
  Conference Series, Vol. 162, Quasars and Cosmology, ed. G.~{Ferland} \&
  J.~{Baldwin},
  \href{https://ui.adsabs.harvard.edu/abs/1999ASPC..162..363F}{363}

\bibitem[{{Fuhrmeister} {et~al.}(2020){Fuhrmeister}, {Czesla}, {Hildebrandt},
  {Nagel}, {Schmitt}, {Jeffers}, {Caballero}, {Hintz}, {Johnson},
  {Sch{\"o}fer}, {Zechmeister}, {Reiners}, {Ribas}, {Amado}, {Quirrenbach},
  {Nortmann}, {Bauer}, {B{\'e}jar}, {Cort{\'e}s-Contreras}, {Dreizler},
  {Galad{\'\i}-Enr{\'\i}quez}, {Hatzes}, {Kaminski}, {K{\"u}rster}, {Lafarga},
  \& {Montes}}]{Fuh20}
{Fuhrmeister}, B., {Czesla}, S., {Hildebrandt}, L., {et~al.} 2020,
  \href{http://dx.doi.org/10.1051/0004-6361/202038279}{\color{blue}\aap},
  \href{https://ui.adsabs.harvard.edu/abs/2020A&A...640A..52F}{640, A52}

\bibitem[{{Fuhrmeister} {et~al.}(2018){Fuhrmeister}, {Czesla}, {Schmitt},
  {Jeffers}, {Caballero}, {Zechmeister}, {Reiners}, {Ribas}, {Amado},
  {Quirrenbach}, {B{\'e}jar}, {Galad{\'\i}-Enr{\'\i}quez}, {Guenther},
  {K{\"u}rster}, {Montes}, \& {Seifert}}]{Fuh18}
{Fuhrmeister}, B., {Czesla}, S., {Schmitt}, J.~H.~M.~M., {et~al.} 2018,
  \href{http://dx.doi.org/10.1051/0004-6361/201732204}{\color{blue}\aap},
  \href{https://ui.adsabs.harvard.edu/abs/2018A&A...615A..14F}{615, A14}

\bibitem[{{Gaia Collaboration} {et~al.}(2018){Gaia Collaboration}, {Brown},
  {Vallenari}, {Prusti}, {de Bruijne}, {Babusiaux}, {Bailer-Jones}, {Biermann},
  {Evans}, {Eyer}, {Jansen}, {Jordi}, {Klioner}, {Lammers}, {Lindegren},
  {Luri}, {Mignard}, {Panem}, {Pourbaix}, {Randich}, {Sartoretti}, {Siddiqui},
  {Soubiran}, {van Leeuwen}, {Walton}, {Arenou}, {Bastian}, {Cropper},
  {Drimmel}, {Katz}, {Lattanzi}, {Bakker}, {Cacciari}, {Casta{\~n}eda},
  {Chaoul}, {Cheek}, {De Angeli}, {Fabricius}, {Guerra}, {Holl}, {Masana},
  {Messineo}, {Mowlavi}, {Nienartowicz}, {Panuzzo}, {Portell}, {Riello},
  {Seabroke}, {Tanga}, {Th{\'e}venin}, {Gracia-Abril}, {Comoretto},
  {Garcia-Reinaldos}, {Teyssier}, {Altmann}, {Andrae}, {Audard},
  {Bellas-Velidis}, {Benson}, {Berthier}, {Blomme}, {Burgess}, {Busso},
  {Carry}, {Cellino}, {Clementini}, {Clotet}, {Creevey}, {Davidson}, {De
  Ridder}, {Delchambre}, {Dell'Oro}, {Ducourant},
  {Fern{\'a}ndez-Hern{\'a}ndez}, {Fouesneau}, {Fr{\'e}mat}, {Galluccio},
  {Garc{\'\i}a-Torres}, {Gonz{\'a}lez-N{\'u}{\~n}ez}, {Gonz{\'a}lez-Vidal},
  {Gosset}, {Guy}, {Halbwachs}, {Hambly}, {Harrison}, {Hern{\'a}ndez},
  {Hestroffer}, {Hodgkin}, {Hutton}, {Jasniewicz}, {Jean-Antoine-Piccolo},
  {Jordan}, {Korn}, {Krone-Martins}, {Lanzafame}, {Lebzelter}, {L{\"o}ffler},
  {Manteiga}, {Marrese}, {Mart{\'\i}n-Fleitas}, {Moitinho}, {Mora}, {Muinonen},
  {Osinde}, {Pancino}, {Pauwels}, {Petit}, {Recio-Blanco}, {Richards},
  {Rimoldini}, {Robin}, {Sarro}, {Siopis}, {Smith}, {Sozzetti}, {S{\"u}veges},
  {Torra}, {van Reeven}, {Abbas}, {Abreu Aramburu}, {Accart}, {Aerts},
  {Altavilla}, {{\'A}lvarez}, {Alvarez}, {Alves}, {Anderson}, {Andrei},
  {Anglada Varela}, {Antiche}, {Antoja}, {Arcay}, {Astraatmadja}, {Bach},
  {Baker}, {Balaguer-N{\'u}{\~n}ez}, {Balm}, {Barache}, {Barata}, {Barbato},
  {Barblan}, {Barklem}, {Barrado}, {Barros}, {Barstow}, {Bartholom{\'e}
  Mu{\~n}oz}, {Bassilana}, {Becciani}, {Bellazzini}, {Berihuete}, {Bertone},
  {Bianchi}, {Bienaym{\'e}}, {Blanco-Cuaresma}, {Boch}, {Boeche}, {Bombrun},
  {Borrachero}, {Bossini}, {Bouquillon}, {Bourda}, {Bragaglia}, {Bramante},
  {Breddels}, {Bressan}, {Brouillet}, {Br{\"u}semeister}, {Brugaletta},
  {Bucciarelli}, {Burlacu}, {Busonero}, {Butkevich}, {Buzzi}, {Caffau},
  {Cancelliere}, {Cannizzaro}, {Cantat-Gaudin}, {Carballo}, {Carlucci},
  {Carrasco}, {Casamiquela}, {Castellani}, {Castro-Ginard}, {Charlot},
  {Chemin}, {Chiavassa}, {Cocozza}, {Costigan}, {Cowell}, {Crifo}, {Crosta},
  {Crowley}, {Cuypers}, {Dafonte}, {Damerdji}, {Dapergolas}, {David}, {David},
  {de Laverny}, {De Luise}, {De March}, {de Martino}, {de Souza}, {de Torres},
  {Debosscher}, {del Pozo}, {Delbo}, {Delgado}, {Delgado}, {Di Matteo},
  {Diakite}, {Diener}, {Distefano}, {Dolding}, {Drazinos}, {Dur{\'a}n},
  {Edvardsson}, {Enke}, {Eriksson}, {Esquej}, {Eynard Bontemps}, {Fabre},
  {Fabrizio}, {Faigler}, {Falc{\~a}o}, {Farr{\`a}s Casas}, {Federici},
  {Fedorets}, {Fernique}, {Figueras}, {Filippi}, {Findeisen}, {Fonti},
  {Fraile}, {Fraser}, {Fr{\'e}zouls}, {Gai}, {Galleti}, {Garabato},
  {Garc{\'\i}a-Sedano}, {Garofalo}, {Garralda}, {Gavel}, {Gavras}, {Gerssen},
  {Geyer}, {Giacobbe}, {Gilmore}, {Girona}, {Giuffrida}, {Glass}, {Gomes},
  {Granvik}, {Gueguen}, {Guerrier}, {Guiraud}, {Guti{\'e}rrez-S{\'a}nchez},
  {Haigron}, {Hatzidimitriou}, {Hauser}, {Haywood}, {Heiter}, {Helmi}, {Heu},
  {Hilger}, {Hobbs}, {Hofmann}, {Holland}, {Huckle}, {Hypki}, {Icardi},
  {Jan{\ss}en}, {Jevardat de Fombelle}, {Jonker}, {Juh{\'a}sz}, {Julbe},
  {Karampelas}, {Kewley}, {Klar}, {Kochoska}, {Kohley}, {Kolenberg},
  {Kontizas}, {Kontizas}, {Koposov}, {Kordopatis}, {Kostrzewa-Rutkowska},
  {Koubsky}, {Lambert}, {Lanza}, {Lasne}, {Lavigne}, {Le Fustec}, {Le
  Poncin-Lafitte}, {Lebreton}, {Leccia}, {Leclerc}, {Lecoeur-Taibi},
  {Lenhardt}, {Leroux}, {Liao}, {Licata}, {Lindstr{\o}m}, {Lister}, {Livanou},
  {Lobel}, {L{\'o}pez}, {Managau}, {Mann}, {Mantelet}, {Marchal}, {Marchant},
  {Marconi}, {Marinoni}, {Marschalk{\'o}}, {Marshall}, {Martino}, {Marton},
  {Mary}, {Massari}, {Matijevi{\v{c}}}, {Mazeh}, {McMillan}, {Messina},
  {Michalik}, {Millar}, {Molina}, {Molinaro}, {Moln{\'a}r}, {Montegriffo},
  {Mor}, {Morbidelli}, {Morel}, {Morris}, {Mulone}, {Muraveva}, {Musella},
  {Nelemans}, {Nicastro}, {Noval}, {O'Mullane}, {Ord{\'e}novic},
  {Ord{\'o}{\~n}ez-Blanco}, {Osborne}, {Pagani}, {Pagano}, {Pailler},
  {Palacin}, {Palaversa}, {Panahi}, {Pawlak}, {Piersimoni}, {Pineau}, {Plachy},
  {Plum}, {Poggio}, {Poujoulet}, {Pr{\v{s}}a}, {Pulone}, {Racero}, {Ragaini},
  {Rambaux}, {Ramos-Lerate}, {Regibo}, {Reyl{\'e}}, {Riclet}, {Ripepi}, {Riva},
  {Rivard}, {Rixon}, {Roegiers}, {Roelens}, {Romero-G{\'o}mez}, {Rowell},
  {Royer}, {Ruiz-Dern}, {Sadowski}, {Sagrist{\`a} Sell{\'e}s}, {Sahlmann},
  {Salgado}, {Salguero}, {Sanna}, {Santana-Ros}, {Sarasso}, {Savietto},
  {Schultheis}, {Sciacca}, {Segol}, {Segovia}, {S{\'e}gransan}, {Shih},
  {Siltala}, {Silva}, {Smart}, {Smith}, {Solano}, {Solitro}, {Sordo}, {Soria
  Nieto}, {Souchay}, {Spagna}, {Spoto}, {Stampa}, {Steele},
  {Steidelm{\"u}ller}, {Stephenson}, {Stoev}, {Suess}, {Surdej}, {Szabados},
  {Szegedi-Elek}, {Tapiador}, {Taris}, {Tauran}, {Taylor}, {Teixeira},
  {Terrett}, {Teyssandier}, {Thuillot}, {Titarenko}, {Torra Clotet}, {Turon},
  {Ulla}, {Utrilla}, {Uzzi}, {Vaillant}, {Valentini}, {Valette}, {van Elteren},
  {Van Hemelryck}, {van Leeuwen}, {Vaschetto}, {Vecchiato}, {Veljanoski},
  {Viala}, {Vicente}, {Vogt}, {von Essen}, {Voss}, {Votruba}, {Voutsinas},
  {Walmsley}, {Weiler}, {Wertz}, {Wevers}, {Wyrzykowski}, {Yoldas},
  {{\v{Z}}erjal}, {Ziaeepour}, {Zorec}, {Zschocke}, {Zucker}, {Zurbach}, \&
  {Zwitter}}]{GaiaDR2}
{Gaia Collaboration}, {Brown}, A.~G.~A., {Vallenari}, A., {et~al.} 2018,
  \href{http://dx.doi.org/10.1051/0004-6361/201833051}{\color{blue}\aap},
  \href{https://ui.adsabs.harvard.edu/abs/2018A&A...616A...1G}{616, A1}

\bibitem[{{Gaia Collaboration} {et~al.}(2021){Gaia Collaboration}, {Brown},
  {Vallenari}, {Prusti}, {de Bruijne}, {Babusiaux}, {Biermann}, {Creevey},
  {Evans}, {Eyer}, \& et~al.}]{GaiaDR3}
{Gaia Collaboration}, {Brown}, A.~G.~A., {Vallenari}, A., {et~al.} 2021,
  \href{http://dx.doi.org/10.1051/0004-6361/202039657}{\color{blue}\aap},
  \href{https://ui.adsabs.harvard.edu/abs/2021A&A...649A...1G}{649, A1}

\bibitem[{{Gaidos} \& {Mann}(2014)}]{Gai14}
{Gaidos}, E. \& {Mann}, A.~W. 2014,
  \href{http://dx.doi.org/10.1088/0004-637X/791/1/54}{\color{blue}\apj},
  \href{https://ui.adsabs.harvard.edu/abs/2014ApJ...791...54G}{791, 54}

\bibitem[{{Garc{\'\i}a P{\'e}rez} {et~al.}(2016){Garc{\'\i}a P{\'e}rez},
  {Allende Prieto}, {Holtzman}, {Shetrone}, {M{\'e}sz{\'a}ros}, {Bizyaev},
  {Carrera}, {Cunha}, {Garc{\'\i}a-Hern{\'a}ndez}, {Johnson}, {Majewski},
  {Nidever}, {Schiavon}, {Shane}, {Smith}, {Sobeck}, {Troup}, {Zamora},
  {Weinberg}, {Bovy}, {Eisenstein}, {Feuillet}, {Frinchaboy}, {Hayden},
  {Hearty}, {Nguyen}, {O'Connell}, {Pinsonneault}, {Wilson}, \&
  {Zasowski}}]{Gar16}
{Garc{\'\i}a P{\'e}rez}, A.~E., {Allende Prieto}, C., {Holtzman}, J.~A.,
  {et~al.} 2016,
  \href{http://dx.doi.org/10.3847/0004-6256/151/6/144}{\color{blue}\aj},
  \href{https://ui.adsabs.harvard.edu/abs/2016AJ....151..144G}{151, 144}

\bibitem[{{Goorvitch}(1994)}]{Goo94}
{Goorvitch}, D. 1994,
  \href{http://dx.doi.org/10.1086/192110}{\color{blue}\apjs},
  \href{https://ui.adsabs.harvard.edu/abs/1994ApJS...95..535G}{95, 535}

\bibitem[{{Gray}(2008)}]{Gra08}
{Gray}, D.~F. 2008, {The Observation and Analysis of Stellar Photospheres}
  ({Cambridge University Press})

\bibitem[{{Gustafsson} {et~al.}(2008){Gustafsson}, {Edvardsson}, {Eriksson},
  {J{\o}rgensen}, {Nordlund}, \& {Plez}}]{Gus08}
{Gustafsson}, B., {Edvardsson}, B., {Eriksson}, K., {et~al.} 2008,
  \href{http://dx.doi.org/10.1051/0004-6361:200809724}{\color{blue}\aap},
  \href{https://ui.adsabs.harvard.edu/abs/2008A&A...486..951G}{486, 951}

\bibitem[{{Heiter} {et~al.}(2021){Heiter}, {Lind}, {Bergemann}, {Asplund},
  {Mikolaitis}, {Barklem}, {Masseron}, {de Laverny}, {Magrini}, {Edvardsson},
  {J{\"o}nsson}, {Pickering}, {Ryde}, {Bayo Ar{\'a}n}, {Bensby}, {Casey},
  {Feltzing}, {Jofr{\'e}}, {Korn}, {Pancino}, {Damiani}, {Lanzafame}, {Lardo},
  {Monaco}, {Morbidelli}, {Smiljanic}, {Worley}, {Zaggia}, {Randich}, \&
  {Gilmore}}]{Hei21}
{Heiter}, U., {Lind}, K., {Bergemann}, M., {et~al.} 2021,
  \href{http://dx.doi.org/10.1051/0004-6361/201936291}{\color{blue}\aap},
  \href{https://ui.adsabs.harvard.edu/abs/2021A&A...645A.106H}{645, A106}

\bibitem[{{Hejazi} {et~al.}(2020){Hejazi}, {L{\'e}pine}, {Homeier}, {Rich}, \&
  {Shara}}]{Hej20}
{Hejazi}, N., {L{\'e}pine}, S., {Homeier}, D., {Rich}, R.~M., \& {Shara}, M.~M.
  2020, \href{http://dx.doi.org/10.3847/1538-3881/ab563c}{\color{blue}\aj},
  \href{https://ui.adsabs.harvard.edu/abs/2020AJ....159...30H}{159, 30}

\bibitem[{{Henry} {et~al.}(2018){Henry}, {Jao}, {Winters}, {Dieterich},
  {Finch}, {Ianna}, {Riedel}, {Silverstein}, {Subasavage}, \&
  {Vrijmoet}}]{Hen18}
{Henry}, T.~J., {Jao}, W.-C., {Winters}, J.~G., {et~al.} 2018,
  \href{http://dx.doi.org/10.3847/1538-3881/aac262}{\color{blue}\aj},
  \href{https://ui.adsabs.harvard.edu/abs/2018AJ....155..265H}{155, 265}

\bibitem[{{Hintz} {et~al.}(2019){Hintz}, {Fuhrmeister}, {Czesla}, {Schmitt},
  {Johnson}, {Schweitzer}, {Caballero}, {Zechmeister}, {Jeffers}, {Reiners},
  {Ribas}, {Amado}, {Quirrenbach}, {Anglada-Escud{\'e}}, {Bauer}, {B{\'e}jar},
  {Cort{\'e}s-Contreras}, {Dreizler}, {Galad{\'\i}-Enr{\'\i}quez}, {Guenther},
  {Hauschildt}, {Kaminski}, {K{\"u}rster}, {Lafarga}, {L{\'o}pez del Fresno},
  {Montes}, {Morales}, {Passegger}, \& {Seifert}}]{Hin19}
{Hintz}, D., {Fuhrmeister}, B., {Czesla}, S., {et~al.} 2019,
  \href{http://dx.doi.org/10.1051/0004-6361/201834788}{\color{blue}\aap},
  \href{https://ui.adsabs.harvard.edu/abs/2019A&A...623A.136H}{623, A136}

\bibitem[{{Hintz} {et~al.}(2020){Hintz}, {Fuhrmeister}, {Czesla}, {Schmitt},
  {Schweitzer}, {Nagel}, {Johnson}, {Caballero}, {Zechmeister}, {Jeffers},
  {Reiners}, {Ribas}, {Amado}, {Quirrenbach}, {Anglada-Escud{\'e}}, {Bauer},
  {B{\'e}jar}, {Cort{\'e}s-Contreras}, {Dreizler}, {Galad{\'\i}-Enr{\'\i}quez},
  {Guenther}, {Hauschildt}, {Kaminski}, {K{\"u}rster}, {Lafarga}, {L{\'o}pez
  del Fresno}, {Montes}, \& {Morales}}]{Hin20}
{Hintz}, D., {Fuhrmeister}, B., {Czesla}, S., {et~al.} 2020,
  \href{http://dx.doi.org/10.1051/0004-6361/202037596}{\color{blue}\aap},
  \href{https://ui.adsabs.harvard.edu/abs/2020A&A...638A.115H}{638, A115}

\bibitem[{{Howard} {et~al.}(2012){Howard}, {Marcy}, {Bryson}, {Jenkins},
  {Rowe}, {Batalha}, {Borucki}, {Koch}, {Dunham}, {Gautier}, {Van Cleve},
  {Cochran}, {Latham}, {Lissauer}, {Torres}, {Brown}, {Gilliland}, {Buchhave},
  {Caldwell}, {Christensen-Dalsgaard}, {Ciardi}, {Fressin}, {Haas}, {Howell},
  {Kjeldsen}, {Seager}, {Rogers}, {Sasselov}, {Steffen}, {Basri},
  {Charbonneau}, {Christiansen}, {Clarke}, {Dupree}, {Fabrycky}, {Fischer},
  {Ford}, {Fortney}, {Tarter}, {Girouard}, {Holman}, {Johnson}, {Klaus},
  {Machalek}, {Moorhead}, {Morehead}, {Ragozzine}, {Tenenbaum}, {Twicken},
  {Quinn}, {Isaacson}, {Shporer}, {Lucas}, {Walkowicz}, {Welsh}, {Boss},
  {Devore}, {Gould}, {Smith}, {Morris}, {Prsa}, {Morton}, {Still}, {Thompson},
  {Mullally}, {Endl}, \& {MacQueen}}]{How12}
{Howard}, A.~W., {Marcy}, G.~W., {Bryson}, S.~T., {et~al.} 2012,
  \href{http://dx.doi.org/10.1088/0067-0049/201/2/15}{\color{blue}\apjs},
  \href{https://ui.adsabs.harvard.edu/abs/2012ApJS..201...15H}{201, 15}

\bibitem[{{Husser} {et~al.}(2013){Husser}, {Wende-von Berg}, {Dreizler},
  {Homeier}, {Reiners}, {Barman}, \& {Hauschildt}}]{Hus13}
{Husser}, T.-O., {Wende-von Berg}, S., {Dreizler}, S., {et~al.} 2013,
  \href{http://dx.doi.org/10.1051/0004-6361/201219058}{\color{blue}\aap},
  \href{http://adsabs.harvard.edu/abs/2013A%26A...553A...6H}{553, A6}

\bibitem[{{Ishikawa} {et~al.}(2020){Ishikawa}, {Aoki}, {Kotani}, {Kuzuhara},
  {Omiya}, {Reiners}, \& {Zechmeister}}]{Tak20}
{Ishikawa}, H.~T., {Aoki}, W., {Kotani}, T., {et~al.} 2020,
  \href{http://dx.doi.org/10.1093/pasj/psaa101}{\color{blue}\pasj},
  \href{https://ui.adsabs.harvard.edu/abs/2020PASJ...72..102I}{72, 102}

\bibitem[{{Johnson} {et~al.}(2010){Johnson}, {Aller}, {Howard}, \&
  {Crepp}}]{Joh10}
{Johnson}, J.~A., {Aller}, K.~M., {Howard}, A.~W., \& {Crepp}, J.~R. 2010,
  \href{http://dx.doi.org/10.1086/655775}{\color{blue}\pasp},
  \href{https://ui.adsabs.harvard.edu/abs/2010PASP..122..905J}{122, 905}

\bibitem[{{Kausch} {et~al.}(2015){Kausch}, {Noll}, {Smette}, {Kimeswenger},
  {Barden}, {Szyszka}, {Jones}, {Sana}, {Horst}, \& {Kerber}}]{Kau15}
{Kausch}, W., {Noll}, S., {Smette}, A., {et~al.} 2015,
  \href{http://dx.doi.org/10.1051/0004-6361/201423909}{\color{blue}\aap},
  \href{https://ui.adsabs.harvard.edu/abs/2015A&A...576A..78K}{576, A78}

\bibitem[{{Khata} {et~al.}(2020){Khata}, {Mondal}, {Das}, {Ghosh}, \&
  {Ghosh}}]{Kha20}
{Khata}, D., {Mondal}, S., {Das}, R., {Ghosh}, S., \& {Ghosh}, S. 2020,
  \href{http://dx.doi.org/10.1093/mnras/staa427}{\color{blue}\mnras},
  \href{https://ui.adsabs.harvard.edu/abs/2020MNRAS.493.4533K}{493, 4533}

\bibitem[{{Kirkpatrick} {et~al.}(1991){Kirkpatrick}, {Henry}, \&
  {McCarthy}}]{Kir91}
{Kirkpatrick}, J.~D., {Henry}, T.~J., \& {McCarthy}, Donald~W., J. 1991,
  \href{http://dx.doi.org/10.1086/191611}{\color{blue}\apjs},
  \href{https://ui.adsabs.harvard.edu/abs/1991ApJS...77..417K}{77, 417}

\bibitem[{{Kopparapu} {et~al.}(2013){Kopparapu}, {Ramirez}, {Kasting}, {Eymet},
  {Robinson}, {Mahadevan}, {Terrien}, {Domagal-Goldman}, {Meadows}, \&
  {Deshpande}}]{Kop13}
{Kopparapu}, R.~K., {Ramirez}, R., {Kasting}, J.~F., {et~al.} 2013,
  \href{http://dx.doi.org/10.1088/0004-637X/765/2/131}{\color{blue}\apj},
  \href{https://ui.adsabs.harvard.edu/abs/2013ApJ...765..131K}{765, 131}

\bibitem[{{Kotani} {et~al.}(2018){Kotani}, {Tamura}, {Nishikawa}, {Ueda},
  {Kuzuhara}, {Omiya}, {Hashimoto}, {Ishizuka}, {Hirano}, {Suto}, {Kurokawa},
  {Kokubo}, {Mori}, {Tanaka}, {Kashiwagi}, {Konishi}, {Kudo}, {Sato},
  {Jacobson}, {Hodapp}, {Hall}, {Aoki}, {Usuda}, {Nishiyama}, {Nakajima},
  {Ikeda}, {Yamamuro}, {Morino}, {Baba}, {Hosokawa}, {Ishikawa}, {Narita},
  {Kokubo}, {Hayano}, {Izumiura}, {Kambe}, {Kusakabe}, {Kwon}, {Ikoma}, {Hori},
  {Genda}, {Fukui}, {Fujii}, {Kawahara}, {Olivier}, {Jovanovic}, {Harakawa},
  {Hayashi}, {Hidai}, {Machida}, {Matsuo}, {Nagata}, {Ogihara}, {Takami},
  {Takato}, {Terada}, \& {Oh}}]{Kot18}
{Kotani}, T., {Tamura}, M., {Nishikawa}, J., {et~al.} 2018, in Society of
  Photo-Optical Instrumentation Engineers (SPIE) Conference Series, Vol. 10702,
  \href{https://ui.adsabs.harvard.edu/abs/2018SPIE10702E..11K}{1070211}

\bibitem[{{Kurucz}(2014)}]{Kur14}
{Kurucz}, R.~L. 2014, {Problems with Atomic and Molecular Data: Including All
  the Lines} (Springer, Cham), 63--73

\bibitem[{{Lafarga} {et~al.}(2020){Lafarga}, {Ribas}, {Lovis}, {Perger},
  {Zechmeister}, {Bauer}, {K{\"u}rster}, {Cort{\'e}s-Contreras}, {Morales},
  {Herrero}, {Rosich}, {Baroch}, {Reiners}, {Caballero}, {Quirrenbach},
  {Amado}, {Alacid}, {B{\'e}jar}, {Dreizler}, {Hatzes}, {Henning}, {Jeffers},
  {Kaminski}, {Montes}, {Pedraz}, {Rodr{\'\i}guez-L{\'o}pez}, \&
  {Schmitt}}]{Laf20}
{Lafarga}, M., {Ribas}, I., {Lovis}, C., {et~al.} 2020,
  \href{http://dx.doi.org/10.1051/0004-6361/201937222}{\color{blue}\aap},
  \href{https://ui.adsabs.harvard.edu/abs/2020A&A...636A..36L}{636, A36}

\bibitem[{{Landi Degl'Innocenti} \& {Landolfi}(2004)}]{Lan04}
{Landi Degl'Innocenti}, E. \& {Landolfi}, M. 2004, {Polarization in Spectral
  Lines}, Vol. 307 (Springer Netherlands)

\bibitem[{{Laughlin} {et~al.}(1997){Laughlin}, {Bodenheimer}, \&
  {Adams}}]{Lau97}
{Laughlin}, G., {Bodenheimer}, P., \& {Adams}, F.~C. 1997,
  \href{http://dx.doi.org/10.1086/304125}{\color{blue}\apj},
  \href{https://ui.adsabs.harvard.edu/abs/1997ApJ...482..420L}{482, 420}

\bibitem[{{Li} {et~al.}(2021){Li}, {Liu}, {Zhang}, {Tian}, {Qiu}, \&
  {Tian}}]{Li21}
{Li}, J., {Liu}, C., {Zhang}, B., {et~al.} 2021,
  \href{http://dx.doi.org/10.3847/1538-4365/abe1c1}{\color{blue}\apjs},
  \href{https://ui.adsabs.harvard.edu/abs/2021ApJS..253...45L}{253, 45}

\bibitem[{{Lohr} {et~al.}(2018){Lohr}, {Negueruela}, {Tabernero}, {Clark},
  {Lewis}, \& {Roche}}]{Loh18}
{Lohr}, M.~E., {Negueruela}, I., {Tabernero}, H.~M., {et~al.} 2018,
  \href{http://dx.doi.org/10.1093/mnras/sty1280}{\color{blue}\mnras},
  \href{https://ui.adsabs.harvard.edu/abs/2018MNRAS.478.3825L}{478, 3825}

\bibitem[{{Luck}(2017)}]{Luc17}
{Luck}, R.~E. 2017,
  \href{http://dx.doi.org/10.3847/1538-3881/153/1/21}{\color{blue}\aj},
  \href{https://ui.adsabs.harvard.edu/abs/2017AJ....153...21L}{153, 21}

\bibitem[{{Luque} {et~al.}(2019){Luque}, {Pall{\'e}}, {Kossakowski},
  {Dreizler}, {Kemmer}, {Espinoza}, {Burt}, {Anglada-Escud{\'e}}, {B{\'e}jar},
  {Caballero}, {Collins}, {Collins}, {Cort{\'e}s-Contreras},
  {D{\'\i}ez-Alonso}, {Feng}, {Hatzes}, {Hellier}, {Henning}, {Jeffers},
  {Kaltenegger}, {K{\"u}rster}, {Madden}, {Molaverdikhani}, {Montes}, {Narita},
  {Nowak}, {Ofir}, {Oshagh}, {Parviainen}, {Quirrenbach}, {Reffert}, {Reiners},
  {Rodr{\'\i}guez-L{\'o}pez}, {Schlecker}, {Stock}, {Trifonov}, {Winn},
  {Zapatero Osorio}, {Zechmeister}, {Amado}, {Anderson}, {Batalha}, {Bauer},
  {Bluhm}, {Burke}, {Butler}, {Caldwell}, {Chen}, {Crane}, {Dragomir},
  {Dressing}, {Dynes}, {Jenkins}, {Kaminski}, {Klahr}, {Kotani}, {Lafarga},
  {Latham}, {Lewin}, {McDermott}, {Monta{\~n}{\'e}s-Rodr{\'\i}guez}, {Morales},
  {Murgas}, {Nagel}, {Pedraz}, {Ribas}, {Ricker}, {Rowden}, {Seager},
  {Shectman}, {Tamura}, {Teske}, {Twicken}, {Vanderspeck}, {Wang}, \&
  {Wohler}}]{Luq19}
{Luque}, R., {Pall{\'e}}, E., {Kossakowski}, D., {et~al.} 2019,
  \href{http://dx.doi.org/10.1051/0004-6361/201935801}{\color{blue}\aap},
  \href{https://ui.adsabs.harvard.edu/abs/2019A&A...628A..39L}{628, A39}

\bibitem[{{Mahadevan} {et~al.}(2012){Mahadevan}, {Ramsey}, {Bender}, {Terrien},
  {Wright}, {Halverson}, {Hearty}, {Nelson}, {Burton}, {Redman}, {Osterman},
  {Diddams}, {Kasting}, {Endl}, \& {Deshpande}}]{Mah12}
{Mahadevan}, S., {Ramsey}, L., {Bender}, C., {et~al.} 2012, in Society of
  Photo-Optical Instrumentation Engineers (SPIE) Conference Series, Vol. 8446,
  \href{https://ui.adsabs.harvard.edu/abs/2012SPIE.8446E..1SM}{84461S}

\bibitem[{{Maldonado} {et~al.}(2015){Maldonado}, {Affer}, {Micela},
  {Scandariato}, {Damasso}, {Stelzer}, {Barbieri}, {Bedin}, {Biazzo},
  {Bignamini}, {Borsa}, {Claudi}, {Covino}, {Desidera}, {Esposito}, {Gratton},
  {Gonz{\'a}lez Hern{\'a}ndez}, {Lanza}, {Maggio}, {Molinari}, {Pagano},
  {Perger}, {Pillitteri}, {Piotto}, {Poretti}, {Prisinzano}, {Rebolo}, {Ribas},
  {Shkolnik}, {Southworth}, {Sozzetti}, \& {Su{\'a}rez Mascare{\~n}o}}]{Mal15}
{Maldonado}, J., {Affer}, L., {Micela}, G., {et~al.} 2015,
  \href{http://dx.doi.org/10.1051/0004-6361/201525797}{\color{blue}\aap},
  \href{https://ui.adsabs.harvard.edu/abs/2015A&A...577A.132M}{577, A132}

\bibitem[{{Maldonado} {et~al.}(2020){Maldonado}, {Micela}, {Baratella},
  {D'Orazi}, {Affer}, {Biazzo}, {Lanza}, {Maggio}, {Gonz{\'a}lez
  Hern{\'a}ndez}, {Perger}, {Pinamonti}, {Scandariato}, {Sozzetti}, {Locci},
  {Di Maio}, {Bignamini}, {Claudi}, {Molinari}, {Rebolo}, {Ribas},
  {Toledo-Padr{\'o}n}, {Covino}, {Desidera}, {Herrero}, {Morales},
  {Su{\'a}rez-Mascare{\~n}o}, {Pagano}, {Petralia}, {Piotto}, \&
  {Poretti}}]{Mal20}
{Maldonado}, J., {Micela}, G., {Baratella}, M., {et~al.} 2020,
  \href{http://dx.doi.org/10.1051/0004-6361/202039478}{\color{blue}\aap},
  \href{https://ui.adsabs.harvard.edu/abs/2020A&A...644A..68M}{644, A68}

\bibitem[{{Mamajek} {et~al.}(2013){Mamajek}, {Bartlett}, {Seifahrt}, {Henry},
  {Dieterich}, {Lurie}, {Kenworthy}, {Jao}, {Riedel}, {Subasavage}, {Winters},
  {Finch}, {Ianna}, \& {Bean}}]{Mam13}
{Mamajek}, E.~E., {Bartlett}, J.~L., {Seifahrt}, A., {et~al.} 2013,
  \href{http://dx.doi.org/10.1088/0004-6256/146/6/154}{\color{blue}\aj},
  \href{https://ui.adsabs.harvard.edu/abs/2013AJ....146..154M}{146, 154}

\bibitem[{{Mann} {et~al.}(2013{\natexlab{a}}){Mann}, {Brewer}, {Gaidos},
  {L{\'e}pine}, \& {Hilton}}]{Man13a}
{Mann}, A.~W., {Brewer}, J.~M., {Gaidos}, E., {L{\'e}pine}, S., \& {Hilton},
  E.~J. 2013{\natexlab{a}},
  \href{http://dx.doi.org/10.1088/0004-6256/145/2/52}{\color{blue}\aj},
  \href{https://ui.adsabs.harvard.edu/abs/2013AJ....145...52M}{145, 52}

\bibitem[{{Mann} {et~al.}(2014){Mann}, {Deacon}, {Gaidos}, {Ansdell}, {Brewer},
  {Liu}, {Magnier}, \& {Aller}}]{Man14}
{Mann}, A.~W., {Deacon}, N.~R., {Gaidos}, E., {et~al.} 2014,
  \href{http://dx.doi.org/10.1088/0004-6256/147/6/160}{\color{blue}\aj},
  \href{https://ui.adsabs.harvard.edu/abs/2014AJ....147..160M}{147, 160}

\bibitem[{{Mann} {et~al.}(2015){Mann}, {Feiden}, {Gaidos}, {Boyajian}, \& {von
  Braun}}]{Man15}
{Mann}, A.~W., {Feiden}, G.~A., {Gaidos}, E., {Boyajian}, T., \& {von Braun},
  K. 2015,
  \href{http://dx.doi.org/10.1088/0004-637X/804/1/64}{\color{blue}\apj},
  \href{https://ui.adsabs.harvard.edu/abs/2015ApJ...804...64M}{804, 64}

\bibitem[{{Mann} {et~al.}(2013{\natexlab{b}}){Mann}, {Gaidos}, \&
  {Ansdell}}]{Man13b}
{Mann}, A.~W., {Gaidos}, E., \& {Ansdell}, M. 2013{\natexlab{b}},
  \href{http://dx.doi.org/10.1088/0004-637X/779/2/188}{\color{blue}\apj},
  \href{https://ui.adsabs.harvard.edu/abs/2013ApJ...779..188M}{779, 188}

\bibitem[{{Marfil} {et~al.}(2020){Marfil}, {Tabernero}, {Montes}, {Caballero},
  {Soto}, {Gonz{\'a}lez Hern{\'a}ndez}, {Kaminski}, {Nagel}, {Jeffers},
  {Reiners}, {Ribas}, {Quirrenbach}, \& {Amado}}]{Mar20}
{Marfil}, E., {Tabernero}, H.~M., {Montes}, D., {et~al.} 2020,
  \href{http://dx.doi.org/10.1093/mnras/staa058}{\color{blue}\mnras},
  \href{https://ui.adsabs.harvard.edu/abs/2020MNRAS.492.5470M}{492, 5470}

\bibitem[{{Masseron} {et~al.}(2016){Masseron}, {Merle}, \& {Hawkins}}]{Mas16}
{Masseron}, T., {Merle}, T., \& {Hawkins}, K. 2016, {BACCHUS: Brussels
  Automatic Code for Characterizing High accUracy Spectra}, Astrophysics Source
  Code Library

\bibitem[{{McKemmish} {et~al.}(2016){McKemmish}, {Yurchenko}, \&
  {Tennyson}}]{kem16}
{McKemmish}, L.~K., {Yurchenko}, S.~N., \& {Tennyson}, J. 2016,
  \href{http://dx.doi.org/10.1093/mnras/stw1969}{\color{blue}\mnras},
  \href{https://ui.adsabs.harvard.edu/abs/2016MNRAS.463..771M}{463, 771}

\bibitem[{{Miret-Roig} {et~al.}(2020){Miret-Roig}, {Galli}, {Brandner}, {Bouy},
  {Barrado}, {Olivares}, {Antoja}, {Romero-G{\'o}mez}, {Figueras}, \&
  {Lillo-Box}}]{Mir20}
{Miret-Roig}, N., {Galli}, P.~A.~B., {Brandner}, W., {et~al.} 2020,
  \href{http://dx.doi.org/10.1051/0004-6361/202038765}{\color{blue}\aap},
  \href{https://ui.adsabs.harvard.edu/abs/2020A&A...642A.179M}{642, A179}

\bibitem[{{Montes} {et~al.}(2018){Montes}, {Gonz{\'a}lez-Peinado}, {Tabernero},
  {Caballero}, {Marfil}, {Alonso-Floriano}, {Cort{\'e}s-Contreras},
  {Gonz{\'a}lez Hern{\'a}ndez}, {Klutsch}, \& {Moreno-J{\'o}dar}}]{Mon18}
{Montes}, D., {Gonz{\'a}lez-Peinado}, R., {Tabernero}, H.~M., {et~al.} 2018,
  \href{http://dx.doi.org/10.1093/mnras/sty1295}{\color{blue}\mnras},
  \href{https://ui.adsabs.harvard.edu/abs/2018MNRAS.479.1332M}{479, 1332}

\bibitem[{{Montes} {et~al.}(2020){Montes}, {L{\'o}pez-Gallifa}, {Labarga},
  {Caballero}, {Marfil}, {Tabernero}, {Lafarga}, {Jeffers}, {Ribas}, {Reiners},
  {Quirrenbach}, {Amado}, \& {CARMENES Consortium}}]{Mon20}
{Montes}, D., {L{\'o}pez-Gallifa}, A., {Labarga}, F., {et~al.} 2020, in
  Contributions to the XIV.0 Scientific Meeting (virtual) of the Spanish
  Astronomical Society,
  \href{https://ui.adsabs.harvard.edu/abs/2020sea..confE.168M}{168}

\bibitem[{{Montes} {et~al.}(2001){Montes}, {L{\'o}pez-Santiago}, {G{\'a}lvez},
  {Fern{\'a}ndez-Figueroa}, {De Castro}, \& {Cornide}}]{Mon01}
{Montes}, D., {L{\'o}pez-Santiago}, J., {G{\'a}lvez}, M.~C., {et~al.} 2001,
  \href{http://dx.doi.org/10.1046/j.1365-8711.2001.04781.x}{\color{blue}\mnras},
  \href{https://ui.adsabs.harvard.edu/abs/2001MNRAS.328...45M}{328, 45}

\bibitem[{{Morales} {et~al.}(2009){Morales}, {Ribas}, {Jordi}, {Torres},
  {Gallardo}, {Guinan}, {Charbonneau}, {Wolf}, {Latham}, {Anglada-Escud{\'e}},
  {Bradstreet}, {Everett}, {O'Donovan}, {Mandushev}, \& {Mathieu}}]{Mor09}
{Morales}, J.~C., {Ribas}, I., {Jordi}, C., {et~al.} 2009,
  \href{http://dx.doi.org/10.1088/0004-637X/691/2/1400}{\color{blue}\apj},
  \href{https://ui.adsabs.harvard.edu/abs/2009ApJ...691.1400M}{691, 1400}

\bibitem[{{Morton}(2000)}]{Mor00}
{Morton}, D.~C. 2000,
  \href{http://dx.doi.org/10.1086/317349}{\color{blue}\apjs},
  \href{https://ui.adsabs.harvard.edu/abs/2000ApJS..130..403M}{130, 403}

\bibitem[{Nagel(2019)}]{Nagelthesis}
Nagel, E. 2019, PhD thesis, Universit\"{a}t Hamburg

\bibitem[{{Neves} {et~al.}(2012){Neves}, {Bonfils}, {Santos}, {Delfosse},
  {Forveille}, {Allard}, {Nat{\'a}rio}, {Fernand es}, \& {Udry}}]{Nev12}
{Neves}, V., {Bonfils}, X., {Santos}, N.~C., {et~al.} 2012,
  \href{http://dx.doi.org/10.1051/0004-6361/201118115}{\color{blue}\aap},
  \href{https://ui.adsabs.harvard.edu/abs/2012A&A...538A..25N}{538, A25}

\bibitem[{{Neves} {et~al.}(2014){Neves}, {Bonfils}, {Santos}, {Delfosse},
  {Forveille}, {Allard}, \& {Udry}}]{Nev14}
{Neves}, V., {Bonfils}, X., {Santos}, N.~C., {et~al.} 2014,
  \href{http://dx.doi.org/10.1051/0004-6361/201424139}{\color{blue}\aap},
  \href{https://ui.adsabs.harvard.edu/abs/2014A&A...568A.121N}{568, A121}

\bibitem[{{Newton} {et~al.}(2014){Newton}, {Charbonneau}, {Irwin},
  {Berta-Thompson}, {Rojas-Ayala}, {Covey}, \& {Lloyd}}]{New14}
{Newton}, E.~R., {Charbonneau}, D., {Irwin}, J., {et~al.} 2014,
  \href{http://dx.doi.org/10.1088/0004-6256/147/1/20}{\color{blue}\aj},
  \href{https://ui.adsabs.harvard.edu/abs/2014AJ....147...20N}{147, 20}

\bibitem[{{Nordstr{\"o}m} {et~al.}(2004){Nordstr{\"o}m}, {Mayor}, {Andersen},
  {Holmberg}, {Pont}, {J{\o}rgensen}, {Olsen}, {Udry}, \& {Mowlavi}}]{Nor04}
{Nordstr{\"o}m}, B., {Mayor}, M., {Andersen}, J., {et~al.} 2004,
  \href{http://dx.doi.org/10.1051/0004-6361:20035959}{\color{blue}\aap},
  \href{https://ui.adsabs.harvard.edu/abs/2004A&A...418..989N}{418, 989}

\bibitem[{{Nortmann} {et~al.}(2018){Nortmann}, {Pall{\'e}}, {Salz},
  {Sanz-Forcada}, {Nagel}, {Alonso-Floriano}, {Czesla}, {Yan}, {Chen},
  {Snellen}, {Zechmeister}, {Schmitt}, {L{\'o}pez-Puertas}, {Casasayas-Barris},
  {Bauer}, {Amado}, {Caballero}, {Dreizler}, {Henning}, {Lamp{\'o}n}, {Montes},
  {Molaverdikhani}, {Quirrenbach}, {Reiners}, {Ribas}, {S{\'a}nchez-L{\'o}pez},
  {Schneider}, \& {Zapatero Osorio}}]{Nor18}
{Nortmann}, L., {Pall{\'e}}, E., {Salz}, M., {et~al.} 2018,
  \href{http://dx.doi.org/10.1126/science.aat5348}{\color{blue}Science},
  \href{https://ui.adsabs.harvard.edu/abs/2018Sci...362.1388N}{362, 1388}

\bibitem[{{Olander} {et~al.}(2021){Olander}, {Heiter}, \& {Kochukhov}}]{Ola21}
{Olander}, T., {Heiter}, U., \& {Kochukhov}, O. 2021,
  \href{http://dx.doi.org/10.1051/0004-6361/202039747}{\color{blue}\aap},
  \href{https://ui.adsabs.harvard.edu/abs/2021A&A...649A.103O}{649, A103}

\bibitem[{{{\"O}nehag} {et~al.}(2012){{\"O}nehag}, {Heiter}, {Gustafsson},
  {Piskunov}, {Plez}, \& {Reiners}}]{One12}
{{\"O}nehag}, A., {Heiter}, U., {Gustafsson}, B., {et~al.} 2012,
  \href{http://dx.doi.org/10.1051/0004-6361/201118101}{\color{blue}\aap},
  \href{https://ui.adsabs.harvard.edu/abs/2012A&A...542A..33O}{542, A33}

\bibitem[{{Passegger} {et~al.}(2021){Passegger}, {Bello-Garc\'{i}a},
  {Ordieres-Mer\'{e}}, {Antoniadis-Karnavas}, {Marfil}, {Duque-Arribas},
  {Amado}, {Delgado-Mena}, {Montes}, {Rojas-Ayala}, {Schweitzer}, {Tabernero},
  {B\'{e}jar}, {Caballero}, {Hatzes}, {Henning}, {Pedraz}, {Quirrenbach},
  {Reiners}, \& {Ribas}}]{Pas20xx}
{Passegger}, V.~M., {Bello-Garc\'{i}a}, A., {Ordieres-Mer\'{e}}, J., {et~al.}
  2021, \aap, \href{none}{submitted}

\bibitem[{{Passegger} {et~al.}(2020){Passegger}, {Bello-Garc{\'\i}a},
  {Ordieres-Mer{\'e}}, {Caballero}, {Schweitzer}, {Gonz{\'a}lez-Marcos},
  {Ribas}, {Reiners}, {Quirrenbach}, {Amado}, {Azzaro}, {Bauer}, {B{\'e}jar},
  {Cort{\'e}s-Contreras}, {Dreizler}, {Hatzes}, {Henning}, {Jeffers},
  {Kaminski}, {K{\"u}rster}, {Lafarga}, {Marfil}, {Montes}, {Morales}, {Nagel},
  {Sarro}, {Solano}, {Tabernero}, \& {Zechmeister}}]{Pas20}
{Passegger}, V.~M., {Bello-Garc{\'\i}a}, A., {Ordieres-Mer{\'e}}, J., {et~al.}
  2020, \href{http://dx.doi.org/10.1051/0004-6361/202038787}{\color{blue}\aap},
  \href{https://ui.adsabs.harvard.edu/abs/2020A&A...642A..22P}{642, A22}

\bibitem[{{Passegger} {et~al.}(2018){Passegger}, {Reiners}, {Jeffers},
  {Wende-von Berg}, {Sch{\"o}fer}, {Caballero}, {Schweitzer}, {Amado},
  {B{\'e}jar}, {Cort{\'e}s-Contreras}, {Hatzes}, {K{\"u}rster}, {Montes},
  {Pedraz}, {Quirrenbach}, {Ribas}, \& {Seifert}}]{Pas18}
{Passegger}, V.~M., {Reiners}, A., {Jeffers}, S.~V., {et~al.} 2018,
  \href{http://dx.doi.org/10.1051/0004-6361/201732312}{\color{blue}\aap},
  \href{http://adsabs.harvard.edu/abs/2018A%26A...615A...6P}{615, A6}

\bibitem[{{Passegger} {et~al.}(2019){Passegger}, {Schweitzer}, {Shulyak},
  {Nagel}, {Hauschildt}, {Reiners}, {Amado}, {Caballero},
  {Cort{\'e}s-Contreras}, {Dom{\'\i}nguez-Fern{\'a}ndez}, {Quirrenbach},
  {Ribas}, {Azzaro}, {Anglada-Escud{\'e}}, {Bauer}, {B{\'e}jar}, {Dreizler},
  {Guenther}, {Henning}, {Jeffers}, {Kaminski}, {K{\"u}rster}, {Lafarga},
  {Mart{\'\i}n}, {Montes}, {Morales}, {Schmitt}, \& {Zechmeister}}]{Pas19}
{Passegger}, V.~M., {Schweitzer}, A., {Shulyak}, D., {et~al.} 2019,
  \href{http://dx.doi.org/10.1051/0004-6361/201935679}{\color{blue}\aap},
  \href{https://ui.adsabs.harvard.edu/abs/2019A&A...627A.161P}{627, A161}

\bibitem[{{Passegger} {et~al.}(2016){Passegger}, {Wende-von Berg}, \&
  {Reiners}}]{Pas16}
{Passegger}, V.~M., {Wende-von Berg}, S., \& {Reiners}, A. 2016,
  \href{http://dx.doi.org/10.1051/0004-6361/201322261}{\color{blue}\aap},
  \href{https://ui.adsabs.harvard.edu/abs/2016A&A...587A..19P}{587, A19}

\bibitem[{{Plez}(2012)}]{Ple12}
{Plez}, B. 2012, {Turbospectrum: Code for spectral synthesis}, Astrophysics
  Source Code Library

\bibitem[{{Quirrenbach} {et~al.}(2020){Quirrenbach}, , {Amado}, {Ribas},
  {Reiners}, {Caballero}, {Seifert}, {Aceituno}, {Azzaro}, {Baroch}, {Barrado},
  {Bauer}, {Becerril}, {B{\`e}jar}, {Ben{\'\i}tez}, {Brinkm{\"o}ller}, {Cardona
  Guill{\'e}n}, {Cifuentes}, {Colom{\'e}}, {Cort{\'e}s-Contreras}, {Czesla},
  {Dreizler}, {Fr{\"o}lich}, {Fuhrmeister}, {Galad{\'\i}-Enr{\'\i}quez},
  {Gonz{\'a}lez Hern{\'a}ndez}, {Gonz{\'a}lez Peinado}, {Guenther}, {de
  Guindos}, {Hagen}, {Hatzes}, {Hauschildt}, {Helmling}, {Henning}, {Herbort},
  {Hern{\'a}ndez Casta{\~n}o}, {Herrero}, {Hintz}, {Jeffers}, {Johnson}, {de
  Juan}, {Kaminski}, {Klahr}, {K{\"u}rster}, {Lafarga}, {Sairam}, {Lamp{\'o}n},
  {Lara}, {Launhardt}, {L{\'o}pez del Fresno}, {L{\'o}pez-Puertas}, {Luque},
  {Mandel}, {Marfil}, {Mart{\'\i}n}, {Mart{\'\i}n-Ruiz}, {Mathar}, {Montes},
  {Morales}, {Nagel}, {Nortmann}, {Nowak}, {Pall{\'e}}, {Passegger}, {Pavlov},
  {Pedraz}, {P{\'e}rez-Medialdea}, {Perger}, {Rebolo}, {Reffert},
  {Rodr{\'\i}guez}, {Rodr{\'\i}guez L{\'o}pez}, {Rosich}, {Sabotta}, {Sadegi},
  {Salz}, {S{\'a}nchez-L{\'o}pez}, {Sanz-Forcada}, {Sarkis}, {Sch{\"a}fer},
  {Schiller}, {Schmitt}, {Sch{\"o}fer}, {Schweitzer}, {Shulyak}, {Solano},
  {Stahl}, {Tala Pinto}, {Trifonov}, {Zapatero Osorio}, {Yan}, {Zechmeister},
  {Abell{\'a}n}, {Abril}, {Alonso-Floriano}, {Ammler-von Eiff},
  {Anglada-Escud{\'e}}, {Anwand-Heerwart}, {Arroyo-Torres}, {Berdi{\~n}as},
  {Bergondy}, {Bl{\"u}mcke}, {del Burgo}, {Cano}, {Carro}, {C{\'a}rdenas},
  {Casal}, {Claret}, {D{\'\i}ez-Alonso}, {Doellinger}, {Dorda}, {Feiz},
  {Fern{\'a}ndez}, {Ferro}, {Gaisn{\'e}}, {Gallardo}, {G{\'a}lvez-Ortiz},
  {Garc{\'\i}a-Piquer}, {Garc{\'\i}a-Vargas}, {Garrido}, {Gesa}, {G{\'o}mez
  Galera}, {Gonz{\'a}lez-{\'A}lvarez}, {Gonz{\'a}lez-Cuesta}, {Grohnert},
  {Gr{\"o}zinger}, {Gu{\`a}rdia}, {Guijarro}, {Hedrosa}, {Hermann}, {Hermelo},
  {Hern{\'a}ndez Arab{\'\i}}, {Hern{\'a}ndez Hernando}, {Hidalgo}, {Holgado},
  {Huber}, {Huber}, {Huke}, {Kehr}, {Kim}, {Klein}, {Kl{\"u}ter}, {Klutsch},
  {Labarga}, {Labiche}, {Lamert}, {Laun}, {L{\'a}zaro}, {Lemke}, {Lenzen},
  {Llamas}, {Lizon}, {Lodieu}, {L{\'o}pez Gonz{\'a}lez}, {L{\'o}pez-Morales},
  {L{\'o}pez Salas}, {L{\'o}pez-Santiago}, {Mag{\'a}n Madinabeitia}, {Mall},
  {Mancini}, {Mar{\'\i}n Molina}, {Mart{\'\i}nez-Rodr{\'\i}guez}, {Maroto
  Fern{\'a}ndez}, {Marvin}, {Mirabet}, {Moreno-Raya}, {Moya}, {Mundt},
  {Naranjo}, {Panduro}, {Pascual}, {P{\'e}rez-Calpena}, {Perryman}, {Pluto},
  {Ram{\'o}n}, {Redondo}, {Reinhart}, {Rhode}, {Rix}, {Rodler}, {Rohloff},
  {S{\'a}nchez-Blanco}, {S{\'a}nchez Carrasco}, {Sarmiento}, {Schmidt},
  {Storz}, {Strachan}, {St{\"u}rmer}, {Su{\'a}rez}, {Tabernero}, {Tal-Or},
  {Tulloch}, {Ulbrich}, {Veredas}, {Vico Linares}, {Vidal-Dasilva},
  {Vilardell}, {Wagner}, {Winkler}, {Wolthoff}, {Xu}, \& {Zhao}}]{Qui20}
{Quirrenbach}, A., , {Amado}, P.~J., {et~al.} 2020, in Society of Photo-Optical
  Instrumentation Engineers (SPIE) Conference Series, Vol. 11447,
  \href{https://ui.adsabs.harvard.edu/abs/2020SPIE11447E..3CQ}{114473C}

\bibitem[{{Rajpurohit} {et~al.}(2018{\natexlab{a}}){Rajpurohit}, {Allard},
  {Rajpurohit}, {Sharma}, {Teixeira}, {Mousis}, \& {Kamlesh}}]{Raj18b}
{Rajpurohit}, A.~S., {Allard}, F., {Rajpurohit}, S., {et~al.}
  2018{\natexlab{a}},
  \href{http://dx.doi.org/10.1051/0004-6361/201833500}{\color{blue}\aap},
  \href{https://ui.adsabs.harvard.edu/abs/2018A&A...620A.180R}{620, A180}

\bibitem[{{Rajpurohit} {et~al.}(2018{\natexlab{b}}){Rajpurohit}, {Allard},
  {Teixeira}, {Homeier}, {Rajpurohit}, \& {Mousis}}]{Raj18a}
{Rajpurohit}, A.~S., {Allard}, F., {Teixeira}, G.~D.~C., {et~al.}
  2018{\natexlab{b}},
  \href{http://dx.doi.org/10.1051/0004-6361/201731507}{\color{blue}\aap},
  \href{https://ui.adsabs.harvard.edu/abs/2018A&A...610A..19R}{610, A19}

\bibitem[{{Rajpurohit} {et~al.}(2014){Rajpurohit}, {Reyl{\'e}}, {Allard},
  {Scholz}, {Homeier}, {Schultheis}, \& {Bayo}}]{Raj14}
{Rajpurohit}, A.~S., {Reyl{\'e}}, C., {Allard}, F., {et~al.} 2014,
  \href{http://dx.doi.org/10.1051/0004-6361/201322881}{\color{blue}\aap},
  \href{https://ui.adsabs.harvard.edu/abs/2014A&A...564A..90R}{564, A90}

\bibitem[{{Rajpurohit} {et~al.}(2012){Rajpurohit}, {Reyl{\'e}}, {Schultheis},
  {Leinert}, {Allard}, {Homeier}, {Ratzka}, {Abraham}, {Moster}, {Witte}, \&
  {Ryde}}]{Raj12}
{Rajpurohit}, A.~S., {Reyl{\'e}}, C., {Schultheis}, M., {et~al.} 2012,
  \href{http://dx.doi.org/10.1051/0004-6361/201219029}{\color{blue}\aap},
  \href{https://ui.adsabs.harvard.edu/abs/2012A&A...545A..85R}{545, A85}

\bibitem[{{Reiners} {et~al.}(2018){Reiners}, {Zechmeister}, {Caballero},
  {Ribas}, {Morales}, {Jeffers}, {Sch{\"o}fer}, {Tal-Or}, {Quirrenbach}, \&
  {Amado}}]{Rei18}
{Reiners}, A., {Zechmeister}, M., {Caballero}, J.~A., {et~al.} 2018,
  \href{http://dx.doi.org/10.1051/0004-6361/201732054}{\color{blue}\aap},
  \href{https://ui.adsabs.harvard.edu/abs/2018A&A...612A..49R}{612, A49}

\bibitem[{{Reyl{\'e}} {et~al.}(2021){Reyl{\'e}}, {Jardine}, {Fouqu{\'e}},
  {Caballero}, {Smart}, \& {Sozzetti}}]{Rey21}
{Reyl{\'e}}, C., {Jardine}, K., {Fouqu{\'e}}, P., {et~al.} 2021,
  \href{http://dx.doi.org/10.1051/0004-6361/202140985}{\color{blue}\aap},
  \href{https://ui.adsabs.harvard.edu/abs/2021A&A...650A.201R}{650, A201}

\bibitem[{{Rojas-Ayala} {et~al.}(2012){Rojas-Ayala}, {Covey}, {Muirhead}, \&
  {Lloyd}}]{Roj12}
{Rojas-Ayala}, B., {Covey}, K.~R., {Muirhead}, P.~S., \& {Lloyd}, J.~P. 2012,
  \href{http://dx.doi.org/10.1088/0004-637X/748/2/93}{\color{blue}\apj},
  \href{https://ui.adsabs.harvard.edu/abs/2012ApJ...748...93R}{748, 93}

\bibitem[{{Rothman} {et~al.}(2009){Rothman}, {Gordon}, {Barbe}, {Benner},
  {Bernath}, {Birk}, {Boudon}, {Brown}, {Campargue}, {Champion}, {Chance},
  {Coudert}, {Dana}, {Devi}, {Fally}, {Flaud}, {Gamache}, {Goldman},
  {Jacquemart}, {Kleiner}, {Lacome}, {Lafferty}, {Mandin}, {Massie},
  {Mikhailenko}, {Miller}, {Moazzen-Ahmadi}, {Naumenko}, {Nikitin}, {Orphal},
  {Perevalov}, {Perrin}, {Predoi-Cross}, {Rinsland}, {Rotger},
  {{\v{S}}ime{\v{c}}kov{\'a}}, {Smith}, {Sung}, {Tashkun}, {Tennyson}, {Toth},
  {Vandaele}, \& {Vander Auwera}}]{Rot09}
{Rothman}, L.~S., {Gordon}, I.~E., {Barbe}, A., {et~al.} 2009,
  \href{http://dx.doi.org/10.1016/j.jqsrt.2009.02.013}{\color{blue}\jqsrt},
  \href{https://ui.adsabs.harvard.edu/abs/2009JQSRT.110..533R}{110, 533}

\bibitem[{{Ryabchikova} {et~al.}(2015){Ryabchikova}, {Piskunov}, {Kurucz},
  {Stempels}, {Heiter}, {Pakhomov}, \& {Barklem}}]{Rya15}
{Ryabchikova}, T., {Piskunov}, N., {Kurucz}, R.~L., {et~al.} 2015,
  \href{http://dx.doi.org/10.1088/0031-8949/90/5/054005}{\color{blue}\physscr},
  \href{http://adsabs.harvard.edu/abs/2015PhyS...90e4005R}{90, 054005}

\bibitem[{{Sabotta} {et~al.}(2021){Sabotta}, {Schlecker}, {Chaturvedi},
  {Guenther}, {Mu{\~n}oz Rodr{\'\i}guez}, {Mu{\~n}oz S{\'a}nchez}, {Caballero},
  {Shan}, {Reffert}, {Ribas}, {Reiners}, {Hatzes}, {Amado}, {Klahr}, {Morales},
  {Quirrenbach}, {Henning}, {Dreizler}, {Pall{\'e}}, {Perger}, {Azzaro},
  {Jeffers}, {Kaminski}, {K{\"u}rster}, {Lafarga}, {Montes}, {Passegger}, \&
  {Zechmeister}}]{Sab21}
{Sabotta}, S., {Schlecker}, M., {Chaturvedi}, P., {et~al.} 2021,
  \href{http://dx.doi.org/10.1051/0004-6361/202140968}{\color{blue}\aap},
  \href{https://ui.adsabs.harvard.edu/abs/2021arXiv210703802S}{653, A114}

\bibitem[{{Sarmento} {et~al.}(2021){Sarmento}, {Rojas-Ayala}, {Delgado Mena},
  \& {Blanco-Cuaresma}}]{Sar21}
{Sarmento}, P., {Rojas-Ayala}, B., {Delgado Mena}, E., \& {Blanco-Cuaresma}, S.
  2021, \href{http://dx.doi.org/10.1051/0004-6361/202039703}{\color{blue}\aap},
  \href{https://ui.adsabs.harvard.edu/abs/2021A&A...649A.147S}{649, A147}

\bibitem[{{Sarro} {et~al.}(2018){Sarro}, {Ordieres-Mer{\'e}},
  {Bello-Garc{\'\i}a}, {Gonz{\'a}lez-Marcos}, \& {Solano}}]{Sar18}
{Sarro}, L.~M., {Ordieres-Mer{\'e}}, J., {Bello-Garc{\'\i}a}, A.,
  {Gonz{\'a}lez-Marcos}, A., \& {Solano}, E. 2018,
  \href{http://dx.doi.org/10.1093/mnras/sty165}{\color{blue}\mnras},
  \href{https://ui.adsabs.harvard.edu/abs/2018MNRAS.476.1120S}{476, 1120}

\bibitem[{{Scalo} {et~al.}(2007){Scalo}, {Kaltenegger}, {Segura}, {Fridlund},
  {Ribas}, {Kulikov}, {Grenfell}, {Rauer}, {Odert}, {Leitzinger}, {Selsis},
  {Khodachenko}, {Eiroa}, {Kasting}, \& {Lammer}}]{Sca07}
{Scalo}, J., {Kaltenegger}, L., {Segura}, A.~G., {et~al.} 2007,
  \href{http://dx.doi.org/10.1089/ast.2006.0125}{\color{blue}Astrobiology},
  \href{https://ui.adsabs.harvard.edu/abs/2007AsBio...7...85S}{7, 85}

\bibitem[{{Sch{\"o}fer} {et~al.}(2019){Sch{\"o}fer}, {Jeffers}, {Reiners},
  {Shulyak}, {Fuhrmeister}, {Johnson}, {Zechmeister}, {Ribas}, {Quirrenbach},
  {Amado}, {Caballero}, {Anglada-Escud{\'e}}, {Bauer}, {B{\'e}jar},
  {Cort{\'e}s-Contreras}, {Dreizler}, {Guenther}, {Kaminski}, {K{\"u}rster},
  {Lafarga}, {Montes}, {Morales}, {Pedraz}, \& {Tal-Or}}]{Scho19}
{Sch{\"o}fer}, P., {Jeffers}, S.~V., {Reiners}, A., {et~al.} 2019,
  \href{http://dx.doi.org/10.1051/0004-6361/201834114}{\color{blue}\aap},
  \href{https://ui.adsabs.harvard.edu/abs/2019A&A...623A..44S}{623, A44}

\bibitem[{{Schwab} {et~al.}(2016){Schwab}, {Rakich}, {Gong}, {Mahadevan},
  {Halverson}, {Roy}, {Terrien}, {Robertson}, {Hearty}, {Levi}, {Monson},
  {Wright}, {McElwain}, {Bender}, {Blake}, {St{\"u}rmer}, {Gurevich},
  {Chakraborty}, \& {Ramsey}}]{Schw16}
{Schwab}, C., {Rakich}, A., {Gong}, Q., {et~al.} 2016, in Ground-based and
  Airborne Instrumentation for Astronomy VI, Vol. 9908,
  \href{https://ui.adsabs.harvard.edu/abs/2016SPIE.9908E..7HS}{99087H}

\bibitem[{{Schweitzer} {et~al.}(2019){Schweitzer}, {Passegger}, {Cifuentes},
  {B{\'e}jar}, {Cort{\'e}s-Contreras}, {Caballero}, {del Burgo}, {Czesla},
  {K{\"u}rster}, {Montes}, {Zapatero Osorio}, {Ribas}, {Reiners},
  {Quirrenbach}, {Amado}, {Aceituno}, {Anglada-Escud{\'e}}, {Bauer},
  {Dreizler}, {Jeffers}, {Guenther}, {Henning}, {Kaminski}, {Lafarga},
  {Marfil}, {Morales}, {Schmitt}, {Seifert}, {Solano}, {Tabernero}, \&
  {Zechmeister}}]{Sch19}
{Schweitzer}, A., {Passegger}, V.~M., {Cifuentes}, C., {et~al.} 2019,
  \href{http://dx.doi.org/10.1051/0004-6361/201834965}{\color{blue}\aap},
  \href{https://ui.adsabs.harvard.edu/abs/2019A&A...625A..68S}{625, A68}

\bibitem[{{Seifahrt} {et~al.}(2020){Seifahrt}, {Bean}, {St{\"u}rmer}, {Kasper},
  {Gers}, {Schwab}, {Zechmeister}, {Stef{\'a}nsson}, {Montet}, {Dos Santos},
  {Peck}, {White}, \& {Tapia}}]{Sei20}
{Seifahrt}, A., {Bean}, J.~L., {St{\"u}rmer}, J., {et~al.} 2020, in Society of
  Photo-Optical Instrumentation Engineers (SPIE) Conference Series, Vol. 11447,
  \href{https://ui.adsabs.harvard.edu/abs/2020SPIE11447E..1FS}{114471F}

\bibitem[{{Selsis} {et~al.}(2007){Selsis}, {Kasting}, {Levrard}, {Paillet},
  {Ribas}, \& {Delfosse}}]{Sel07}
{Selsis}, F., {Kasting}, J.~F., {Levrard}, B., {et~al.} 2007,
  \href{http://dx.doi.org/10.1051/0004-6361:20078091}{\color{blue}\aap},
  \href{https://ui.adsabs.harvard.edu/abs/2007A&A...476.1373S}{476, 1373}

\bibitem[{{Shan} {et~al.}(2021){Shan}, {Reiners}, {Fabbian}, {Marfil},
  {Montes}, {Tabernero}, {Ribas}, {Caballero}, {Quirrenbach}, {Amado},
  {Aceituno}, {Bejar}, {Cortes-Contreras}, {Dreizler}, {Hatzes}, {Henning},
  {Jeffers}, {Kaminski}, {Kurster}, {Lafarga}, {Morales}, {Nagel}, {Palle},
  {Passegger}, {Rodriguez Lopez}, {Schweitzer}, \& {Zechmeister}}]{Sha21}
{Shan}, Y., {Reiners}, A., {Fabbian}, D., {et~al.} 2021,
  \href{http://dx.doi.org/10.1051/0004-6361/202141530}{\color{blue}\aap},
  \href{none}{\href{https://ui.adsabs.harvard.edu/abs/2021arXiv210812442S}{in
  press}}

\bibitem[{{Shulyak} {et~al.}(2019){Shulyak}, {Reiners}, {Nagel}, {Tal-Or},
  {Caballero}, {Zechmeister}, {B{\'e}jar}, {Cort{\'e}s-Contreras}, {Martin},
  {Kaminski}, {Ribas}, {Quirrenbach}, {Amado}, {Anglada-Escud{\'e}}, {Bauer},
  {Dreizler}, {Guenther}, {Henning}, {Jeffers}, {K{\"u}rster}, {Lafarga},
  {Montes}, {Morales}, \& {Pedraz}}]{Shu19}
{Shulyak}, D., {Reiners}, A., {Nagel}, E., {et~al.} 2019,
  \href{http://dx.doi.org/10.1051/0004-6361/201935315}{\color{blue}\aap},
  \href{https://ui.adsabs.harvard.edu/abs/2019A&A...626A..86S}{626, A86}

\bibitem[{{Smette} {et~al.}(2015){Smette}, {Sana}, {Noll}, {Horst}, {Kausch},
  {Kimeswenger}, {Barden}, {Szyszka}, {Jones}, {Gallenne}, {Vinther},
  {Ballester}, \& {Taylor}}]{Sme15}
{Smette}, A., {Sana}, H., {Noll}, S., {et~al.} 2015,
  \href{http://dx.doi.org/10.1051/0004-6361/201423932}{\color{blue}\aap},
  \href{https://ui.adsabs.harvard.edu/abs/2015A&A...576A..77S}{576, A77}

\bibitem[{{Souto} {et~al.}(2020){Souto}, {Cunha}, {Smith}, {Allende Prieto},
  {Burgasser}, {Covey}, {Garc{\'\i}a-Hern{\'a}ndez}, {Holtzman}, {Johnson},
  {J{\"o}nsson}, {Mahadevan}, {Majewski}, {Masseron}, {Shetrone},
  {Rojas-Ayala}, {Sobeck}, {Stassun}, {Terrien}, {Teske}, {Wanderley}, \&
  {Zamora}}]{Sou20}
{Souto}, D., {Cunha}, K., {Smith}, V.~V., {et~al.} 2020,
  \href{http://dx.doi.org/10.3847/1538-4357/ab6d07}{\color{blue}\apj},
  \href{https://ui.adsabs.harvard.edu/abs/2020ApJ...890..133S}{890, 133}

\bibitem[{{Tabernero} {et~al.}(2018){Tabernero}, {Dorda}, {Negueruela}, \&
  {Gonz{\'a}lez-Fern{\'a}ndez}}]{Tab18}
{Tabernero}, H.~M., {Dorda}, R., {Negueruela}, I., \&
  {Gonz{\'a}lez-Fern{\'a}ndez}, C. 2018,
  \href{http://dx.doi.org/10.1093/mnras/sty399}{\color{blue}\mnras},
  \href{http://adsabs.harvard.edu/abs/2018MNRAS.476.3106T}{476, 3106}

\bibitem[{{Tabernero} {et~al.}(2019){Tabernero}, {Marfil}, {Montes}, \&
  {Gonz{\'a}lez Hern{\'a}ndez}}]{Tab19}
{Tabernero}, H.~M., {Marfil}, E., {Montes}, D., \& {Gonz{\'a}lez
  Hern{\'a}ndez}, J.~I. 2019,
  \href{http://dx.doi.org/10.1051/0004-6361/201935465}{\color{blue}\aap},
  \href{https://ui.adsabs.harvard.edu/abs/2019A&A...628A.131T}{628, A131}

\bibitem[{{Tabernero} {et~al.}(2021){Tabernero}, {Marfil}, {Montes}, \&
  {Gonz{\'a}lez Hern{\'a}ndez}}]{Tab20xx}
{Tabernero}, H.~M., {Marfil}, E., {Montes}, D., \& {Gonz{\'a}lez
  Hern{\'a}ndez}, J.~I. 2021,
  \href{https://ui.adsabs.harvard.edu/abs/2021arXiv211000444T}{arXiv e-prints,
  arXiv:2110.00444}

\bibitem[{{Tal-Or} {et~al.}(2018){Tal-Or}, {Zechmeister}, {Reiners}, {Jeffers},
  {Sch{\"o}fer}, {Quirrenbach}, {Amado}, {Ribas}, {Caballero}, {Aceituno},
  {Bauer}, {B{\'e}jar}, {Czesla}, {Dreizler}, {Fuhrmeister}, {Hatzes},
  {Johnson}, {K{\"u}rster}, {Lafarga}, {Montes}, {Morales}, {Reffert},
  {Sadegi}, {Seifert}, \& {Shulyak}}]{Tal18}
{Tal-Or}, L., {Zechmeister}, M., {Reiners}, A., {et~al.} 2018,
  \href{http://dx.doi.org/10.1051/0004-6361/201732362}{\color{blue}\aap},
  \href{https://ui.adsabs.harvard.edu/abs/2018A&A...614A.122T}{614, A122}

\bibitem[{{Terrien} {et~al.}(2012){Terrien}, {Mahadevan}, {Bender},
  {Deshpande}, {Ramsey}, \& {Bochanski}}]{Ter12}
{Terrien}, R.~C., {Mahadevan}, S., {Bender}, C.~F., {et~al.} 2012,
  \href{http://dx.doi.org/10.1088/2041-8205/747/2/L38}{\color{blue}\apjl},
  \href{https://ui.adsabs.harvard.edu/abs/2012ApJ...747L..38T}{747, L38}

\bibitem[{{Tinney} \& {Reid}(1998)}]{Tin98}
{Tinney}, C.~G. \& {Reid}, I.~N. 1998,
  \href{http://dx.doi.org/10.1046/j.1365-8711.1998.02079.x}{\color{blue}\mnras},
  \href{https://ui.adsabs.harvard.edu/abs/1998MNRAS.301.1031T}{301, 1031}

\bibitem[{{Trifonov} {et~al.}(2018){Trifonov}, {K{\"u}rster}, {Zechmeister},
  {Tal-Or}, {Caballero}, {Quirrenbach}, {Amado}, {Ribas}, {Reiners}, {Reffert},
  {Dreizler}, {Hatzes}, {Kaminski}, {Launhardt}, {Henning}, {Montes},
  {B{\'e}jar}, {Mundt}, {Pavlov}, {Schmitt}, {Seifert}, {Morales}, {Nowak},
  {Jeffers}, {Rodr{\'\i}guez-L{\'o}pez}, {del Burgo}, {Anglada-Escud{\'e}},
  {L{\'o}pez-Santiago}, {Mathar}, {Ammler-von Eiff}, {Guenther}, {Barrado},
  {Gonz{\'a}lez Hern{\'a}ndez}, {Mancini}, {St{\"u}rmer}, {Abril}, {Aceituno},
  {Alonso-Floriano}, {Antona}, {Anwand-Heerwart}, {Arroyo-Torres}, {Azzaro},
  {Baroch}, {Bauer}, {Becerril}, {Ben{\'\i}tez}, {Berdi{\~n}as}, {Bergond},
  {Bl{\"u}mcke}, {Brinkm{\"o}ller}, {Cano}, {C{\'a}rdenas V{\'a}zquez},
  {Casal}, {Cifuentes}, {Claret}, {Colom{\'e}}, {Cort{\'e}s-Contreras},
  {Czesla}, {D{\'\i}ez-Alonso}, {Feiz}, {Fern{\'a}ndez}, {Ferro},
  {Fuhrmeister}, {Galad{\'\i}-Enr{\'\i}quez}, {Garcia-Piquer}, {Garc{\'\i}a
  Vargas}, {Gesa}, {G{\'o}mez Galera}, {Gonz{\'a}lez-Peinado}, {Gr{\"o}zinger},
  {Grohnert}, {Gu{\`a}rdia}, {Guijarro}, {de Guindos}, {Guti{\'e}rrez-Soto},
  {Hagen}, {Hauschildt}, {Hedrosa}, {Helmling}, {Hermelo}, {Hern{\'a}ndez
  Arab{\'\i}}, {Hern{\'a}ndez Casta{\~n}o}, {Hern{\'a}ndez Hernando},
  {Herrero}, {Huber}, {Huke}, {Johnson}, {de Juan}, {Kim}, {Klein},
  {Kl{\"u}ter}, {Klutsch}, {Lafarga}, {Lamp{\'o}n}, {Lara}, {Laun}, {Lemke},
  {Lenzen}, {L{\'o}pez del Fresno}, {L{\'o}pez-Gonz{\'a}lez},
  {L{\'o}pez-Puertas}, {L{\'o}pez Salas}, {Luque}, {Mag{\'a}n Madinabeitia},
  {Mall}, {Mandel}, {Marfil}, {Mar{\'\i}n Molina}, {Maroto Fern{\'a}ndez},
  {Mart{\'\i}n}, {Mart{\'\i}n-Ruiz}, {Marvin}, {Mirabet}, {Moya},
  {Moreno-Raya}, {Nagel}, {Naranjo}, {Nortmann}, {Ofir}, {Oreiro}, {Pall{\'e}},
  {Panduro}, {Pascual}, {Passegger}, {Pedraz}, {P{\'e}rez-Calpena}, {P{\'e}rez
  Medialdea}, {Perger}, {Perryman}, {Pluto}, {Rabaza}, {Ram{\'o}n}, {Rebolo},
  {Redondo}, {Reinhardt}, {Rhode}, {Rix}, {Rodler}, {Rodr{\'\i}guez},
  {Rodr{\'\i}guez Trinidad}, {Rohloff}, {Rosich}, {Sadegi},
  {S{\'a}nchez-Blanco}, {S{\'a}nchez Carrasco}, {S{\'a}nchez-L{\'o}pez},
  {Sanz-Forcada}, {Sarkis}, {Sarmiento}, {Sch{\"a}fer}, {Schiller},
  {Sch{\"o}fer}, {Schweitzer}, {Solano}, {Stahl}, {Strachan}, {Su{\'a}rez},
  {Tabernero}, {Tala}, {Tulloch}, {Veredas}, {Vico Linares}, {Vilardell},
  {Wagner}, {Winkler}, {Wolthoff}, {Xu}, {Yan}, \& {Zapatero Osorio}}]{Tri18}
{Trifonov}, T., {K{\"u}rster}, M., {Zechmeister}, M., {et~al.} 2018,
  \href{http://dx.doi.org/10.1051/0004-6361/201731442}{\color{blue}\aap},
  \href{https://ui.adsabs.harvard.edu/abs/2018A&A...609A.117T}{609, A117}

\bibitem[{{Trifonov} {et~al.}(2020){Trifonov}, {Lee}, {K{\"u}rster}, {Henning},
  {Grishin}, {Stock}, {Tjoa}, {Caballero}, {Wong}, {Bauer}, {Quirrenbach},
  {Zechmeister}, {Ribas}, {Reffert}, {Reiners}, {Amado}, {Kossakowski},
  {Azzaro}, {B{\'e}jar}, {Cort{\'e}s-Contreras}, {Dreizler}, {Hatzes},
  {Jeffers}, {Kaminski}, {Lafarga}, {Montes}, {Morales}, {Pavlov},
  {Rodr{\'\i}guez-L{\'o}pez}, {Schmitt}, {Solano}, \& {Barnes}}]{Tri20}
{Trifonov}, T., {Lee}, M.~H., {K{\"u}rster}, M., {et~al.} 2020,
  \href{http://dx.doi.org/10.1051/0004-6361/201936987}{\color{blue}\aap},
  \href{https://ui.adsabs.harvard.edu/abs/2020A&A...638A..16T}{638, A16}

\bibitem[{{Tsuji} {et~al.}(1996){Tsuji}, {Ohnaka}, \& {Aoki}}]{Tsu96a}
{Tsuji}, T., {Ohnaka}, K., \& {Aoki}, W. 1996, \aap,
  \href{https://ui.adsabs.harvard.edu/abs/1996A&A...305L...1T}{305, L1}

\bibitem[{{Valenti} \& {Fischer}(2005)}]{Val05}
{Valenti}, J.~A. \& {Fischer}, D.~A. 2005,
  \href{http://dx.doi.org/10.1086/430500}{\color{blue}\apjs},
  \href{https://ui.adsabs.harvard.edu/abs/2005ApJS..159..141V}{159, 141}

\bibitem[{Van~Rossum(2020)}]{Van20}
Van~Rossum, G. 2020, The Python Library Reference, release 3.8.2 (Python
  Software Foundation)

\bibitem[{{Veyette} {et~al.}(2016){Veyette}, {Muirhead}, {Mann}, \&
  {Allard}}]{Vey16}
{Veyette}, M.~J., {Muirhead}, P.~S., {Mann}, A.~W., \& {Allard}, F. 2016,
  \href{http://dx.doi.org/10.3847/0004-637X/828/2/95}{\color{blue}\apj},
  \href{https://ui.adsabs.harvard.edu/abs/2016ApJ...828...95V}{828, 95}

\bibitem[{{Veyette} {et~al.}(2017){Veyette}, {Muirhead}, {Mann}, {Brewer},
  {Allard}, \& {Homeier}}]{Vey17}
{Veyette}, M.~J., {Muirhead}, P.~S., {Mann}, A.~W., {et~al.} 2017,
  \href{http://dx.doi.org/10.3847/1538-4357/aa96aa}{\color{blue}\apj},
  \href{https://ui.adsabs.harvard.edu/abs/2017ApJ...851...26V}{851, 26}

\bibitem[{{Virtanen} {et~al.}(2020){Virtanen}, {Gommers}, {Oliphant},
  {Haberland}, {Reddy}, {Cournapeau}, {Burovski}, {Peterson}, {Weckesser},
  {Bright}, {van der Walt}, {Brett}, {Wilson}, {Millman}, {Mayorov}, {Nelson},
  {Jones}, {Kern}, {Larson}, {Carey}, {Polat}, {Feng}, {Moore}, {VanderPlas},
  {Laxalde}, {Perktold}, {Cimrman}, {Henriksen}, {Quintero}, {Harris},
  {Archibald}, {Ribeiro}, {Pedregosa}, {van Mulbregt}, \& {SciPy 1. 0
  Contributors}}]{Vir20}
{Virtanen}, P., {Gommers}, R., {Oliphant}, T.~E., {et~al.} 2020,
  \href{http://dx.doi.org/10.1038/s41592-019-0686-2}{\color{blue}Nature
  Methods}, \href{https://ui.adsabs.harvard.edu/abs/2020NatMe..17..261V}{17,
  261}

\bibitem[{{von Braun} {et~al.}(2014){von Braun}, {Boyajian}, {van Belle},
  {Kane}, {Jones}, {Farrington}, {Schaefer}, {Vargas}, {Scott}, {ten
  Brummelaar}, {Kephart}, {Gies}, {Ciardi}, {L{\'o}pez-Morales}, {Mazingue},
  {McAlister}, {Ridgway}, {Goldfinger}, {Turner}, \& {Sturmann}}]{vBra14}
{von Braun}, K., {Boyajian}, T.~S., {van Belle}, G.~T., {et~al.} 2014,
  \href{http://dx.doi.org/10.1093/mnras/stt2360}{\color{blue}\mnras},
  \href{https://ui.adsabs.harvard.edu/abs/2014MNRAS.438.2413V}{438, 2413}

\bibitem[{{Wildi} {et~al.}(2017){Wildi}, {Blind}, {Reshetov}, {Hernandez},
  {Genolet}, {Conod}, {Sordet}, {Segovilla}, {Rasilla}, {Brousseau},
  {Thibault}, {Delabre}, {Bandy}, {Sarajlic}, {Cabral}, {Bovay}, {Vall{\'e}e},
  {Bouchy}, {Doyon}, {Artigau}, {Pepe}, {Hagelberg}, {Melo}, {Delfosse},
  {Figueira}, {Santos}, {Gonz{\'a}lez Hern{\'a}ndez}, {de Medeiros}, {Rebolo},
  {Broeg}, {Benz}, {Boisse}, {Malo}, {K{\"a}ufl}, \& {Saddlemyer}}]{Wil17}
{Wildi}, F., {Blind}, N., {Reshetov}, V., {et~al.} 2017, in Society of
  Photo-Optical Instrumentation Engineers (SPIE) Conference Series, Vol. 10400,
  \href{https://ui.adsabs.harvard.edu/abs/2017SPIE10400E..18W}{1040018}

\bibitem[{{Wilson} {et~al.}(2010){Wilson}, {Hearty}, {Skrutskie}, {Majewski},
  {Schiavon}, {Eisenstein}, {Gunn}, {Blank}, {Henderson}, {Smee}, {Barkhouser},
  {Harding}, {Fitzgerald}, {Stolberg}, {Arns}, {Nelson}, {Brunner}, {Burton},
  {Walker}, {Lam}, {Maseman}, {Barr}, {Leger}, {Carey}, {MacDonald}, {Horne},
  {Young}, {Rieke}, {Rieke}, {O'Brien}, {Hope}, {Krakula}, {Crane}, {Zhao},
  {Carr}, {Harrison}, {Stoll}, {Vernieri}, {Holtzman}, {Shetrone},
  {Allende-Prieto}, {Johnson}, {Frinchaboy}, {Zasowski}, {Bizyaev},
  {Gillespie}, \& {Weinberg}}]{Wil10}
{Wilson}, J.~C., {Hearty}, F., {Skrutskie}, M.~F., {et~al.} 2010, in Society of
  Photo-Optical Instrumentation Engineers (SPIE) Conference Series, Vol. 7735,
  \href{https://ui.adsabs.harvard.edu/abs/2010SPIE.7735E..1CW}{77351C}

\bibitem[{{Winn} \& {Fabrycky}(2015)}]{Win15}
{Winn}, J.~N. \& {Fabrycky}, D.~C. 2015,
  \href{http://dx.doi.org/10.1146/annurev-astro-082214-122246}{\color{blue}\araa},
  \href{https://ui.adsabs.harvard.edu/abs/2015ARA&A..53..409W}{53, 409}

\bibitem[{{Yan} {et~al.}(2019){Yan}, {Casasayas-Barris}, {Molaverdikhani},
  {Alonso-Floriano}, {Reiners}, {Pall{\'e}}, {Henning}, {Molli{\`e}re}, {Chen},
  {Nortmann}, {Snellen}, {Ribas}, {Quirrenbach}, {Caballero}, {Amado},
  {Azzaro}, {Bauer}, {Cort{\'e}s Contreras}, {Czesla}, {Khalafinejad}, {Lara},
  {L{\'o}pez-Puertas}, {Montes}, {Nagel}, {Oshagh}, {S{\'a}nchez-L{\'o}pez},
  {Stangret}, \& {Zechmeister}}]{Yan19}
{Yan}, F., {Casasayas-Barris}, N., {Molaverdikhani}, K., {et~al.} 2019,
  \href{http://dx.doi.org/10.1051/0004-6361/201936396}{\color{blue}\aap},
  \href{https://ui.adsabs.harvard.edu/abs/2019A&A...632A..69Y}{632, A69}

\bibitem[{{Zechmeister} {et~al.}(2019){Zechmeister}, {Dreizler}, {Ribas},
  {Reiners}, {Caballero}, {Bauer}, {B{\'e}jar}, {Gonz{\'a}lez-Cuesta},
  {Herrero}, {Lalitha}, {L{\'o}pez-Gonz{\'a}lez}, {Luque}, {Morales},
  {Pall{\'e}}, {Rodr{\'\i}guez}, {Rodr{\'\i}guez L{\'o}pez}, {Tal-Or},
  {Anglada-Escud{\'e}}, {Quirrenbach}, {Amado}, {Abril}, {Aceituno},
  {Aceituno}, {Alonso-Floriano}, {Ammler-von Eiff}, {Antona Jim{\'e}nez},
  {Anwand-Heerwart}, {Arroyo-Torres}, {Azzaro}, {Baroch}, {Barrado},
  {Becerril}, {Ben{\'\i}tez}, {Berdi{\~n}as}, {Bergond}, {Bluhm},
  {Brinkm{\"o}ller}, {del Burgo}, {Calvo Ortega}, {Cano}, {Cardona
  Guill{\'e}n}, {Carro}, {C{\'a}rdenas V{\'a}zquez}, {Casal},
  {Casasayas-Barris}, {Casanova}, {Chaturvedi}, {Cifuentes}, {Claret},
  {Colom{\'e}}, {Cort{\'e}s-Contreras}, {Czesla}, {D{\'\i}ez-Alonso}, {Dorda},
  {Fern{\'a}ndez}, {Fern{\'a}ndez-Mart{\'\i}n}, {Fuhrmeister}, {Fukui},
  {Galad{\'\i}-Enr{\'\i}quez}, {Gallardo Cava}, {Garcia de la Fuente},
  {Garcia-Piquer}, {Garc{\'\i}a Vargas}, {Gesa}, {G{\'o}ngora Rueda},
  {Gonz{\'a}lez-{\'A}lvarez}, {Gonz{\'a}lez Hern{\'a}ndez},
  {Gonz{\'a}lez-Peinado}, {Gr{\"o}zinger}, {Gu{\`a}rdia}, {Guijarro}, {de
  Guindos}, {Hatzes}, {Hauschildt}, {Hedrosa}, {Helmling}, {Henning},
  {Hermelo}, {Hern{\'a}ndez Arabi}, {Hern{\'a}ndez Casta{\~n}o}, {Hern{\'a}ndez
  Otero}, {Hintz}, {Huke}, {Huber}, {Jeffers}, {Johnson}, {de Juan},
  {Kaminski}, {Kemmer}, {Kim}, {Klahr}, {Klein}, {Kl{\"u}ter}, {Klutsch},
  {Kossakowski}, {K{\"u}rster}, {Labarga}, {Lafarga}, {Llamas}, {Lamp{\'o}n},
  {Lara}, {Launhardt}, {L{\'a}zaro}, {Lodieu}, {L{\'o}pez del Fresno},
  {L{\'o}pez-Puertas}, {L{\'o}pez Salas}, {L{\'o}pez-Santiago}, {Mag{\'a}n
  Madinabeitia}, {Mall}, {Mancini}, {Mand el}, {Marfil}, {Mar{\'\i}n Molina},
  {Maroto Fern{\'a}ndez}, {Mart{\'\i}n}, {Mart{\'\i}n-Fern{\'a}ndez},
  {Mart{\'\i}n-Ruiz}, {Marvin}, {Mirabet}, {Monta{\~n}{\'e}s-Rodr{\'\i}guez},
  {Montes}, {Moreno-Raya}, {Nagel}, {Naranjo}, {Narita}, {Nortmann}, {Nowak},
  {Ofir}, {Oshagh}, {Panduro}, {Parviainen}, {Pascual}, {Passegger}, {Pavlov},
  {Pedraz}, {P{\'e}rez-Calpena}, {P{\'e}rez Medialdea}, {Perger}, {Perryman},
  {Rabaza}, {Ram{\'o}n Ballesta}, {Rebolo}, {Redondo}, {Reffert}, {Reinhardt},
  {Rhode}, {Rix}, {Rodler}, {Rodr{\'\i}guez Trinidad}, {Rosich}, {Sadegi},
  {S{\'a}nchez-Blanco}, {S{\'a}nchez Carrasco}, {S{\'a}nchez-L{\'o}pez},
  {Sanz-Forcada}, {Sarkis}, {Sarmiento}, {Sch{\"a}fer}, {Schmitt},
  {Sch{\"o}fer}, {Schweitzer}, {Seifert}, {Shulyak}, {Solano}, {Sota}, {Stahl},
  {Stock}, {Strachan}, {Stuber}, {St{\"u}rmer}, {Su{\'a}rez}, {Tabernero},
  {Tala Pinto}, {Trifonov}, {Veredas}, {Vico Linares}, {Vilardell}, {Wagner},
  {Wolthoff}, {Xu}, {Yan}, \& {Zapatero Osorio}}]{Zec19}
{Zechmeister}, M., {Dreizler}, S., {Ribas}, I., {et~al.} 2019,
  \href{http://dx.doi.org/10.1051/0004-6361/201935460}{\color{blue}\aap},
  \href{https://ui.adsabs.harvard.edu/abs/2019A&A...627A..49Z}{627, A49}

\bibitem[{{Zechmeister} {et~al.}(2018){Zechmeister}, {Reiners}, {Amado},
  {Azzaro}, {Bauer}, {B{\'e}jar}, {Caballero}, {Guenther}, {Hagen}, {Jeffers},
  {Kaminski}, {K{\"u}rster}, {Launhardt}, {Montes}, {Morales}, {Quirrenbach},
  {Reffert}, {Ribas}, {Seifert}, {Tal-Or}, \& {Wolthoff}}]{Zec18}
{Zechmeister}, M., {Reiners}, A., {Amado}, P.~J., {et~al.} 2018,
  \href{http://dx.doi.org/10.1051/0004-6361/201731483}{\color{blue}\aap},
  \href{https://ui.adsabs.harvard.edu/abs/2018A&A...609A..12Z}{609, A12}

\bibitem[{{Zechmeister} {et~al.}(2020){Zechmeister}, {Reiners}, {Amado},
  {Azzaro}, {Bauer}, {B{\'e}jar}, {Caballero}, {Guenther}, {Hagen}, {Jeffers},
  {Kaminski}, {K{\"u}rster}, {Launhardt}, {Montes}, {Morales}, {Quirrenbach},
  {Reffert}, {Ribas}, {Seifert}, \& {Tal-Or}}]{Zec20}
{Zechmeister}, M., {Reiners}, A., {Amado}, P.~J., {et~al.} 2020, {SERVAL:
  SpEctrum Radial Velocity AnaLyser}, Astrophysics Source Code Library

\end{thebibliography}

\begin{appendix}
\section{Additional figures}

\begin{figure}
    \centering
    \includegraphics[width=0.49\textwidth]{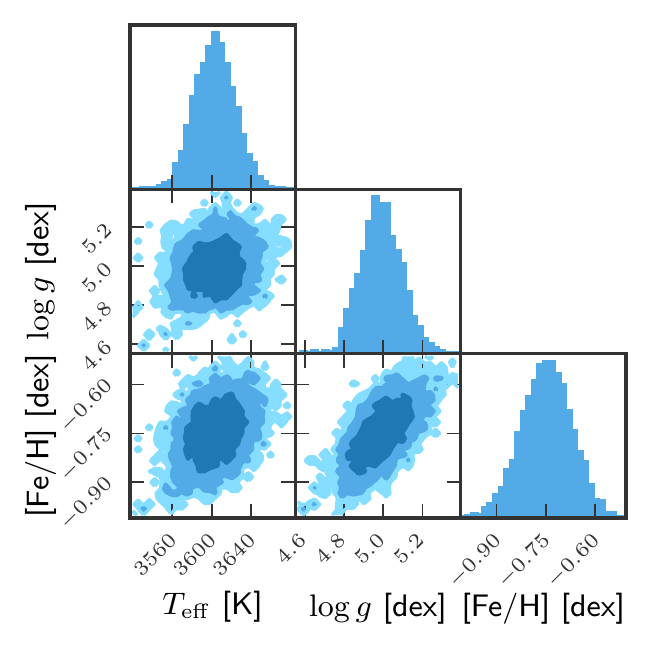}
    \caption{Same as Fig.~\ref{fig:cornerplot}, but for GX~And (M1.0\,V, J00183$+$440). The retrieved parameters are $T_{\rm eff}=3603\pm24$\,K, $\log{g}=4.99\pm0.14$\,dex, and ${\rm [Fe/H]=-0.75\pm0.11}$\,dex.}
    \label{fig:refGXAndcorner}
\end{figure}

\begin{figure}
    \centering
    \includegraphics[width=0.49\textwidth]{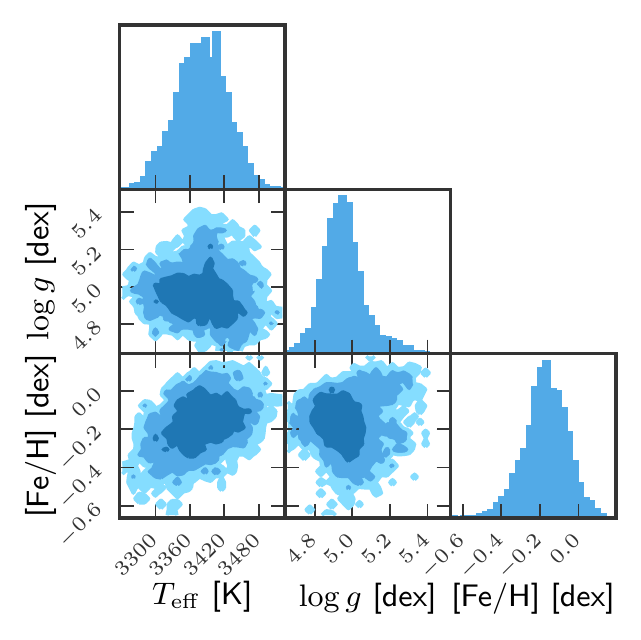}
    \caption{Same as Fig.~\ref{fig:cornerplot}, but for Luyten's~star (M3.5\,V, J07274$+$052). The retrieved parameters are $T_{\rm eff}=3380\pm43$\,K, $\log{g}=4.96\pm0.11$\,dex, and ${\rm [Fe/H]=-0.17\pm0.11}$\,dex.}
    \label{fig:refLuytencorner}
\end{figure}

\begin{figure}
    \centering
    \includegraphics[width=0.49\textwidth]{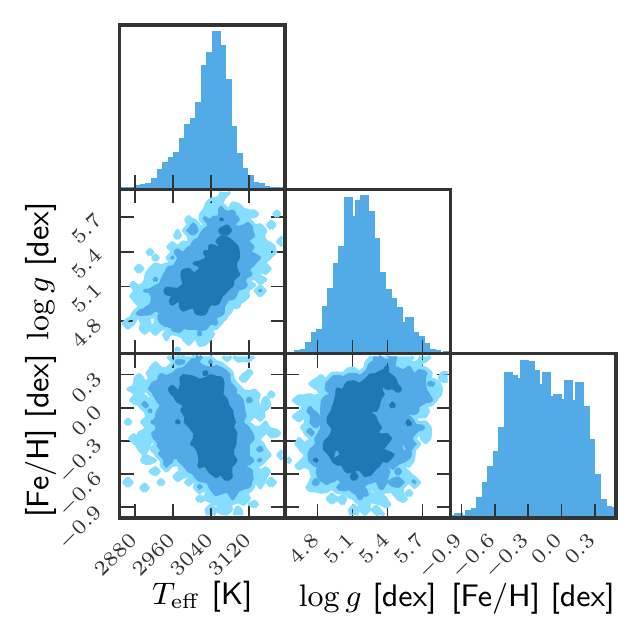}
    \caption{Same as Fig.~\ref{fig:cornerplot}, but for Teegarden's~star (M7.0\,V, J02530$+$168). The retrieved parameters are $T_{\rm eff}=3034\pm45$\,K, $\log{g}=5.19\pm0.20$\,dex, and ${\rm [Fe/H]=-0.17\pm0.28}$\,dex.}
    \label{fig:refTeegardencorner}
\end{figure}

\begin{figure*}
    \centering
    \includegraphics[width=\textwidth]{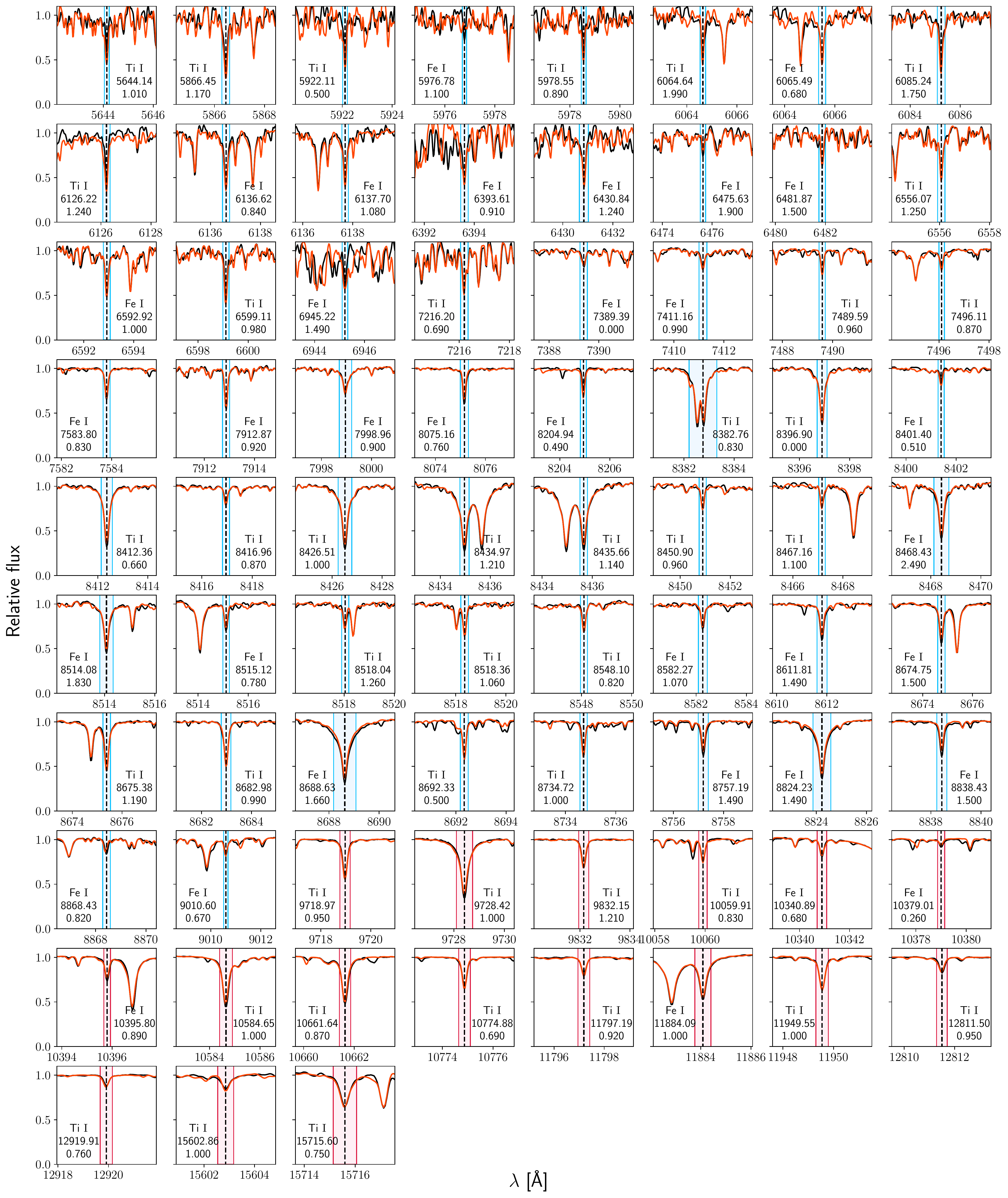}
    \caption{Same as Fig.~\ref{fig:lineplot}, but for GX~And (M1.0\,V, J00183$+$440).}
    \label{fig:refGXAndlines}
\end{figure*}

\begin{figure*}
    \centering
    \includegraphics[width=\textwidth]{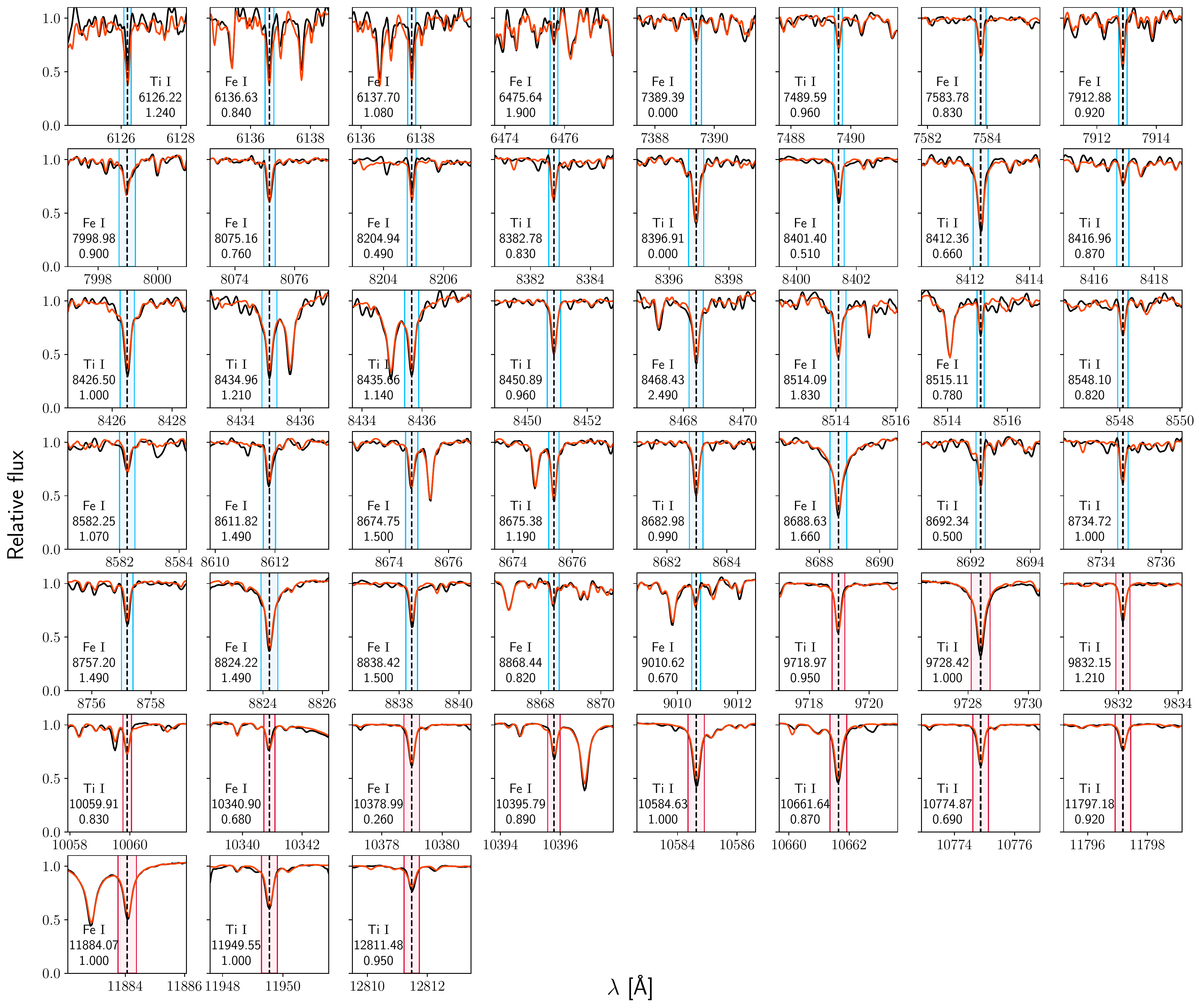}
    \caption{Same as Fig.~\ref{fig:lineplot}, but for Luyten's~star (M3.5\,V, J07274$+$052).}
    \label{fig:refLuytenlines}
\end{figure*}

\begin{figure*}
    \centering
    \includegraphics[width=\textwidth]{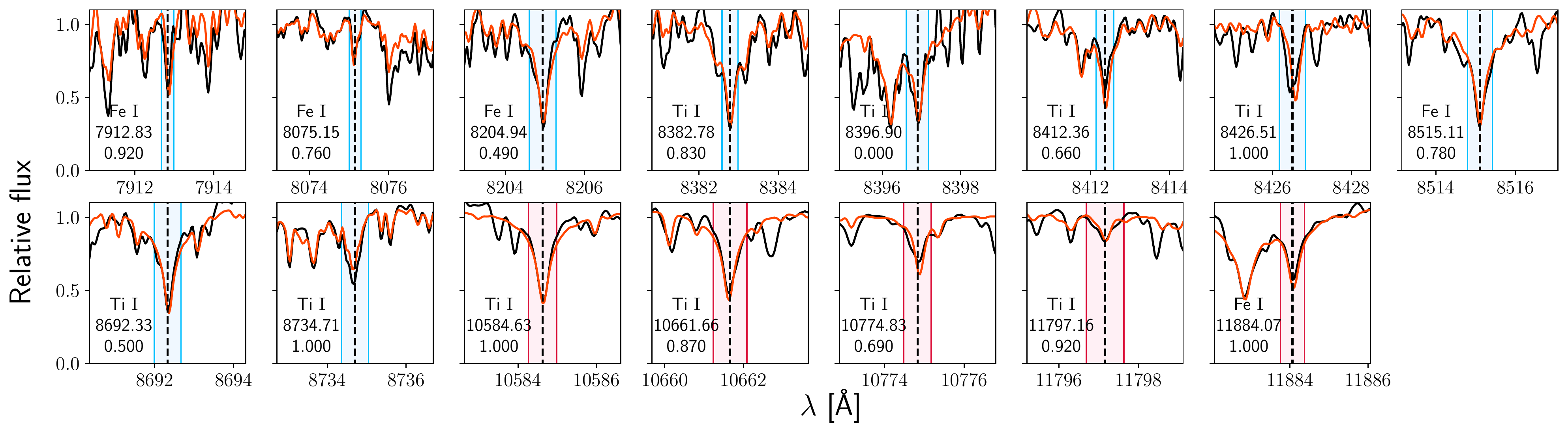}
    \caption{Same as Fig.~\ref{fig:lineplot}, but for Teegarden's~star (M7.0\,V, J02530$+$168).}
    \label{fig:refTeegardenlines}
\end{figure*}

\begin{figure}
    \centering
    \includegraphics[width=0.49\textwidth]{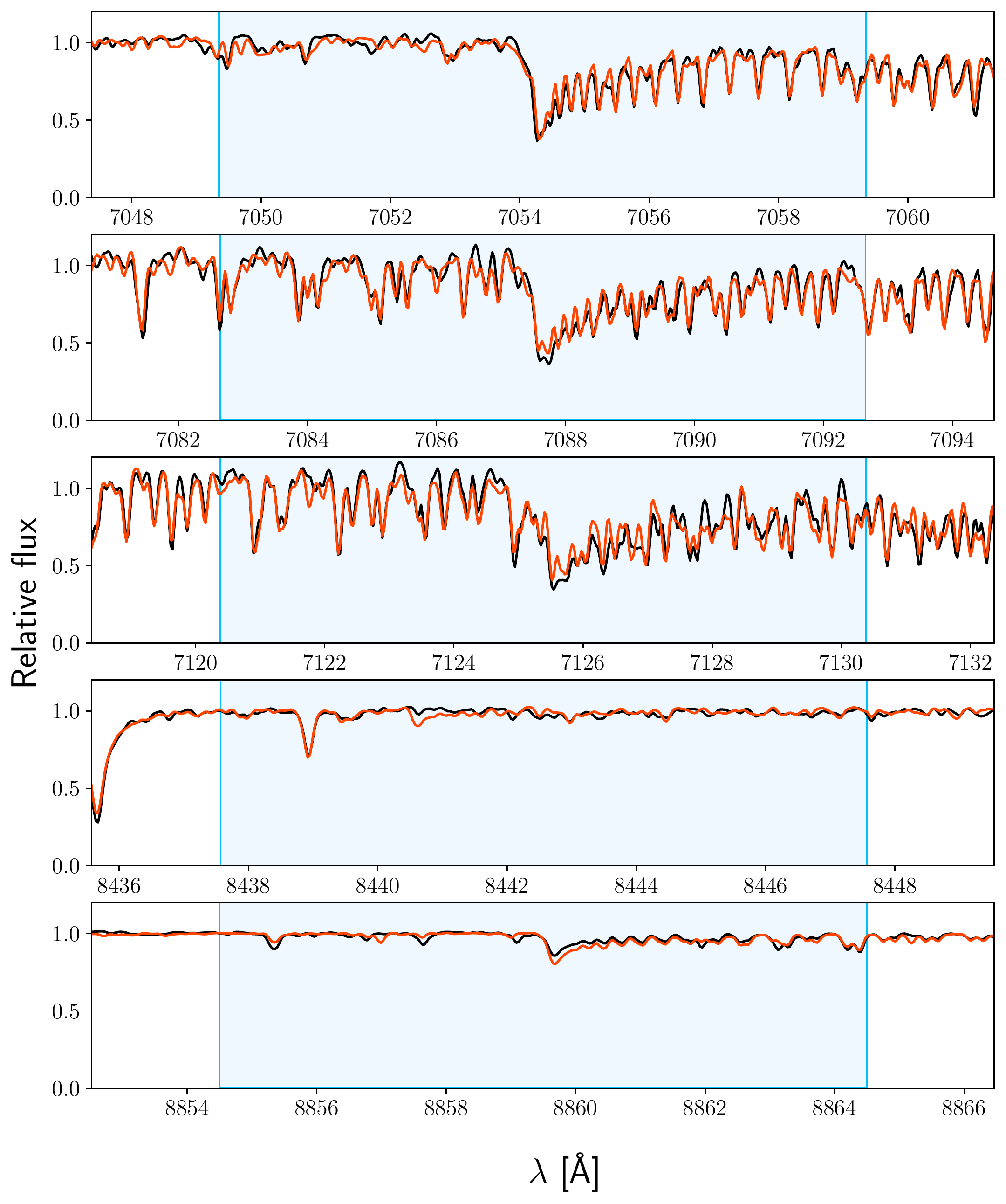}
    \caption{Same as Fig.~\ref{fig:bandplot}, but for GX~And (M1.0\,V, J00183$+$440).}
    \label{fig:refGXAndbands}
\end{figure}

\begin{figure}
    \centering
    \includegraphics[width=0.49\textwidth]{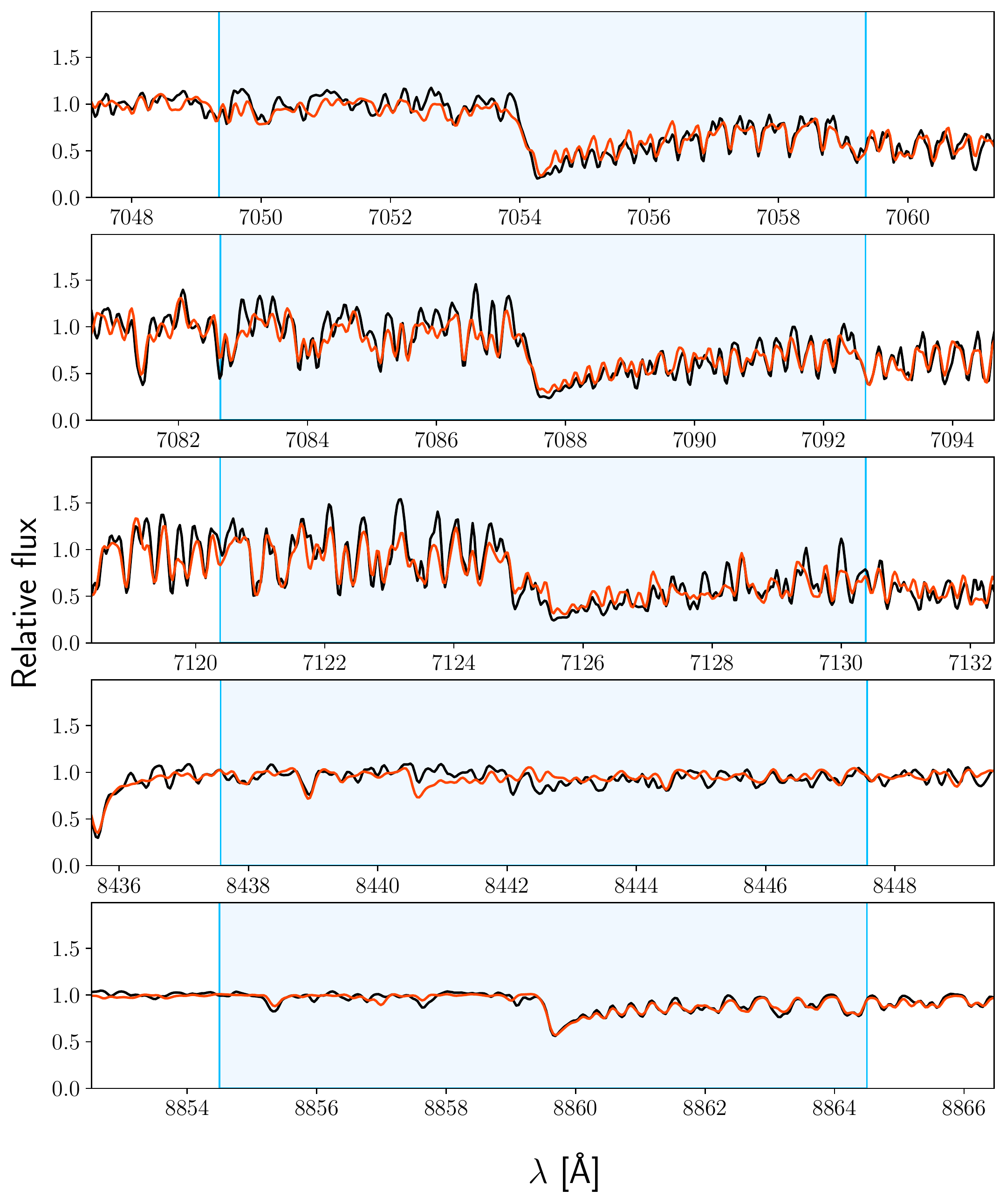}
    \caption{Same as Fig.~\ref{fig:bandplot}, but for Luyten's~star (M3.5\,V, J07274$+$052).}
    \label{fig:refLuytenbands}
\end{figure}

\begin{figure}
    \centering
    \includegraphics[width=0.49\textwidth]{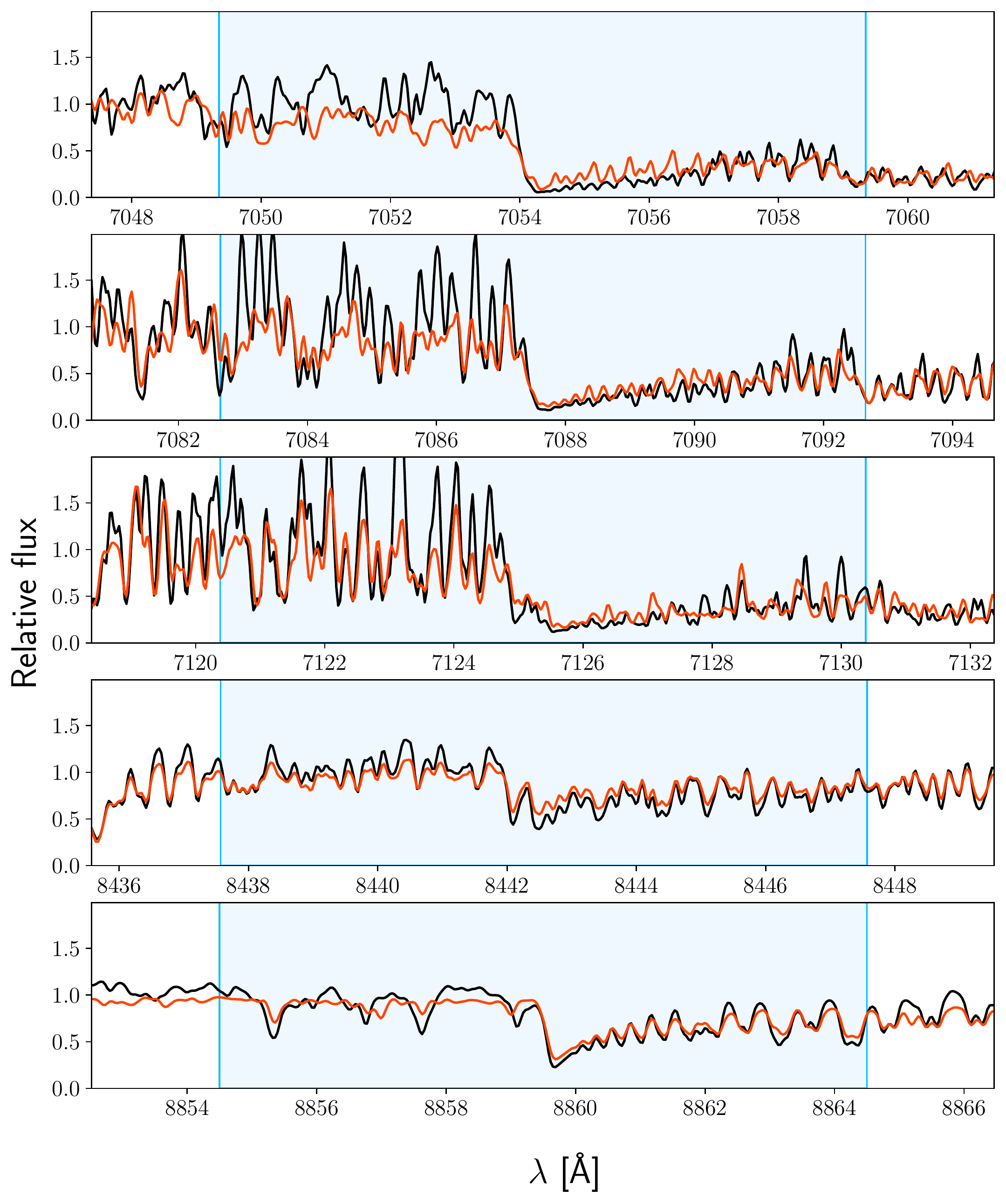}
    \caption{Same as Fig.~\ref{fig:bandplot}, but for Teegarden's~star (M7.0\,V, J02530$+$168).}
    \label{fig:refTeegardenbands}
\end{figure}

Figures~\ref{fig:refGXAndcorner}, \ref{fig:refLuytencorner}, and \ref{fig:refTeegardencorner} show the marginalised posterior distributions in $T_{\rm eff}$, $\log{g}$, and [Fe/H] for the reference stars, namely GX~And, Luyten's star, and Teegarden's star, respectively (see Table~\ref{tab:div}). The best synthetic fits for the atomic lines and molecular bands for these stars can be found in Figs.~\ref{fig:refGXAndlines}, \ref{fig:refLuytenlines}, \ref{fig:refTeegardenlines}, \ref{fig:refGXAndbands}, \ref{fig:refLuytenbands}, and \ref{fig:refTeegardenbands}.

\begin{figure*}
\centering
\includegraphics[width=0.7\textwidth]{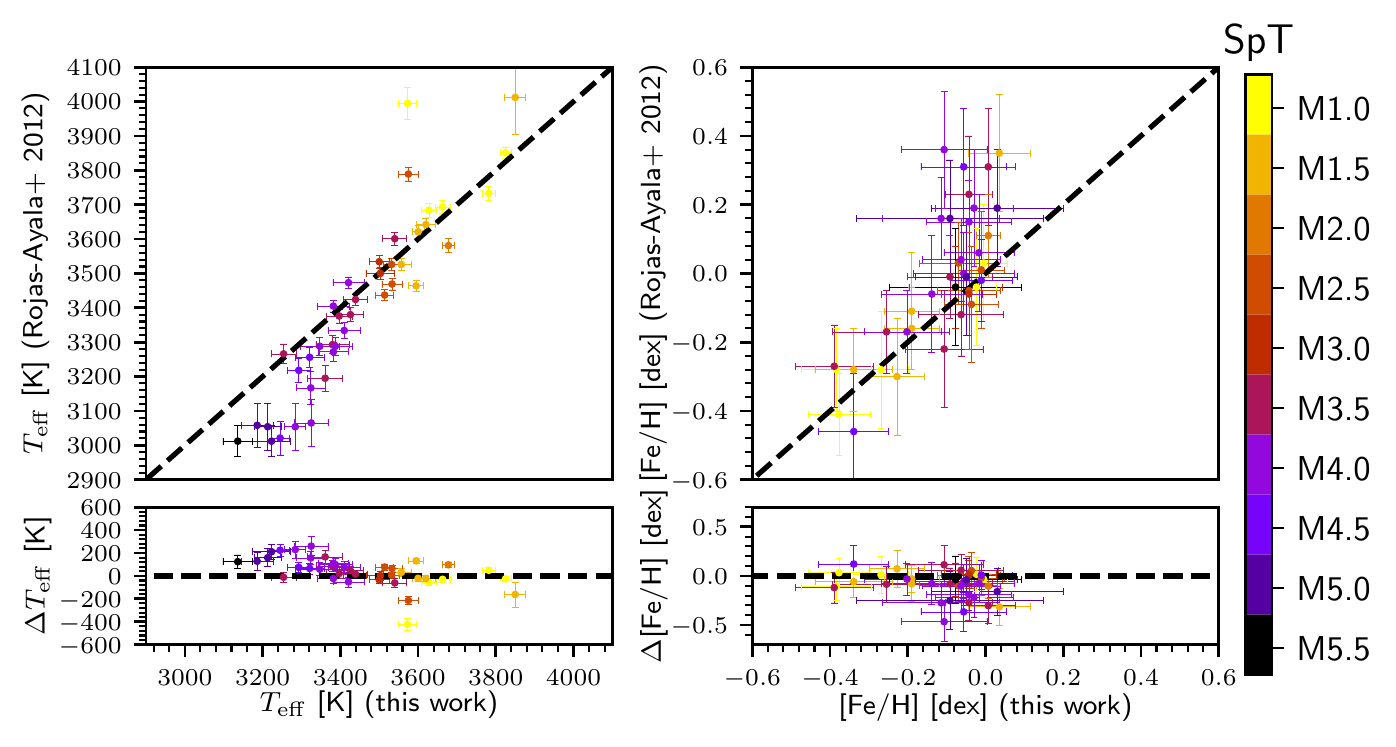}
\caption{\label{fig:compRA12} Comparison between this work and \citet{Roj12}.}
\end{figure*}

\begin{figure*}
\centering
\includegraphics[width=\textwidth]{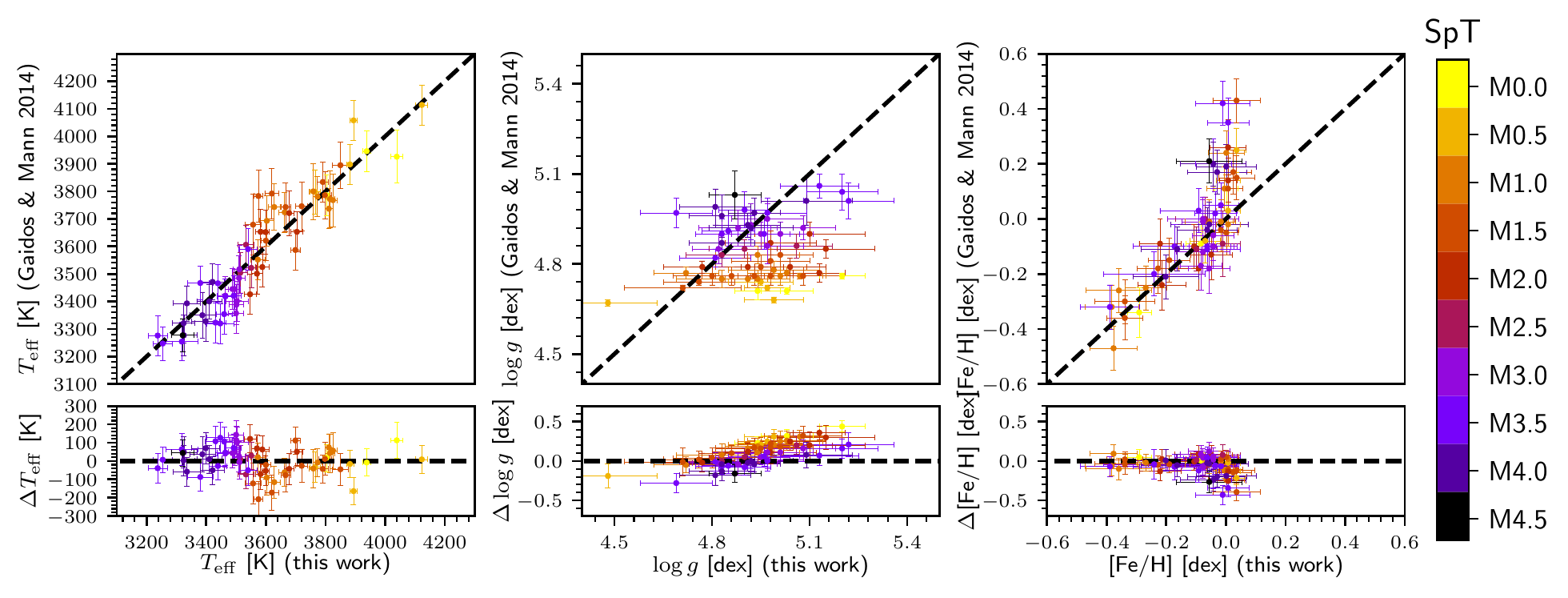}
\caption{\label{fig:compGM14} Comparison between this work and \citet{Gai14}.}
\end{figure*}

\begin{figure*}
\centering
\includegraphics[width=\textwidth]{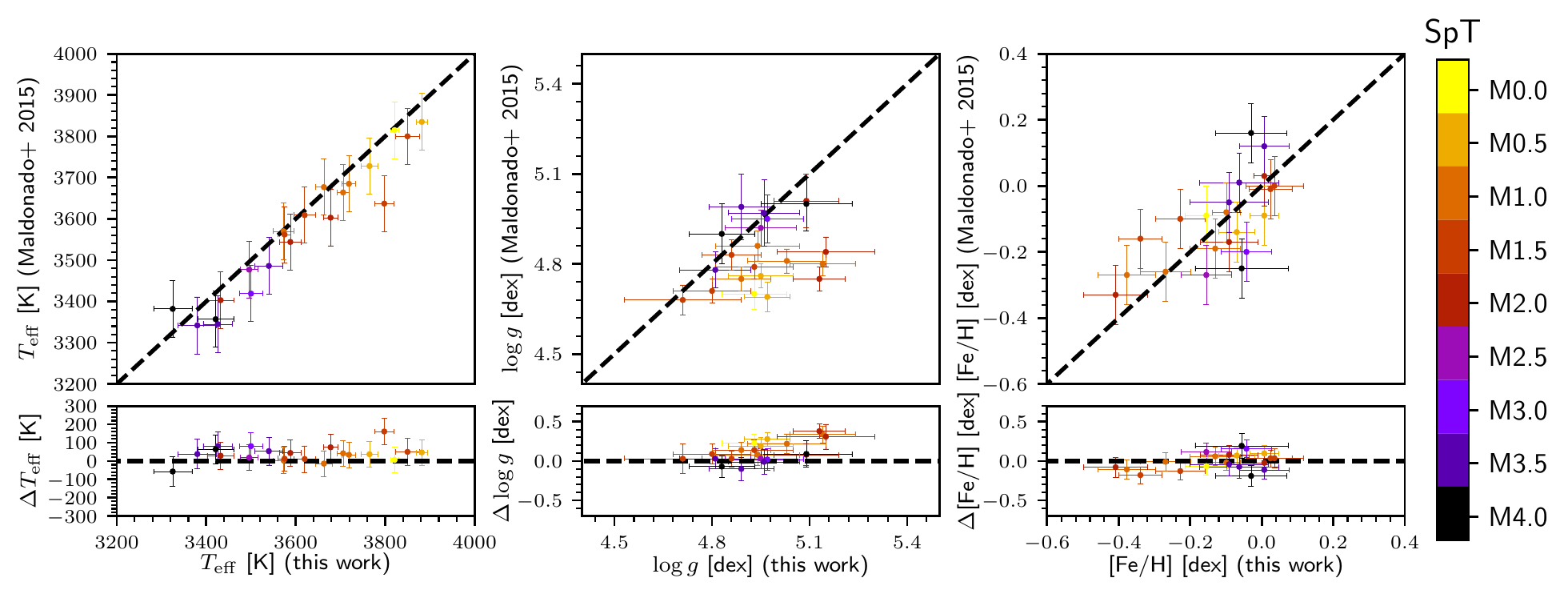}
\caption{\label{fig:compMald15} Comparison between this work and \citet{Mal15}.}
\end{figure*}

\begin{figure*}
\centering
\includegraphics[width=\textwidth]{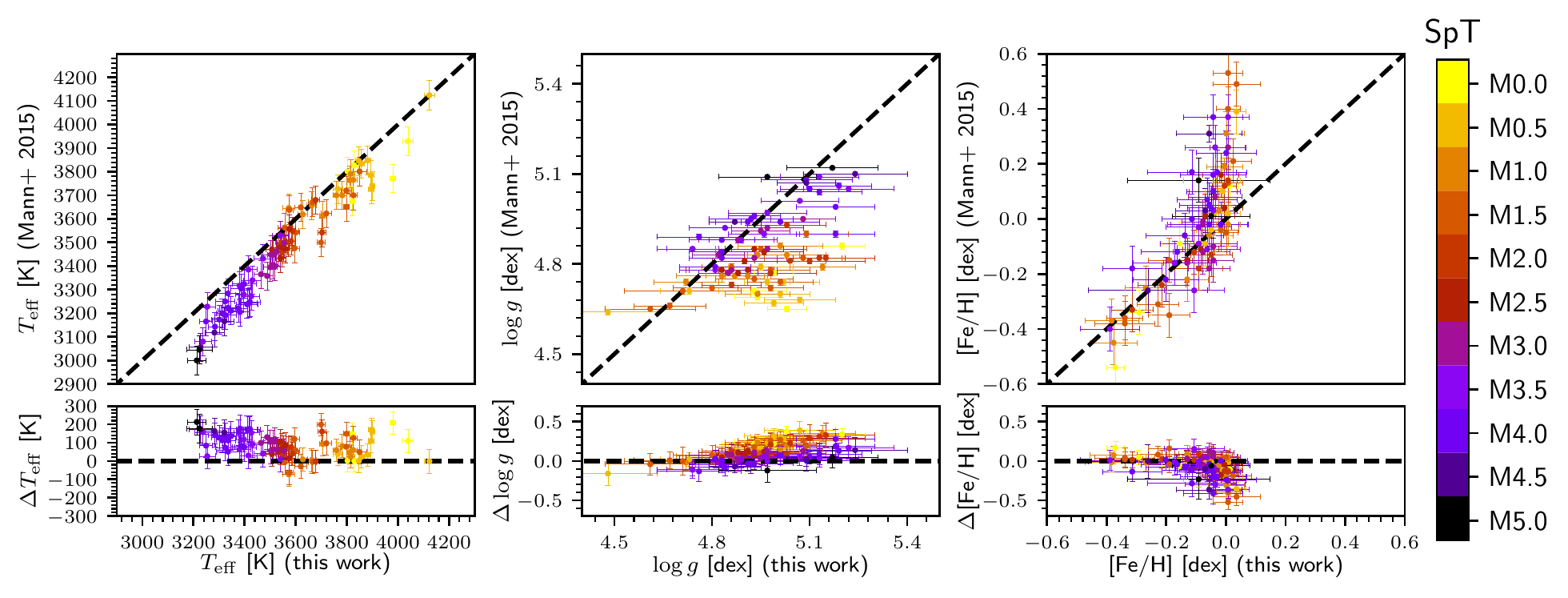}
\caption{\label{fig:compMann15} Comparison between this work and \citet{Man15}.}
\end{figure*}

\begin{figure*}
\centering
\includegraphics[width=\textwidth]{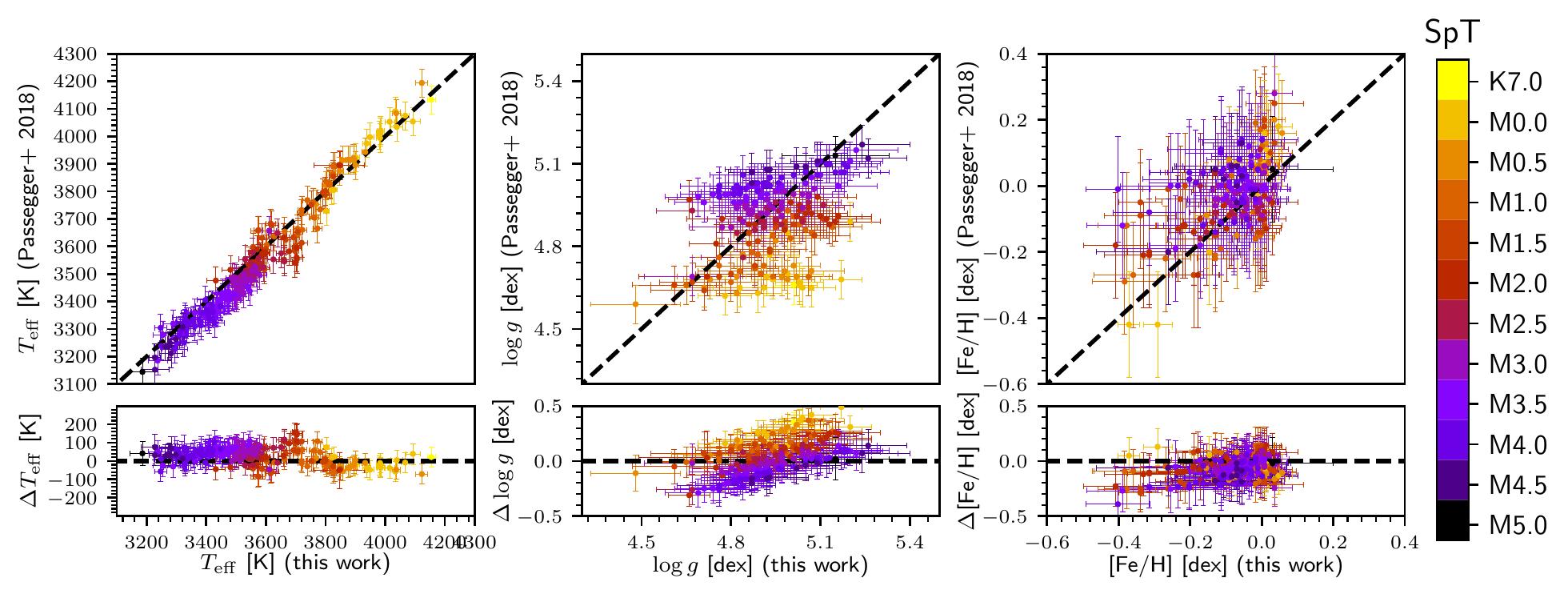}
\caption{\label{fig:compPas18} Comparison between this work and \citet{Pas18}.}
\end{figure*}

\begin{figure*}
\centering
\includegraphics[width=\textwidth]{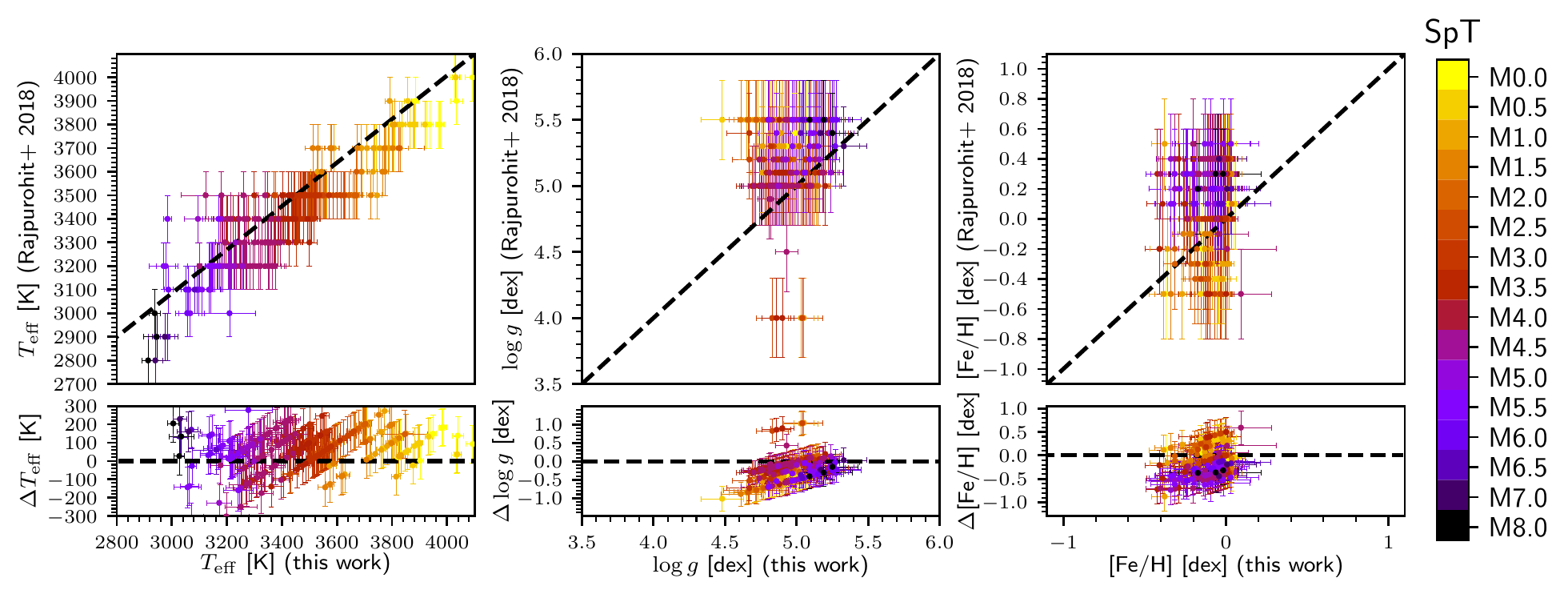}
\caption{\label{fig:compRaj18} Comparison between this work and \citet{Raj18b}.}
\end{figure*}

\begin{figure*}
\centering
\includegraphics[width=\textwidth]{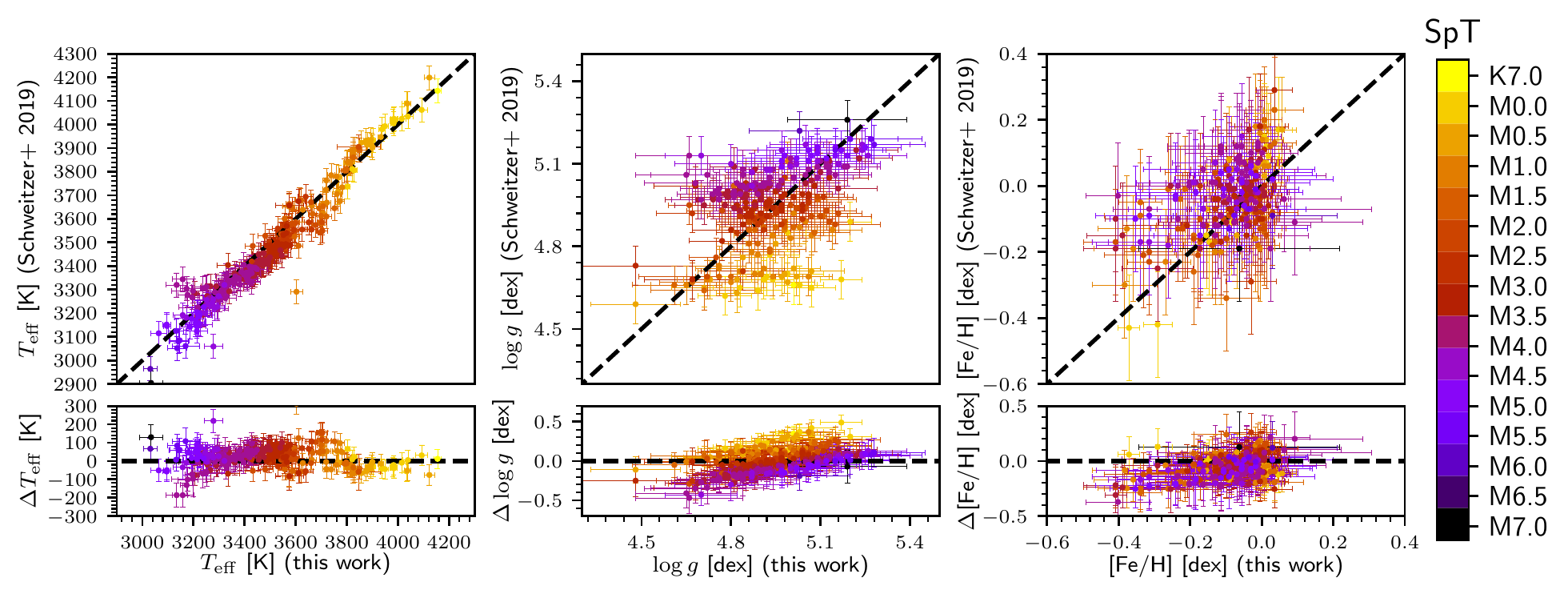}
\caption{\label{fig:compSch19} Comparison between this work and \citet{Sch19}.}
\end{figure*}

\begin{figure*}
\centering
\includegraphics[width=\textwidth]{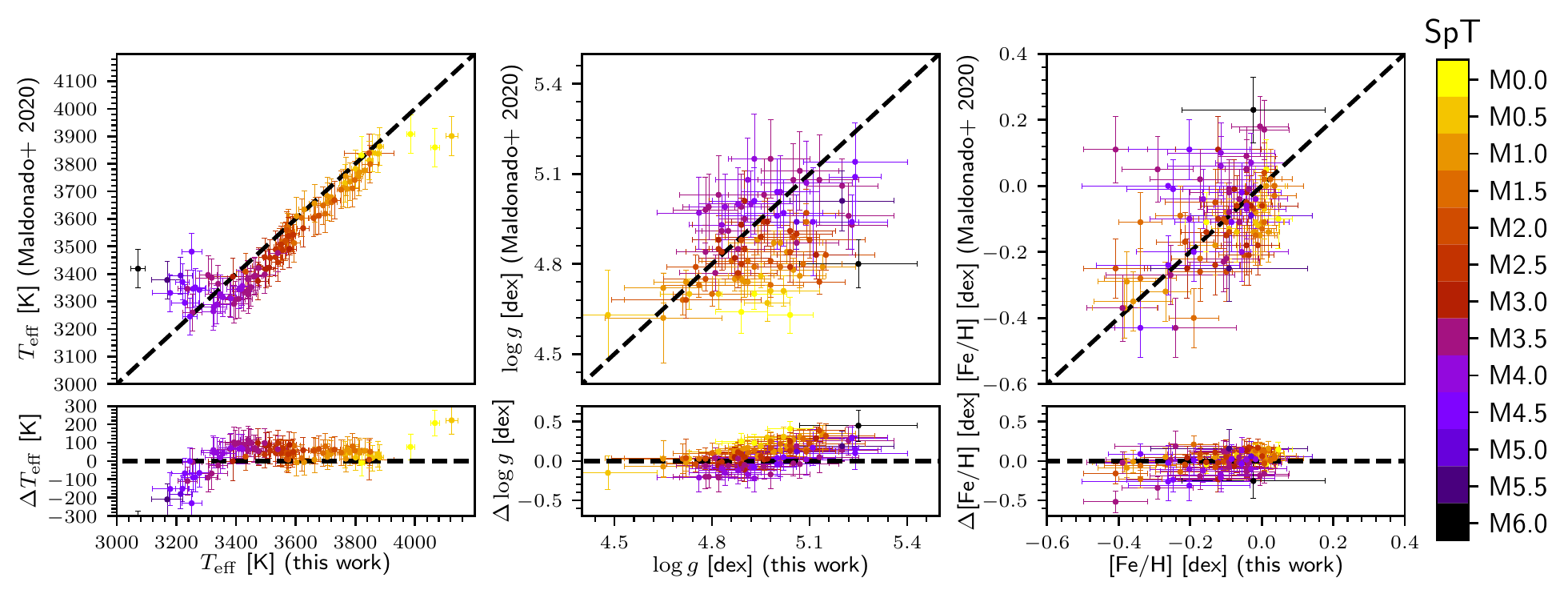}
\caption{\label{fig:compMald20} Comparison between this work and \citet{Mal20}.}
\end{figure*}

\begin{figure*}
\centering
\includegraphics[width=\textwidth]{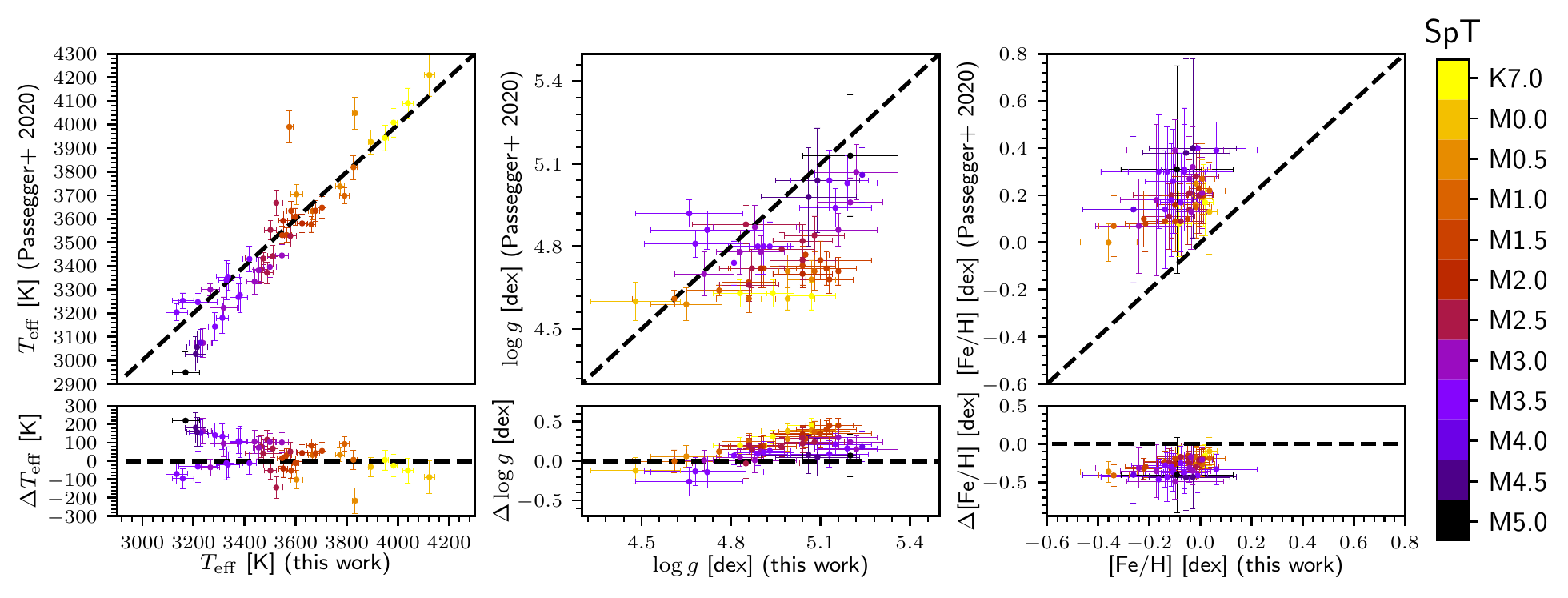}
\caption{\label{fig:compPas20} Comparison between this work and \citet{Pas20}.}
\end{figure*}

Figures~\ref{fig:compRA12}, \ref{fig:compGM14}, \ref{fig:compMald15}, \ref{fig:compMann15}, \ref{fig:compPas18}, \ref{fig:compRaj18}, \ref{fig:compSch19}, \ref{fig:compMald20}, and \ref{fig:compPas20} show the comparison between our results and those of \citet{Roj12}, \citet{Gai14}, \citet{Mal15}, \citet{Man15}, \citet{Pas18}, \citet{Raj18b}, \citet{Sch19}, \citet{Mal20}, and \citet{Pas20}, respectively, as discussed in Sect.~\ref{sec:discussion}.

\clearpage
\onecolumn

\section{Additional tables}

Table~\ref{tab:gtosample} contains the sample of 343 M dwarfs from the CARMENES GTO survey analysed in this work. Table~\ref{tab:linelist} displays the spectral ranges synthesised around the selected \ion{Fe}{i} and \ion{Ti}{i} lines according to the template spectra of GX~And, Luyten's star, and Teegarden's star, as explained in Sect.~\ref{subsec:selection}. Table~\ref{tab:par_stars_stepar} shows the adopted prior distributions in $T_{\rm eff}$ and $\log{g}$ along with the stellar atmospheric parameters of the sample computed with {\sc SteParSyn}, as discussed in Sects.~\ref{subsec:steparsyn} and \ref{sec:discussion}.

\begin{longtable}{llcclccc}

\caption{\label{tab:gtosample} Sample of M dwarfs from the CARMENES radial velocity survey analysed in this work.}\\
\hline\hline\noalign{\smallskip}
Karmn & Name & SpT\,$^{(a)}$ & $\varv_{\rm r}$      & Ref.\,$^{(b)}$ & $\varv\sin{i}$\,$^{(c)}$ & S/N & H$\alpha$ flag\,$^{(d)}$ \\
      &      &               & [km\,s$^{-1}$] &                & [km\,s$^{-1}$]            &     &                          \\
\noalign{\smallskip}\hline\noalign{\smallskip}
\endfirsthead

\caption{continued.}\\
\hline\hline\noalign{\smallskip}
Karmn & Name & SpT\,$^{(a)}$ & $\varv_{\rm r}$      & Ref.\,$^{(b)}$ & $\varv\sin{i}$\,$^{(c)}$ & S/N & H$\alpha$ flag\,$^{(d)}$ \\
      &      &               & [km\,s$^{-1}$] &                & [km\,s$^{-1}$]            &     &                          \\
\noalign{\smallskip}\hline\noalign{\smallskip}
\endhead

\noalign{\smallskip}\hline
\endfoot

\hline\noalign{\smallskip}
\multicolumn{8}{p{17.2cm}}{{\bf Notes.} $^{(a)}$Spectral types from Carmencita \citep{Alo15, Cab16a}. $^{(b)}$Radial velocities from \citet[][Laf20]{Laf20} if available, or computed using {\tt serval} \citep{Zec18} otherwise. $^{(c)}$Projected rotational velocities computed as in \citet{Rei18}. $^{(d)}$H$\alpha$-active stars following \citet{Scho19} (1 if active). $^{(e)}$ M8.0\,V or later (see Table~\ref{tab:excluded}).}\\
\endlastfoot

J04167$-$120  & LP~714-47                 & K7.0\,V &   $-7.45\pm0.05$ & This work &     $\leq2.0$  &  258 & 0 \\
J11110$+$304E & HD~97101                  & K7.0\,V &  $-16.00\pm0.03$ & This work &     $\leq2.0$  &  422 & 0 \\
J18198$-$019  & HD~168442                 & K7.0\,V &  $-14.03\pm0.03$ & Laf20     &     $\leq3.0$  & 1447 & 0 \\
J03463$+$262  & HD~23453                  & M0.0\,V &   $35.59\pm0.02$ & Laf20     &   $3.3\pm4.0$  &  728 & 0 \\
J04376$+$528  & BD$+$52~857               & M0.0\,V &   $33.91\pm0.01$ & Laf20     &   $3.4\pm1.5$  & 1167 & 0 \\
J04588$+$498  & BD$+$49~1280              & M0.0\,V &  $-34.35\pm0.02$ & Laf20     &     $\leq2.0$  &  877 & 0 \\
J05365$+$113  & V2689~Ori                 & M0.0\,V &   $21.51\pm0.02$ & Laf20     &   $3.8\pm1.5$  & 1291 & 1 \\
J06371$+$175  & HD~260655                 & M0.0\,V &  $-58.75\pm0.02$ & Laf20     &     $\leq2.0$  &  780 & 0 \\
\end{longtable}

{\tiny
\begin{longtable}{cccccccclccc}
\caption{\label{tab:linelist} Selection of \ion{Ti}{i} and \ion{Fe}{i} lines in the spectra of GX~And (M1.0\,V), Luyten's star (M3.5\,V), and Teegarden's star (M7.0\,V).}\\
\hline\hline\noalign{\smallskip}
\multicolumn{2}{c}{Range}                       & \multicolumn{5}{c}{Atomic line\,$^{(a)}$}                                                            & Magnetically & Ref.\,$^{(c)}$ & GX~And & Luyten's & Teegarden's \\
${\lambda}_{\rm min}$ [\AA] & ${\lambda}_{\rm max}$ [\AA] & ${\lambda}_{\rm line}$ [\AA] & Element & $\chi$ [eV] & $\log{gf}$ & $g_{\rm eff}$ &  sensitive\,$^{(b)}$ &        &        & star   & star      \\
\noalign{\smallskip}\hline\noalign{\smallskip}
\endfirsthead

\caption{continued.}\\
\hline\hline\noalign{\smallskip}
\multicolumn{2}{c}{Range}                       & \multicolumn{5}{c}{Atomic line\,$^{(a)}$}                                        & Magnetically  & Ref.\,$^{(c)}$ & GX~And & Luyten's & Teegarden's \\
${\lambda}_{\rm min}$ [\AA] & ${\lambda}_{\rm max}$ [\AA] & ${\lambda}_{\rm line}$ [\AA] & Element & $\chi$ [eV] & $\log{gf}$ & $g_{\rm eff}$ & sensitive\,$^{(b)}$ &        &        & star    & star      \\

\noalign{\smallskip}\hline\noalign{\smallskip}
\endhead

\hline
\endfoot

\hline
\noalign{\smallskip}
\multicolumn{12}{p{17.45cm}}{{\bf Notes.} The corresponding wavelength ranges synthesised in this work are listed in the first two columns. $\chi$, $\log{gf}$, and $g_{\rm eff}$ are the excitation potential, the oscillator strength, and the effective Land\'{e} factors of the lines, respectively. $^{(a)}$Atomic line data from the VALD3 database \citep{Rya15}. $^{(b)}$Lines particularly affected by the stellar magnetic field ({\color{blue} L\'{o}pez-Gallifa et al. in prep.)}. $^{(c)}$Lines employed in \citet[][Tab18]{Tab18}, and \citet[][Pas19]{Pas19}.}\\
\endlastfoot

  5636.61 &  5651.65 &  5644.134 & \ion{Ti}{i} & 2.2674 & $+$0.120 & 1.010 &             & This work & \textbullet &             &             \\
  5859.04 &  5874.01 &  5866.451 & \ion{Ti}{i} & 1.0666 & $-$0.790 & 1.170 &             & This work & \textbullet &             &             \\
  5914.95 &  5929.51 &  5922.110 & \ion{Ti}{i} & 1.0460 & $-$1.380 & 0.500 &             & This work & \textbullet &             &             \\
  5969.39 &  5985.75 &  5976.776 & \ion{Fe}{i} & 3.9433 & $-$1.236 & 1.100 &             & This work & \textbullet &             &             \\
          &          &  5978.541 & \ion{Ti}{i} & 1.8732 & $-$0.312 & 0.890 &             & This work & \textbullet &             &             \\
  6056.60 &  6072.73 &  6064.626 & \ion{Ti}{i} & 1.0460 & $-$1.670 & 1.990 &             & This work & \textbullet &             &             \\
          &          &  6065.481 & \ion{Fe}{i} & 2.6085 & $-$1.530 & 0.680 &             & This work & \textbullet &             &             \\
  6077.90 &  6092.38 &  6085.226 & \ion{Ti}{i} & 1.0529 & $-$1.331 & 1.750 &             & This work & \textbullet &             &             \\
\end{longtable}
}

\begin{landscape}
\begin{longtable}{lcccccccccl}

\caption{\label{tab:par_stars_stepar} Prior $T_{\rm eff}$ and $\log{g}$ distributions adopted from \citet{Cif20} and stellar atmospheric parameters of the sample computed with \textsc{SteParSyn}.}\\
\hline\hline\noalign{\smallskip}
Karmn & Prior $T_{\rm eff}$ & Prior $\log{g}$ & $T_{\rm eff}$ & $\log{g}$ & [Fe/H] & [Fe/H]$_{\rm corr}$ & $T_{\rm eff,\,fixed}$ & [Fe/H]$_{\rm fixed}$ & [Fe/H]$_{\rm corr,\,fixed}$ & Pop.\,$^{(a)}$ \\
      & [K]                 & [dex]           & [K]           & [dex]     & [dex]  & [dex]               & [K]             & [dex] & [dex] & \\
\noalign{\smallskip}\hline\noalign{\smallskip}
\endfirsthead

\caption{continued.}\\
\hline\hline\noalign{\smallskip}
Karmn & Prior $T_{\rm eff}$ & Prior $\log{g}$ & $T_{\rm eff}$ & $\log{g}$ & [Fe/H] & [Fe/H]$_{\rm corr}$ & $T_{\rm eff,\,fixed}$ & [Fe/H]$_{\rm fixed}$ & [Fe/H]$_{\rm corr,\,fixed}$ & Pop.\,$^{(a)}$ \\
      & [K]                 & [dex]           & [K]           & [dex]     & [dex]  & [dex]               & [K]            & [dex] & [dex] & \\
\noalign{\smallskip}\hline\noalign{\smallskip}
\endhead

\noalign{\smallskip}\hline
\endfoot

\hline\noalign{\smallskip}
\multicolumn{11}{p{22.5cm}}{{\bf Notes.} $^{(a)}$Galactic populations, including the thick disc (TD), the thick disc-thin disc transition (TD-D), the thin disc (D), and the young disc (YD) ({\color{blue} Cort\'{e}s-Contreras et al. in prep.}).} \\
\endlastfoot

\multicolumn{11}{c}{K7.0\,V}\\
J04167$-$120  & $3700\pm200$ & $4.62\pm0.20$ & $3961\pm13$ & $5.01\pm0.09$ &  $0.15\pm0.04$ &  $0.15\pm0.04$ & $3922\pm14$ &  $0.00\pm0.02$ &  $0.00\pm0.02$ & D    \\
J11110$+$304E & $4100\pm200$ & $4.60\pm0.20$ & $4211\pm13$ & $4.98\pm0.07$ &  $0.04\pm0.03$ &  $0.04\pm0.03$ & $4195\pm15$ & $-0.07\pm0.02$ & $-0.04\pm0.02$ & D    \\
J18198$-$019  & $4000\pm200$ & $4.64\pm0.20$ & $4155\pm14$ & $5.01\pm0.06$ & $-0.21\pm0.04$ & $-0.14\pm0.04$ & $4215\pm5$  & $-0.33\pm0.03$ & $-0.22\pm0.03$ & YD   \\
\hline\noalign{\smallskip}
\multicolumn{11}{c}{M0.0\,V}\\
J03463$+$262  & $3900\pm200$ & $4.67\pm0.20$ & $3949\pm12$ & $4.83\pm0.09$ &  $0.01\pm0.04$ &  $0.01\pm0.04$ & $3941\pm12$ & $-0.08\pm0.02$ & $-0.04\pm0.02$ & YD   \\
J04376$+$528  & $3900\pm200$ & $4.66\pm0.20$ & $4039\pm20$ & $4.94\pm0.07$ & $-0.21\pm0.05$ & $-0.14\pm0.05$ & $4005\pm9$  & $-0.42\pm0.03$ & $-0.28\pm0.03$ & D    \\
J04588$+$498  & $3900\pm200$ & $4.65\pm0.20$ & $3983\pm13$ & $5.07\pm0.08$ &  $0.02\pm0.03$ &  $0.02\pm0.03$ & $3981\pm15$ & $-0.21\pm0.03$ & $-0.14\pm0.03$ & D    \\
\end{longtable}
\end{landscape}

\end{appendix}

\end{document}